\documentclass[12pt]{iopart}
\usepackage{epsfig,cite,latexsym,axodraw,psfrag}

\newcommand{\amu}{a_\mu}
\newcommand{\amuSMTL}{a_\mu^{\rm SM,2L}}

\newcommand{\amuSUSY}{a_\mu^{\rm SUSY}}
\newcommand{\amuSU}{a_\mu^{\rm SUSY}}
\newcommand{\amuSUOL}{a_\mu^{\rm SUSY,1L}}

\newcommand{\amuSUTLa}{a_\mu^{\rm SUSY,2L(a)}}
\newcommand{\amuSUTLb}{a_\mu^{\rm SUSY,2L(b)}}

\newcommand{\amuSUTLferm}{a_\mu^{\rm SUSY,ferm,2L}}
\newcommand{\amuSUTLbos}{a_\mu^{\rm SUSY,bos,2L}}
\newcommand{\amuneu}{a_\mu^{\chi^0}}
\newcommand{\amucha}{a_\mu^{\chi^\pm}}
\newcommand{\amuchiTL}{a_\mu^{\chi,\rm 2L}}
\newcommand{\amusfTL}{a_\mu^{\tilde{f},\rm 2L}}

\newcommand{\tunit}{\times10^{-10}}

\newcommand{\Usmu}{U^{\tilde{\mu}}}
\newcommand{\MSUSY}{M_{\rm SUSY}}

\newcommand{\gsim}
{\;\raisebox{-.3em}{$\stackrel{\displaystyle >}{\sim}$}\;}

\begin{document}
\def\thefootnote{\fnsymbol{footnote}}
\def\thefootnote{\arabic{footnote}}
\setcounter{page}{0}
\setcounter{footnote}{0}
\null
\vspace{5em}
\topical{\null\\[-1ex]The Muon Magnetic Moment and Supersymmetry\\[1em]}

\author{Dominik St\"ockinger\\[0.5em]}

\address{School of Physics, \\University of Edinburgh, \\Edinburgh EH9 3JZ, UK\\[0.5em]}

\ead{dominik.stockinger@ed.ac.uk}

\vspace{3em}
\begin{abstract}
\null\\[1em]
The present review is devoted to the muon magnetic moment and its role
in supersymmetry 
phenomenology. Analytical results for 
the leading supersymmetric one- and two-loop  
contributions are provided, numerical examples are given and the
dominant $\tan\beta\,\mbox{sign}(\mu)/M_{\rm SUSY}^2$ behaviour is
qualitatively explained. The consequences of the Brookhaven measurement
are discussed. The $2\sigma$ deviation from the
Standard Model prediction implies preferred ranges for supersymmetry
parameters, in particular upper and lower mass bounds. Correlations
with other observables from collider physics and cosmology are
reviewed.
We give, wherever possible, an intuitive understanding of
each result before providing a detailed discussion.
\end{abstract}
\vspace{2em}

\newpage
\tableofcontents
\markboth{The Muon Magnetic Moment and Supersymmetry}{The Muon Magnetic Moment and Supersymmetry}

\section{Introduction}

The magnetic moment of the muon is one of the most precisely measured
and calculated quantities in elementary particle physics. It has been
measured recently at Brookhaven National Laboratory to a precision of
0.54 parts per 
million (ppm) \cite{BNL1,BNL2,BNL3,BNL4,BNL5,BNL6}. 
The Standard Model (SM) theory prediction has reached a
comparable level. Thanks to this fantastic precision the
comparison of theory and experiment is not only a sensitive test of
all SM interactions but also of possible new physics at the
electroweak scale, which is typically suppressed by a factor
$(m_\mu/M_{\rm new\ ph.})^2$ and can be expected to contribute at the ppm-level. 

Indeed, the experimental result \cite{BNL3} published in 2001 showed a 4 ppm
deviation with a statistical significance of almost $3\sigma$ from the SM
prediction. This caused a lot of enthusiasm and could be nicely
explained by a variety of new physics scenarios at the weak scale, in
particular by weak-scale supersymmetry (SUSY). In
the meantime several errors in the SM prediction have been corrected,
but due to smaller error bars on the theoretical and the experimental
side, the current 2 ppm deviation  still has a significance of
more than $2\sigma$ (if the SM prediction is based on $e^+e^-$ data,
see below). Of course, this deviation does not rule out the
SM. However, it is intriguing that SUSY, which is widely regarded as
one of the most compelling ideas for physics beyond the SM, would
naturally lead to a deviation of the observed magnitude.

In the present review we describe the rich interplay between SUSY and
the magnetic moment of the muon. In section \ref{sec:SUSYcontrib} we
present the analytical results for the SUSY contributions at the one-
and two-loop level and discuss their main features. In section
\ref{sec:Numerics} the numerical values of the SUSY contributions as
functions of the SUSY parameters are analyzed. In section
\ref{sec:Impact} we discuss the implications of the experimental result
for the possible values of SUSY parameters and the correlations
with other observables. In the remainder of this
introduction we review the status of the comparison of experiment and SM
theory and provide some necessary background on SUSY and on the theory
of magnetic moments.

\subsection{Comparison of experiment and Standard Model theory}

The magnetic moment $\vec{\mu}$ of a particle with mass $m$ and charge
$e$ is related to the particle spin $\vec{S}$ by the gyromagnetic
ratio $g$:
\begin{eqnarray}
\vec{\mu} &=& g \left(\frac{e}{2m}\right) \vec{S}.
\end{eqnarray}
At tree level, QED predicts the exact result $g=2$ for elementary
spin-$\frac12$-particles such as electron and muon. Quantum effects
from QED loop diagrams, from strong or weak interactions, or from
hypothetical new particles lead to a deviation
\begin{eqnarray}
a &=& \frac{1}{2}(g-2),
\end{eqnarray}
the so-called anomalous magnetic moment.
The theoretical prediction of the anomalous magnetic moment of a
lepton with mass $m$ is
dominated by the one-loop QED contribution, the famous Schwinger term
$\alpha/2\pi$ \cite{Schwinger}, followed by higher-order QED and
strong interaction effects. Loop contributions from heavy particles
with mass $M$ are generally suppressed by a factor $m^2/M^2$ as
explained below in section \ref{sec:gm2}. 
Therefore, the anomalous magnetic moment of the muon is a
factor $(m_\mu/m_e)^2\approx 40\,000$  more sensitive to such
contributions than the one of the electron.

The anomalous magnetic moment of the muon $\amu$ has been determined
to an unprecedented precision 
of 0.54 parts per million (ppm) at the Brookhaven $g-2$ experiment
E821 \cite{BNL1,BNL2,BNL3,BNL4,BNL5,BNL6}:
\begin{eqnarray}
\amu^{\rm exp} &=& 11\, 659\, 208.0 \,(6.3) \tunit.
\label{exp}
\end{eqnarray}
This is the first magnetic moment measurement that is sensitive to
effects from physics at the electroweak scale, and it has the potential
to constrain and discriminate between models of physics beyond the
Standard Model (SM).

Inspired by the success of this experiment, the SM theory prediction
of $\amu$ has been considerably refined and scrutinized in the last
few years (see e.g.\ \cite{MRRev,MelnRev,KnechtRev,DM,Passera} for
recent reviews and references). The SM prediction can be decomposed
into QED, hadronic and weak contributions. The QED contribution is the
largest, but it is theoretically well under control. 
The weak contributions are the smallest SM contributions and  amount
to $15.4(0.2)\tunit$, but they are relevant at the current level 
of experimental sensitivity. Their structure and numerical value are
comparable to the ones of potential new physics contributions. 

Currently, the hadronic contributions are the main source of the SM
theory uncertainty. They can be decomposed according to the hadronic
subdiagrams into vacuum polarization and light-by-light scattering
contributions. The hadronic vacuum polarization contributions can
be inferred from experimental data on the hadronic $e^+e^-$
annihilation cross section using a dispersion relation. Thus
ultimately the precision of these contributions is related to the
precision of experimental data, which is constantly increasing due to
ongoing measurements at Novosibirsk, $B$~factories and the
$\phi$~factory DA$\Phi$NE. The alternative
method of determining the hadronic vacuum polarization from data on
hadronic $\tau$ decays suffers from theory uncertainties that are
difficult to assess and is not in perfect agreement with the
$e^+e^-$-based results. The current relative accuracy of these
contributions is about $1\%$, corresponding to $7.2\tunit$ error in
$\amu$. The hadronic light-by-light contributions cannot be related to
experimental data and are notoriously difficult to evaluate. Early
evaluations had a sign error, identified in \cite{KneN1,KneN2}, which
led to a seemingly large deviation 
between the experimental result published in 2001 and the then current
SM prediction \cite{BNL3}. Current estimates vary between
$8.6(3.5)\tunit$ \cite{DEHZ} and $13.6(2.5)\tunit$ \cite{MelnV}.

For the purpose of the present review we will use the SM theory
prediction given in \cite{DM}, based on the most recent
$e^+e^-$-based evaluations of the hadronic contributions
\cite{DEHZ,HMNT} (see \cite{TrocY} for another recent evaluation,
leading to a similar result):
\begin{eqnarray}
\eqalign{
\amu^{\rm SM} &= 11 \, 659 \, 184.1\, (7.2)^{\rm Vac.Pol.}\,
(3.5)^{\rm LBL}\,(0.3)^{\rm QED/weak}\tunit\\
&=
11 \, 659 \, 184.1 \,(8.0) \tunit.
}
\end{eqnarray}
Thus there is a deviation of about 2 ppm between the experimental
result and the SM theory prediction:
\begin{eqnarray}
\label{expSM}
\Delta\amu(\mbox{exp}-\mbox{SM}) &=& 23.9\,(9.9)
\tunit.
\end{eqnarray}

In order to appreciate the result further, we put it into historical
context and 
compare it to the situation after the first measurement of $\amu$ at
CERN in the 1970's, with the result \cite{CERNfinal}
\begin{eqnarray}
\amu^{\rm exp,1978} &=&
11\, 659\, 240\,(85) \tunit.
\end{eqnarray}
This experiment was the first to be sensitive to hadronic
contributions to $\amu$, so it was interesting to compare its result
to the theory prediction without hadronic contributions and to the
hadronic contributions individually,
\begin{eqnarray}
\Delta\amu([\mbox{exp,1978}] - [\mbox{SM without had}]) &=&
720\,(85)\tunit,\\
\amu^{{\rm had,1978}} &=&667\,(81)\tunit,
\end{eqnarray}
where the central values and errors from Ref.\ \cite{CERNfinal} have
been used. Without including the hadronic contributions there
was a gap of more than $8\sigma$ between theory and experiment, but
this gap was beautifully closed by the hadronic contributions. Hence
the CERN experiment confirmed the existence of 
hadronic contributions to $\amu$ and the correctness of the SM with a
high significance.

Likewise, the new Brookhaven experiment is the first to be
sensitive to weak interaction effects, and it is instructive to
compare its result to the SM prediction without weak contributions and
the weak contributions individually,
\begin{eqnarray}
\label{gapweak}
\Delta\amu([\mbox{exp}] - [\mbox{SM without weak}]) &=& 39.3\,(9.9)
\tunit,\\
\amu^{{\rm weak}} &=& 15.4\,(0.2)\tunit
\label{SMweak}
\end{eqnarray}
Again, without including the weak contributions there is a gap of
about $4\sigma$, which establishes the existence of contributions
beyond the QED and hadronic effects. However, in this case the gap
(\ref{gapweak}) is about $2.5$ times larger than the SM weak
contribution (\ref{SMweak}).

The deviation (\ref{expSM}) or the difference between (\ref{gapweak})
and (\ref{SMweak}) has a significance of about $2.4\sigma$. This is
clearly not sufficient to prove the existence of physics beyond the
SM. The deviation could be due to e.g.\ statistical fluctuations in
the experimental result of $\amu$ itself or the experimental data
leading to the prediction of $\amu^{\rm Vac.Pol.}$, or the imperfect
understanding of $\amu^{\rm LBL}$, or a combination of these effects. 

Nevertheless, the result is tantalizing in view of new physics at the
weak scale \cite{Czarnecki:2001pv}. The existence of new physics at
the weak scale has been 
suspected long before the $\amu$ measurement of (\ref{exp}), for
reasons related 
e.g.\ to the naturalness problem of the SM Higgs sector, grand
unification, or cosmology and dark matter. Generically, new
physics at a scale $M$ can be expected to contribute at the order
$m_\mu^2/M^2$ to $\amu$, up to some numerical prefactors. For $M\sim
M_W$ such a contribution 
could easily amount to 2~ppm, and this is just the magnitude of 
the observed deviation (\ref{expSM}). In the following we focus on
supersymmetry as a particularly well-motivated and predictive idea
for physics beyond the SM.

\subsection{Relevant properties of the MSSM}

The deviation between experimental and theoretical
value of $a_\mu$ could be due 
to contributions from supersymmetry. SUSY
at the electroweak scale is one of the most compelling ideas of
physics beyond the SM (see e.g.\
\cite{Nilles,HaKa,MartinRev,Chung:2003fi,Aitchison:2005cf} for 
reviews). SUSY is the unique symmetry that relates fermions and bosons
in relativistic quantum field theories. It eliminates 
the quadratic divergences associated with the Higgs boson mass and
thus stabilizes the weak scale against quantum corrections from
ultra-high scales. SUSY at the weak scale also automatically leads to
gauge coupling unification, and the
lightest SUSY particle (LSP) can be neutral and stable and constitutes
a natural candidate for cold 
dark matter. Moreover, in contrast to many other scenarios for physics
beyond the SM, the minimal supersymmetric standard model (MSSM) is a
weakly coupled, renormalizable gauge theory \cite{HKRRSS}, such that
quantum effects are computable and well-defined, and it has survived
many non-trivial electroweak precision tests \cite{HHW}.

The MSSM is the appropriate framework for a general discussion of
$a_\mu$ and SUSY. The unknown  supersymmetry breaking mechanism is
parametrized in terms of a set of in principle arbitrary soft SUSY
breaking parameters. Specific models of supersymmetry breaking can
be accommodated within the MSSM by suitable restrictions on these
parameters.

\begin{table}
\caption{\label{tab:MSSM}The field/particle content of the MSSM. Only
  2nd generation (s)leptons and 3rd generation (s)quarks are listed
  explicitly. The mass
  eigenstates corresponding to the electroweak gauge
  and Higgs bosons and their superpartners are indicated.
}
\lineup
\begin{tabular}{@{}lcccc}
\br
& (s)leptons & (s)quarks 
&
\parbox{2.75cm}{\null\hfill Higgs\hfill\null\\\null\hfill
  Higgsinos\hfill\null}
&\parbox{2.65cm}{\null\hfill gauge bosons\hfill\null\\\null\hfill
  gauginos\hfill\null}
\\
\mr
SM/THDM & \null\ ${\displaystyle{{\nu_\mu}\choose{\mu_L}}}, \mu_R,\ldots$ \ \null
&
\null\ ${\displaystyle{t_L\choose b_L}}, t_R, b_R, \ldots$\ \null
&
${\cal H}_1, {\cal H}_2$ 
& 
$B^\mu, W^{a\mu};\quad G^{a\mu}\hspace{-1ex}\null$
\\[-2.0ex]
&&&
\multicolumn{2}{c}{
\parbox{4.6cm}{
$\underbrace{\mbox{\parbox{4.1cm}{\ }}}$\\
\null\hfill$\gamma, Z, W^\pm, G^{0,\pm},$\ \ \hfill\null\\
\null\hfill
$h^0, H^0, A^0, H^\pm$\ \ \hfill\null
}
}
\\[1.5em]
SUSY partners\hspace{-.5em}\null
&$ {\displaystyle{\tilde\nu_\mu\choose\tilde\mu_L}},
\tilde\mu_R,\ldots$
&
${\displaystyle{\tilde{t}_L\choose \tilde{b}_L}}, \tilde{t}_R, \tilde{b}_R, \ldots$
&
${\tilde H}_1, {\tilde H}_2 $
& 
$\tilde{B}, \tilde{W}^{a};\quad \tilde{g}^{a}$\\[-2.0ex]
&&&
\multicolumn{2}{c}{
\parbox{4.6cm}{
$\underbrace{\mbox{\parbox{4.1cm}{\ }}}$\\
\null\hfill$\chi^0_{1,2,3,4}$, $\chi^\pm_{1,2}$\ \ \hfill\null}
}
\\
\br
\end{tabular}
\end{table}

The MSSM as the minimal supersymmetric extension of the SM
contains all SM particles and corresponding SUSY partners, see table
\ref{tab:MSSM}. In addition
it also contains a second Higgs doublet with associated SUSY partner;
hence the MSSM can actually be 
regarded as the SUSY version of the two-Higgs-doublet model
(THDM). Two Higgs doublets ${\cal H}_{1,2}$ of opposite hypercharge
$\mp1$ are required for the cancellation of chiral gauge anomalies
caused by the corresponding Higgsinos. Thus the field content of the
MSSM comprises the THDM fields, including five physical Higgs bosons,
see (\ref{Higgsbegin})--(\ref{Higgsend}) below, scalar SUSY partners
of each chiral 
SM fermion, called sfermions $\tilde{f}_{L,R}$, Higgsino doublets 
$\tilde{H}_{1,2}$, and U(1), SU(2) and SU(3) gauginos (called bino,
winos and gluinos) $\tilde{B}$, $\tilde{W}^{\pm,3}$,
$\tilde{g}$. Right-handed (s)neutrinos as well as  
non-vanishing neutrino masses are not relevant for
this review and are ignored.

Two central MSSM parameters that are of particular importance for
$\amu$ are related to the two Higgs doublets. The first of these is
the ratio of the two vacuum expectation values,
\begin{eqnarray}
\tan\beta &=& \frac{v_2}{v_1}.
\end{eqnarray}
SUSY and gauge invariance require that the doublet ${\cal H}_1$ gives
masses to down-type fermions, while ${\cal H}_2$ gives masses to
up-type fermions. As a result, e.g.\ the top- and bottom-Yukawa
couplings in the MSSM are enhanced by factors $1/\sin\beta$ and
$1/\cos\beta$, respectively. In order to avoid non-perturbative values
of these Yukawa couplings, $\tan\beta$  is commonly restricted to the
range between about 1 and 50. High values $\tan\beta={\cal O}(50)$
lead to similar top and bottom Yukawa couplings and are therefore
favoured by the idea of top--bottom Yukawa coupling unification
\cite{Ananthanarayan:1991xp}.

The second important parameter relating the two Higgs doublets is the
$\mu$-parameter, which appears in the MSSM Lagrangian in the terms
\begin{eqnarray}
\mu\tilde{H}_1\tilde{H}_2-\mu F_{H_1}{\cal H}_2-\mu F_{H_2}{\cal H}_1 +
h.c. .
\label{mudef}
\end{eqnarray}
The first term describes a Higgsino mass term, while in the other terms
$F_{H_{1,2}}$ are auxiliary fields whose elimination gives rise
to interactions of ${\cal H}_{1,2}$ with sfermions of the opposite type
compared to the Yukawa couplings, e.g.\ to ${\cal
  H}_1^0\tilde{t}_L\tilde{t}_R^\dagger$ and ${\cal
  H}_2^0\tilde{\mu}_L\tilde{\mu}_R^\dagger$. 

In addition the MSSM contains a large number of parameters that
parametrize soft SUSY breaking. Except where explicitly stated we will
restrict the number of these parameters by neglecting generation
mixing in the sfermion sectors.
Furthermore, we restrict ourselves to the case
of $R$-parity conservation, since $R$-parity violating interactions
have not much impact on $\amu$.

The SUSY particles of particular importance to $\amu$ are the
smuons, muon-sneutrino,
gauginos and Higgsinos since they appear in the  SUSY one-loop
contributions. At higher order, also other sectors of the MSSM become
relevant, most notably the third generation squarks and the Higgs
sector. Since in the MSSM all particles of equal quantum numbers can
mix and these mixings have an important influence on the
$\amu$-prediction, we briefly discuss the mixing of the individual
sectors in the following.

The sfermions $\tilde{f}_{L,R}$ for each flavour can mix, and the
mass matrices corresponding to the 
$\tilde{f}_L, \tilde{f}_R$ basis read
\begin{eqnarray}
M_{\tilde{f}}^2 &=& \left(\begin{array}{lr}
M_{LL}^2
&
m_f X_f^*
\\
m_f X_f
&
M_{RR}^2
			\end{array}
\right),
\label{sfmassmat}
\end{eqnarray}
where  
\begin{eqnarray}
M_{LL}^2 &=&
m_f^2 + m_{L,\tilde{f}}^2 + M_Z^2\cos2\beta(I_3^f-Q^fs_W^2),\\
M_{RR}^2 &=&
m_f^2 + m_{R,\tilde{f}}^2 + M_Z^2\cos2\beta\ Q^fs_W^2,\\
X_f &=&
A_f-\mu^*\{\cot\beta,\tan\beta\}
\end{eqnarray}
with $\{\cot\beta,\tan\beta\}$ for up- and down-type sfermions,
respectively. 
$m_f$, $I^f_3$ and $Q^f$ denote the mass, weak isospin and
electric charge of the corresponding fermion;
$s_W^2\equiv\sin^2\theta_W=1-M_W^2/M_Z^2$, where $\theta_W$ 
denotes the weak mixing angle and $M_{W,Z}$ the $W$ and $Z$ boson
masses. The quantities  $A_{f}$ are soft SUSY breaking parameters for
trilinear interactions of sfermions with Higgs bosons of the form $\tilde{f}_L$--$\tilde{f}_R$--Higgs. The remaining
entries $m_{L,R}$ of the diagonal elements are 
governed by the five independent soft SUSY-breaking parameters for
each generation:
\begin{eqnarray}
m_{L,\tilde{t}} = 
m_{L,\tilde{b}} \equiv M_{Q3},\qquad&
m_{R,\tilde{b}} \equiv M_{D3},\qquad&
m_{R,\tilde{t}} \equiv M_{U3},
\\
m_{L,\tilde{\mu}} =
m_{L,\tilde{\nu}_\mu} \equiv M_{L2},\qquad&
m_{R,\tilde{\mu}} \equiv  M_{R2}.\qquad&
\end{eqnarray}
We have given these relations for 3rd generation squarks and 2nd
generation sleptons as these are most important for our
purposes. Analogous formulas hold for the other generations. The mass
matrices can be diagonalized by unitary matrices 
$U^{\tilde{f}}$ in the form
\begin{eqnarray}
U^{\tilde{f}} M_{\tilde{f}}^2 U^{\tilde{f}}{}^\dagger
 = {\rm diag}(m_{\tilde{f}_1}^2,
m_{\tilde{f}_2}^2),
\end{eqnarray}
and sfermion mass eigenstates can be defined by
\begin{eqnarray}
\left(\begin{array}{c}\tilde{f}_1\\\tilde{f}_2\end{array}\right)
&=& U^{\tilde{f}}
\left(\begin{array}{c}\tilde{f}_L\\\tilde{f}_R\end{array}\right)
.
\end{eqnarray}
The mass of the sneutrino $\tilde{\nu}_\mu$ is given by
\begin{eqnarray}
m_{\tilde{\nu}_\mu}^2 &=
m_{L,\tilde{\mu}}^2 + \frac12 M_Z^2\cos2\beta
\end{eqnarray}
and similar for the other generations.

The superpartners of the charged gauge and Higgs bosons also mix, and
the mass and mixing terms can be easiest expressed in terms of the
Weyl spinor combinations
$\psi^- = {
(\tilde{W}^-,\tilde{H}_1^-})
$, 
$\psi^+ = 
(\tilde{W}^+, \tilde{H}_2^+).
$
The mass term for these fields is given by $\psi^- X \psi^++h.c.$ with
the mass matrix
\begin{eqnarray}
X &= 
\left(\begin{array}{cc}
M_2 & M_W\sqrt2 \sin\beta\\
 M_W\sqrt2 \cos\beta & \mu
      \end{array}
\right)
,
\end{eqnarray}
where $M_2$ is the soft SUSY breaking parameter corresponding to the
SU(2) gaugino mass.
This matrix can be diagonalized using two unitary matrices $U,V$ in
the form
\begin{eqnarray}
U^* X V^{-1} &= {\rm diag}(m_{\chi^\pm_1},m_{\chi^\pm_2}).
\end{eqnarray}
Similarly, the mass matrix $Y$ corresponding to the superpartners of
the neutral gauge and Higgs bosons in the basis 
$\psi^0=(\tilde{B},\tilde{W}^3,\tilde{H}_1^0,\tilde{H}_2^0)$
is given by
\begin{eqnarray}
\fl \qquad Y =
\left(\begin{array}{cccc}
M_1 & 0   & -M_Z s_W\cos\beta & M_Z s_W\sin\beta \\
0   & M_2 & M_Z c_W \cos\beta & -M_Z c_W\sin\beta \\
-M_Z s_W\cos\beta &M_Z c_W \cos\beta & 0 & -\mu \\
M_Z s_W\sin\beta & -M_Z c_W\sin\beta& -\mu & 0
      \end{array}
\right),
\end{eqnarray}
where $M_1$ is a U(1) gaugino (bino) soft SUSY breaking parameter and
$c_W=M_W/M_Z$. It can be diagonalized with the help of one unitary
matrix $N$ in the form
\begin{eqnarray}
N^* Y N^{-1} &= {\rm diag}(m_{\chi^0_1},\ldots, m_{\chi^0_4}).
\end{eqnarray}
The mass eigenstates corresponding to the charged and neutral gauginos
and Higgsinos are called charginos and neutralinos, and
they are related to the interaction eigenstates by
\begin{eqnarray}
\chi^+_i&=V_{ij}\psi^+_j,\\
\chi^-_i&=U_{ij}\psi^-_j,\\
\chi^0_i&=N_{ij}\psi^0_j.
\end{eqnarray}
The gluinos do not mix, and their tree-level mass is given by $|M_3|$
in terms of the SU(3) gaugino mass parameter $M_3$.

The SUSY parameters $\mu$, $A_f$, $M_{1,2,3}$ can be complex. However,
not all complex phases can appear in observables. The only physical
phases of the MSSM (beyond the phase in the CKM matrix) are the ones
of the combinations
\begin{eqnarray}
\mu A_f,\quad
\mu M_{1,2,3}.
\end{eqnarray}
Hence the frequently adopted convention that $M_2$ is real and
positive constitutes no restriction. Below, we will adopt this
convention only in section \ref{sec:Impact} and remain general
elsewhere.

After spontaneous symmetry breaking the two MSSM Higgs doublets lead
to 5 physical Higgs bosons and 3 unphysical Goldstone
bosons. Parametrizing the two doublets in the form
\begin{eqnarray}
{\cal H}_1 &=
\left(\begin{array}{c}
v_1+\frac{1}{\sqrt2}(\phi_1^0-i\chi_1^0)\\
-\phi_1^-
      \end{array}\right)
,\qquad
{\cal H}_2 &=
\left(\begin{array}{c}
\phi_2^+\\
v_2+\frac{1}{\sqrt2}(\phi_2^0+i\chi_2^0)
      \end{array}\right),
\end{eqnarray}
the mass eigenstates are given by
\begin{eqnarray}
\label{Higgsbegin}
\left(\begin{array}{c}
H^0\\ h^0
      \end{array}\right)
&=
\left(\begin{array}{cc}
\cos\alpha & \sin\alpha \\ -\sin\alpha & \cos\alpha
      \end{array}\right)
\left(\begin{array}{c}
\phi_1^0\\ \phi_2^0
      \end{array}\right)
,\\
\left(\begin{array}{c}
G^0\\ A^0
      \end{array}\right)
&=
\left(\begin{array}{cc}
\cos\beta & \sin\beta \\ -\sin\beta & \cos\beta
      \end{array}\right)
\left(\begin{array}{c}
\chi_1^0\\ \chi_2^0
      \end{array}\right)
,\\
\left(\begin{array}{c}
G^\pm\\ H^\pm
      \end{array}\right)
&=
\left(\begin{array}{cc}
\cos\beta & \sin\beta \\ -\sin\beta & \cos\beta
      \end{array}\right)
\left(\begin{array}{c}
\phi_1^\pm\\ \phi_2^\pm
      \end{array}\right),
\label{Higgsend}
\end{eqnarray}
where the mixing angle $\alpha$ is related to $\beta$ and the mass
$M_A$ of the CP-odd scalar $A^0$ by
\begin{eqnarray}
\tan2\alpha &= \tan2\beta\ \frac{M_A^2+M_Z^2}{M_A^2-M_Z^2},\qquad
-\frac{\pi}{2}<\alpha<0.
\end{eqnarray}
The physical Higgs degrees of freedom are the light and heavy CP-even
scalars $h^0$, $H^0$, the CP-odd scalar $A^0$ and the charged Higgs
bosons $H^\pm$. For $M_A\gg M_Z$ the masses of $H^0$ and $H^\pm$ are
of the order $M_A$. The three unphysical Goldstone bosons $G^{0,\pm}$
are eaten to give masses to the $Z$ and $W^\pm$ bosons.

For the discussion of two-loop contributions to $\amu$ with
Higgs exchange it is important to note that the
muon receives its mass from the doublet 
${\cal H}_1$. In the case that $M_A\gg M_Z$ and $\tan\beta$
is large the heavy Higgs bosons $H^0$, $A^0$ and $H^\pm$ are
predominantly composed of ${\cal H}_1$-components. In this case they
have larger couplings to the muon than the light Higgs boson $h^0$.

\subsection{Magnetic moment of the muon and chirality flips}
\label{sec:gm2}

The magnetic moment is a property of the muon in presence of an
electromagnetic field. In quantum field theory it is related to the
muon--photon vertex function $\Gamma_{\mu\bar\mu A^\rho}(p,-p',q)$,
which has the covariant decomposition
\begin{eqnarray}
\label{covdecomp}
\fl\qquad
\bar u(p')\Gamma_{\mu\bar\mu A^\rho}(p,-p',q) u(p) & = 
e\,
\bar u(p')\left[\gamma_\rho F_V(q^2) + (p+p')_\rho F_M(q^2) +
  \ldots\right] u(p)
\end{eqnarray}
for on-shell momenta $p$, $p'$ and spinors $u$, $\bar u$ that satisfy
the Dirac equation. The charge renormalization condition implies
$F_V(0)+2m_\mu F_M(0)=1$, and rearranging (\ref{covdecomp}) leads to a
term
\begin{eqnarray}
\frac{i\, e}{2 m_\mu}[1-2m_\mu F_M(0)]\sigma_{\rho\nu}q^\nu
\end{eqnarray}
for $q^2\to0$ in the covariant decomposition. This term describes the
interaction of the muon dipole moment with a magnetic field, and the
corresponding gyromagnetic ratio is given by $g=2[1-2m_\mu F_M(0)]$,
or equivalently
\begin{eqnarray}
\amu &= -2m_\mu F_M(0).
\label{amuFM}
\end{eqnarray}
In practical calculations it is often useful to extract the form
factor $F_M$ before evaluating the full vertex function with the help
of projection operators \cite{RoskiesProjOp,CKProjOp}.

\begin{figure}
\begin{center}
\begin{tabular}{ll}
\\
\setlength{\unitlength}{1pt}
\begin{picture}(60,30)
\Vertex(30,5){2}\Line(0,5)(60,5)
\put(-3,17){$\mu_L$}\put(45,17){$\mu_R$}
\end{picture}
\ $\propto\ {m_\mu}$\ 
&
\setlength{\unitlength}{1pt}
$\mu_L,\nu_\mu$
\begin{picture}(60,30)
\Line(0,5)(60,5)
{\Vertex(30,5){2}}\DashLine(30,5)(30,30){4}
\put(30,30){${\cal H}_1$}
\end{picture}
$\mu_R$
\ $\propto\  {m_\mu\,\tan\beta}$\ 
\\
\\
\setlength{\unitlength}{1pt}
\begin{picture}(60,30)
\Vertex(30,5){2}\DashLine(0,5)(60,5){4}
\put(-3,17){$\tilde\mu_L$}\put(45,17){$\tilde{\mu}_R$}
\end{picture}
\ $\propto\ {m_\mu\,\tan\beta}\,{\mu}$\ \qquad
&
\setlength{\unitlength}{1pt}
$\mu_L,\nu_\mu$
\begin{picture}(60,30)
\Line(0,5)(30,5)\DashLine(30,5)(60,5){4}
{\Vertex(30,5){2}}\Line(30,5)(30,30)
\put(30,30){$\tilde{H}_1$}
\end{picture}
$\tilde\mu_R$
\ $\propto\ {m_\mu\,\tan\beta}$\ 
\end{tabular}
\end{center}
\caption{\label{fig:chiralityflips}
Possibilities for chirality-flips along the line carrying the
$\mu$-lepton number.}
\end{figure}
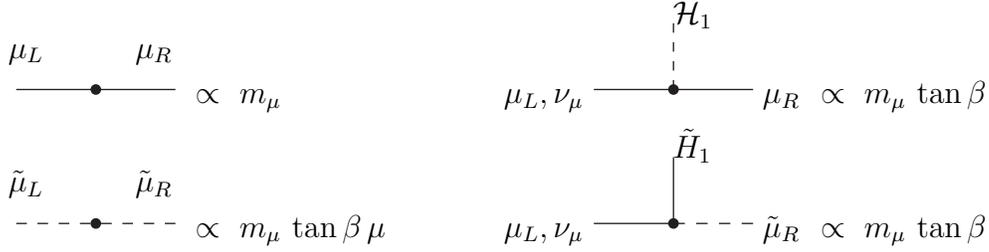

It is important to understand where the generic behaviour
$\amuSUSY\propto m_\mu^2/\MSUSY^2$ comes from and how it can be
modified by additional factors.\footnote{
  Logarithmic factors like $\log M_{\rm SUSY}/m_\mu$, which can appear
  at the two-loop level, are disregarded here.}
The $1/M_{\rm SUSY}^2$-behaviour for SUSY masses of order
$M_{\rm SUSY}$ reflects the decoupling properties of SUSY. The
$m_\mu^2$-behaviour of $\amu$, or equivalently the relation
$F_M\propto m_\mu$ is related to chiral symmetry, and crucial
modifications by additional factors are possible.
The form factor $F_M$ in (\ref{covdecomp}) corresponds to a
chirality-flipping interaction between the left- and right-handed
muon. If the MSSM were invariant under the discrete chiral
transformation
\begin{eqnarray}
\eqalign{
\left.
\begin{array}{ll}
{\displaystyle{\nu_\mu\choose \mu_L},} &
{\displaystyle{\tilde{\nu}_\mu\choose \tilde{\mu}_L}}
\\[2ex]
\ \ \mu_R, & \ \ \tilde{\mu}_R 
\end{array}
\right\}
&\to
\left\{
\begin{array}{ll}
+{\displaystyle{\nu_\mu\choose \mu_L},} &
+{\displaystyle{\tilde{\nu}_\mu\choose \tilde{\mu}_L} ,}\\[2ex]
-\ \ \mu_R, &
-\ \ \tilde{\mu}_R
\end{array}
\right.
}
\label{ChiralTransf}
\end{eqnarray}
of the left-handed doublets and right-handed singlets, $F_M$ and thus
$\amu$ would vanish in the MSSM. $F_M$ is proportional to $m_\mu$
because the invariance of the MSSM under (\ref{ChiralTransf}) is broken
by the muon mass, or more precisely by all terms in the MSSM
Lagrangian that are proportional to the muon Yukawa coupling. In each
Feynman diagram that contributes to $\amu$, the $\mu$-chirality has to
be flipped by one of these terms. The main possibilities for the
chirality flip are illustrated in figure \ref{fig:chiralityflips} and
are the following:
\begin{itemize}
\item at a $\mu$-line through a muon mass term,
  contributing a factor $m_\mu$,
\item at a Yukawa coupling of ${\cal H}_1$ to $\mu_R$
  and $\mu_L$ or $\nu_\mu$, contributing a factor
  $y_\mu$,
\item at a $\tilde{\mu}$-line, corresponding to the transition
  $\tilde{\mu}_L$--$\tilde{\mu}_R$, contributing a factor $m_\mu
  X_\mu\approx m_\mu\,\tan\beta\,\mu$ for large $\tan\beta$ from the
  smuon mass matrix,
\item at a SUSY Yukawa coupling of a Higgsino to $\mu$ and
  $\tilde{\mu}$ or $\tilde{\nu}_\mu$, contributing a factor $y_\mu$.
\end{itemize}
The muon Yukawa coupling $y_\mu$ is given by
\begin{eqnarray}
y_\mu=\frac{m_\mu}{v_1}=\frac{m_\mu g_2}{\sqrt2
M_W\cos\beta},
\label{ymu}
\end{eqnarray}
where $g_2=e/s_W$, and is thus enhanced by the factor
$1/\cos\beta\approx\tan\beta$ for 
large $\tan\beta$ compared to its SM value. Hence, while all four
possibilities are proportional to the muon mass $m_\mu$, the last
three are enhanced by a factor $\tan\beta$ for large $\tan\beta$.
SUSY contributions to $\amu$ that make use of these enhanced chirality
flips are themselves enhanced compared to the generic estimate
$m_\mu^2/M_{\rm SUSY}^2$.

\section{Analytic results}
\label{sec:SUSYcontrib}

In a theory with unbroken SUSY, the gyromagnetic ratio of a charged
fermion is exactly 2 and thus $\amu=0$ \cite{FeRe,FePo},  that is the
SUSY contribution exactly cancels the SM contribution. In
the MSSM, SUSY is softly broken, and SUSY contributions to $\amu$ were
studied already in the early 1980's
\cite{Fayet1L,Grifols1L,Ellis1L,Barbieri1L,Kosower1L,Yuan1L,Romao1L},
when it was still conceivable that SUSY particles were significantly
lighter than the $W$ and $Z$ bosons. Although very
light SUSY particles are now experimentally excluded and SUSY
contributions to $\amu$ are suppressed as $1/\MSUSY^2$, it is still
possible that these contributions are significant.  In
the following we first discuss the generic size of the SUSY
one- and two-loop contributions on a more intuitive level;
then we present the analytic results for the most important
contributions. 

\subsection{How large can the SUSY contributions be?}
\label{sec:massinsertions}

Before presenting the full SUSY one-loop contributions to $\amu$
\cite{Fayet1L,Grifols1L,Ellis1L,Barbieri1L,Kosower1L,Yuan1L,Romao1L,Vendramin1L,Abel1L,Lopez1L,Chatt1L,Moroi1L},
it is instructive to discuss the dominant parameter 
dependence on an intuitive level and to obtain useful
estimates. As shown in section \ref{sec:gm2}
the contributions from SUSY particles of a generic mass $\MSUSY$ are
of the order $m_\mu^2/\MSUSY^2$, and hence suppressed by a factor
$M_W^2/\MSUSY^2$ compared to the SM electroweak contributions.

However, it has been observed in \cite{Kosower1L} and further stressed
and discussed in \cite{Lopez1L,Chatt1L,Moroi1L} that the SUSY
contributions can be significantly enhanced if $\tan\beta$ is
large. Moreover, for large $\tan\beta$ the sign of the one-loop
contributions is mainly determined by the sign of the $\mu$-parameter
introduced in  (\ref{mudef}).
We will see here that not only the one-loop but also the leading
two-loop contributions behave in this way.
\begin{figure}[tb]
\null\vspace{1ex}
\setlength{\unitlength}{1pt}
\begin{picture}(190,65)(-65,-12)
\Text(-65,55)[l]{(C)}
\Vertex(25,5){2}
\Vertex(100,5){2}
\Vertex(62.5,42.5){2}
\Vertex(36.06,31.43){2}
\Vertex(88.94,31.43){2}
\CArc(62.5,5)(37.5,0,180)
\Line(0,5)(25,5)
\Line(100,5)(125,5)
\DashLine(25,5)(100,5){4}
\Text(0,-6)[l]{$\mu_R$}\Text(125,-6)[r]{${\mu}_L$}
\Text(62.5,-6)[c]{$\tilde{\nu}_\mu$}
\Text(26,20)[r]{$\tilde{H}_1^+$}\Text(99,20)[l]{$\tilde{W}^+$}
\Text(44,42)[b]{$\tilde{H}_2^+$}\Text(81,42)[b]{$\tilde{W}^+$}
\end{picture}
\begin{picture}(190,65)(-15,-12)
\Text(0,5)[l]{$\quad\propto\quad m_\mu^2\, \tan\beta\ \mu\, M_2\,
  F(\mu,M_2,m_{\tilde{\mu}_L})$}
\end{picture}
\\[3em]
\begin{picture}(190,65)(-65,-12)
\Text(-65,55)[tl]{(N1)}
\Vertex(25,5){2}
\Vertex(100,5){2}
\Vertex(62.5,5){2}
\Vertex(62.5,42.5){2}
\CArc(62.5,5)(37.5,0,180)
\Line(0,5)(25,5)
\Line(100,5)(125,5)
\DashLine(25,5)(100,5){4}
\Text(0,-6)[l]{$\mu_R$}\Text(125,-6)[r]{${\mu}_L$}
\Text(44,-6)[c]{$\tilde{\mu}_R$}\Text(81,-6)[c]{$\tilde{\mu}_L$}
\Text(44,42)[b]{$\tilde{B}$}
\Text(81,42)[b]{$\tilde{B}$}
\end{picture}
\begin{picture}(190,65)(-15,-12)
\Text(0,5)[l]{$\quad\propto\quad m_\mu^2\, \tan\beta\ \mu\, M_1\,
  F(M_1,m_{\tilde{\mu}_{L,R}})$}
\end{picture}
\\[3em]
\begin{picture}(190,65)(-65,-12)
\Text(-65,55)[tl]{(N2)}
\Vertex(25,5){2}
\Vertex(100,5){2}
\Vertex(62.5,42.5){2}
\Vertex(36.06,31.43){2}
\Vertex(88.94,31.43){2}
\CArc(62.5,5)(37.5,0,180)
\Line(0,5)(25,5)
\Line(100,5)(125,5)
\DashLine(25,5)(100,5){4}
\Text(0,-6)[l]{$\mu_R$}\Text(125,-6)[r]{${\mu}_L$}
\Text(62.5,-6)[c]{$\tilde{\mu}_R$}
\Text(26,20)[r]{$\tilde{B}$}\Text(99,20)[l]{$\tilde{H}_1^0$}
\Text(44,42)[b]{$\tilde{B}$}\Text(81,42)[b]{$\tilde{H}_2^0$}
\end{picture}
\begin{picture}(190,65)(-15,-12)
\Text(0,5)[l]{$\quad\propto\quad m_\mu^2\, \tan\beta\ \mu\, M_1\,
  F(\mu,M_1,m_{\tilde{\mu}_R})$}
\end{picture}
\\[3em]
\begin{picture}(190,100)(-65,-12)
\Text(-65,90)[l]{(C$_{\rm 2L}$)}
\Vertex(25,5){2}
\Vertex(100,5){2}
\Vertex(45,40){2}
\Vertex(80,40){2}
\Vertex(40,72.3){2}
\Vertex(85,72.3){2}
\Vertex(62.5,85.67){2}
\DashLine(25,5)(45,40){4}
\Photon(100,5)(80,40){4}{4}
\CArc(62.5,60)(25.67,0,360)
\Line(0,5)(25,5)
\Line(100,5)(125,5)
\Line(25,5)(100,5)
\Text(0,-6)[l]{$\mu_R$}\Text(125,-6)[r]{${\mu}_L$}
\Text(62.5,-6)[c]{${\mu}_L$}
\Text(28,25)[r]{${\cal H}_1^0$}\Text(97,25)[l]{$\gamma$}
\Text(46,88)[r]{$\tilde{H}_2^+$}\Text(79,88)[l]{$\tilde{W}^+$}
\Text(35,55)[r]{$\tilde{H}_1^+$}\Text(90,55)[l]{$\tilde{W}^+$}
\end{picture}
\begin{picture}(190,100)(-15,-12)
\Text(0,5)[l]{$\quad\propto\quad m_\mu^2\, \tan\beta\ \mu\, M_2\,
  F(\mu,M_2,M_{{\cal H}_1})$}
\end{picture}
\\[3em]
\begin{picture}(190,100)(-65,-12)
\Text(-65,90)[l]{($\tilde{t}_{\rm 2L}$)}
\Vertex(25,5){2}
\Vertex(100,5){2}
\Vertex(45,40){2}
\Vertex(80,40){2}
\Vertex(62.5,85.67){2}
\DashLine(25,5)(45,40){4}
\Photon(100,5)(80,40){4}{4}
\DashCArc(62.5,60)(25.67,0,360){4}
\Line(0,5)(25,5)
\Line(100,5)(125,5)
\Line(25,5)(100,5)
\Text(0,-6)[l]{$\mu_R$}\Text(125,-6)[r]{${\mu}_L$}
\Text(62.5,-6)[c]{${\mu}_L$}
\Text(28,25)[r]{${\cal H}_1^0$}\Text(97,25)[l]{$\gamma$}
\Text(40,85)[r]{$\tilde{t}_L$}\Text(85,85)[l]{$\tilde{t}_R$}
\end{picture}
\begin{picture}(190,100)(-15,-12)
\Text(0,5)[l]{$\quad\propto \quad
m_\mu^2\, \tan\beta\ \frac{\mu\, m_t}{M_W^2}\ (m_t X_t)\, 
F(m_{\tilde{t}_{L,R}},M_{{\cal H}_1^0})
$}
\end{picture}
\caption{Five sample mass-insertion diagrams. Vertices and
  mass insertions are denoted by dots, and the interaction eigenstates
  corresponding to each line are displayed explicitly. The external
  photon has to be attached in all possible ways to the charged
  internal lines. The one-loop diagrams (C), (N1), (N2) have been
  discussed also in \cite{Moroi1L}. The loop functions $F$ in the
  results are different in the five cases and depend on different
  masses.
  }
\label{fig:massinsertiondiagrams}
\end{figure}

It is easiest to understand the leading behaviour with the help of
diagrams that are written in terms of interaction eigenstates, where
the  insertions of mass and mixing terms and chirality flips are
explicitly shown \cite{Moroi1L}. The five diagrams in figure
\ref{fig:massinsertiondiagrams} exemplify the
main enhancement mechanisms. The basic reason for the
$\tan\beta$-enhancement is the fact that the muon Yukawa coupling in
the MSSM is larger by a factor $1/\cos\beta\approx \tan\beta$ for
large $\tan\beta$ than its SM counterpart. This Yukawa coupling enters
the diagrams in figure \ref{fig:massinsertiondiagrams} in the vertices
where the muon chirality is flipped, i.e.\ in the couplings
of the muon to the Higgsino or Higgs boson in cases
(C,N2,C$_{\rm 2L}$,$\tilde{t}_{\rm 2L}$), and in the
$\tilde{\mu}_L$--$\tilde{\mu}_R$ transition, given by 
$(M_{\tilde{\mu}}^2)_{12}$, in case (N1).

The second important parameter entering all five diagrams is the
$\mu$-parameter, which governs the mixing between the two Higgs
doublets. In all cases, the enhancement due to this mixing can be
traced back to the fact that ${\cal H}_2$ has the larger vacuum
expectation value and strongly couples to top quarks, while only
${\cal H}_1$ couples to muons. 

In diagrams (C,N2,C$_{\rm 2L}$) the $\mu$-parameter enters via the
Higgsino $\tilde{H}_1$--$\tilde{H}_2$ transitions. These transitions
enhance the diagrams because the following $\tilde{H}_2$--gaugino
transitions are by a factor $v_2:v_1=\tan\beta$ larger than
$\tilde{H}_1$--gaugino transitions. In diagram (N1) $\mu$ enters
via the dominant part of the smuon mixing. This mass insertion is obtained from
the $F$-term $F_{H_1}{\cal H}_2$, see (\ref{mudef}), by replacing
${\cal H}_2$ by its large vacuum expectation value. Finally, in
diagram ($\tilde{t}_{\rm 2L}$) the dominant part of the Higgs--stop
coupling originates from $F_{H_2}{\cal H}_1$ and thus enables ${\cal
  H}_1$ to couple with the top-Yukawa coupling.

The remaining mass insertions in the diagrams provide additional
factors of the gaugino mass  $M_{1,2}$ and stop mixing parameter
$X_t$. They are necessary in order
to obtain an even number of $\gamma$-matrices in the fermion line and
in order to connect $\tilde{t}_L$ and $\tilde{t}_R$,
respectively. As an illustration, the relevant
factors of diagram (C) are given by
\begin{eqnarray}
y_\mu\,X_{22}\,X_{12}\,X_{22} & = 
\frac{m_\mu}{v_1}\,\mu\,(g_2v_2)\,M_2 =
g_2\, m_\mu \, \tan\beta\,\mu\,M_2.
\end{eqnarray}
Combining the enhancement factors of all diagrams leads to the
estimates given in figure \ref{fig:massinsertiondiagrams}. They all
have a similar form,
\begin{eqnarray}
\amu^{\rm (C,C_{\rm 2L};N1,N2)} &\propto
m_\mu^2\, \tan\beta\ \mu\, M_{2;1}\, F
,\\
\amu^{(\tilde{t}_{\rm 2L})} &\propto
m_\mu^2\, \tan\beta\ \frac{\mu\, m_t}{M_W^2}\ (m_t X_t)\, F
,
\end{eqnarray}
where $M_{2;1}$ corresponds to ${\rm (C,C_{\rm 2L})}$ and
${\rm(N1,N2)}$, respectively. The loop functions $F$ are different in
the five cases, depend on the masses appearing in the respective
diagrams and generally  behave as $F\propto M_{\rm SUSY}^{-4}$ for
large SUSY masses. In these formulas one power of $m_\mu$ is due to
the $\amu$--$F_M$ relation (\ref{amuFM}) and gauge couplings have been
suppressed.

Therefore all leading one- and two-loop contributions are
approximately linear in $\tan\beta$, and their sign is given by the
sign of $\mu$, together with the sign of $M_{1,2}$ or
$X_t$. Generally, all diagrams  
are suppressed by two powers of the SUSY mass scale. Hence the basic
behaviour of diagrams (C,C$_{\rm 2L}$,N1,N2) is given by
\begin{eqnarray}
\amu^{\rm (C,C_{\rm 2L};N1,N2)} &\propto
\frac{m_\mu^2}{\MSUSY^2}\ \tan\beta\ \mbox{sign}(\mu M_{2;1})
\label{Diagabcd}
\end{eqnarray}
if all SUSY masses are set equal to a
common scale $\MSUSY$.
However, it is important to keep in mind that the relevant SUSY masses
are different in the five diagrams. 

In particular, diagrams (N1) and ($\tilde{t}_{\rm 2L}$) are special
because they increase linearly with $\mu$,
while the all other diagrams are suppressed for large $\mu$ by their
$\mu$-dependent loop functions $F$. Likewise, only the one-loop
diagrams are sensitive to 
the smuon and sneutrino masses. If these are large, the one-loop
diagrams can be suppressed and the two-loop diagrams can become
dominant. The chargino-loop diagram (C$_{\rm 2L}$) can be large if the
chargino and Higgs masses are small. 

The discussion of the stop-loop diagram
($\tilde{t}_{\rm 2L}$) is complicated by the fact that the seemingly
linear dependence on 
$(m_t X_t)$ is cut off by the requirement that both stop mass
eigenvalues are positive.
This diagram is largest for maximal stop mixing, i.e.\ if $(m_t
X_t)$ is large but both eigenvalues are positive and  $m_{\tilde{t}_1}\ll
m_{\tilde{t}_2}$, and if $m_{\tilde{t}_1}$ and the Higgs boson mass
are small. In this case, diagram
($\tilde{t}_{\rm 2L}$) has the 
behaviour 
\begin{eqnarray}
\amu^{(\tilde{t}_{\rm 2L})} &\propto
\frac{m_\mu^2}{\MSUSY^2}\ 
\frac{\mu\, m_t}{M_W^2}\ \tan\beta\ \mbox{sign}(X_t),
\label{Diage}
\end{eqnarray}
where $\MSUSY$ denotes here the common mass scale of the appearing
Higgs boson and the lightest stop. Thus the diagram is linearly
enhanced by large $\mu$,
and its sign is determined by sign($X_t$).
In the following subsections we
will provide the 
exact analytical formulas for all these diagrams and also derive the
numerical prefactors in the proportionalities (\ref{Diagabcd}) and
(\ref{Diage}).

\subsection{One-loop contributions}

Each diagram that contributes to $\amu$ contains one line carrying the
$\mu$-lepton number. This fact allows to divide the MSSM one-loop
diagrams into two classes:
\begin{enumerate}
\item SM-like diagrams, where the $\mu$-lepton number is carried only
  by $\mu$ and/or $\nu_\mu$.
\item SUSY diagrams, where the $\mu$-lepton number is carried also by
  $\tilde{\mu}$ and/or $\tilde{\nu}_\mu$.
\end{enumerate}
The diagrams of the first class involve only SM-particles, and they
are essentially identical in the SM and the MSSM. The only
non-identical diagrams involve two couplings of physical SM or MSSM
Higgs bosons to the muon line. Owing to the additional suppression
factor $m_\mu^2/M_W^2$ such diagrams are entirely negligible 
both in the SM and the MSSM.

\begin{figure}[tb]
\null\hfill
\setlength{\unitlength}{1pt}
\begin{picture}(125,80)(0,-12)
\Vertex(25,5){2}
\Vertex(100,5){2}
\CArc(62.5,5)(37.5,0,180)
\Line(0,5)(25,5)
\Line(100,5)(125,5)
\DashLine(25,5)(100,5){4}
\Text(0,-6)[l]{$\mu$}\Text(125,-6)[r]{${\mu}$}
\Text(62.5,-6)[c]{$\tilde{\nu}_\mu$}
\Text(62.5,45)[b]{$\chi_k^+$}
\end{picture}
\hfill
\begin{picture}(125,80)(0,-12)
\Vertex(25,5){2}
\Vertex(100,5){2}
\CArc(62.5,5)(37.5,0,180)
\Line(0,5)(25,5)
\Line(100,5)(125,5)
\DashLine(25,5)(100,5){4}
\Text(0,-6)[l]{$\mu$}\Text(125,-6)[r]{${\mu}$}
\Text(62.5,-6)[c]{$\tilde{\mu}_m$}
\Text(62.5,45)[b]{$\chi^0_i$}
\end{picture}
\hfill\null
\caption{\label{fig:oneloop}
The two SUSY one-loop diagrams, written in terms of mass
eigenstates. The external photon line has to be attached to the
charged internal lines.
  }
\end{figure}
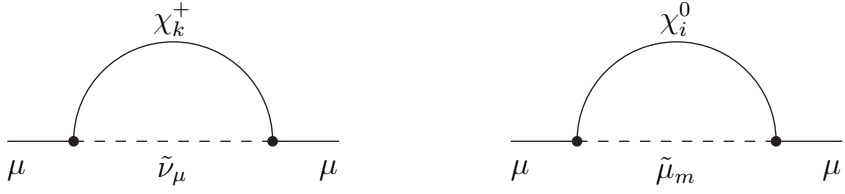

Therefore the SUSY one-loop contribution, i.e.\ the difference
between $\amu$ in the MSSM and the SM, is given entirely by the
diagrams of the second class. They are displayed in figure
\ref{fig:oneloop} and 
involve either a chargino--sneutrino or a neutralino--smuon loop. In
contrast to the diagrams in figure \ref{fig:massinsertiondiagrams}
they are written in terms of interaction eigenstates, which is more
appropriate for an exact evaluation. The
diagrams have been evaluated in 
Refs.\ \cite{Fayet1L,Ellis1L,Grifols1L,Barbieri1L} with various
restrictions on the masses and mixings. These restrictions have been
dropped in Refs.\ \cite{Kosower1L,Yuan1L,Romao1L,Vendramin1L}, and exact
results have been derived. Later, more comprehensive and general
evaluations of these diagrams have been presented in the context of
particular supersymmetric models \cite{Abel1L,Lopez1L,Chatt1L} and 
the unconstrained MSSM \cite{Moroi1L} (see also \cite{HIRS1,HIRS2} for
related results on weak dipole moments in the MSSM). We present the
general result in the form given in \cite{MartinWells}:
\begin{eqnarray}
\label{amuSUOL}
\amuSUOL &=\amuneu+\amucha,
\end{eqnarray}
with
\begin{eqnarray}
\fl \amuneu&=\frac{m_\mu}{16\pi^2}\sum_{i,m}
\Big\{
-\frac{m_\mu}{12m_{\tilde{\mu}_m}^2}(|n_{im}^L|^2+|n_{im}^R|^2)
F_1^N(x_{im})
+\frac{m_{\chi_i^0}}{3m_{\tilde{\mu}_m}^2}
{\rm Re}[n_{im}^Ln_{im}^R] F_2^N(x_{im})
\Big\}
\label{amuchi0}
,\\
\fl
\amucha&=\frac{m_\mu}{16\pi^2}\sum_{k}
\Big\{
\frac{m_\mu}{12m_{\tilde{\nu}_\mu}^2}(|c_{k}^L|^2+|c_{k}^R|^2)
F_1^C(x_{k})
+\frac{2m_{\chi_k^\pm}}{3m_{\tilde{\nu}_\mu}^2}
{\rm Re}[c_{k}^Lc_{k}^R] F_2^C(x_{k})
\Big\}
,
\label{amuchipm}
\end{eqnarray}
where $i=1\ldots4$ and $k=1,2$ denote the neutralino and chargino
indices, $m=1,2$ denotes the smuon index, and the couplings are given
by
\begin{eqnarray}
n_{im}^L &= \frac{1}{\sqrt2}(g_1 N_{i1}+g_2 N_{i2})\Usmu_{m1}{}^*
-y_\mu N_{i3}\Usmu_{m2}{}^*,\\
n_{im}^R &= \sqrt2 g_1 N_{i1} \Usmu_{m2} + y_\mu N_{i3}\Usmu_{m1},\\
c_{k}^L &= - g_2 V_{k1},\\
c_{k}^R &= y_\mu U_{k2}.
\end{eqnarray}
The kinematic variables are defined as the mass ratios
$x_{im}=m_{\chi_i^0}^2/m_{\tilde{\mu}_m}^2$,
$x_k=m_{\chi_k^\pm}^2/m_{\tilde{\nu}_\mu}^2$, and the loop functions
are given by
\begin{eqnarray}
F_1^N(x) &= \frac{ 2 }{(1-x)^4 }\big[
1-6x+3x^2+2x^3-6x^2\log x
\big],\\
F_2^N(x) &= \frac{ 3 }{(1-x)^3 }\big[
1-x^2+2x\log x
\big],\\
F_1^C(x) &= \frac{ 2 }{(1-x)^4 }\big[
2+3x-6x^2+x^3+6x\log x
\big],\\
F_2^C(x) &= \frac{ 3 }{(1-x)^3 }\big[
-3+4x-x^2-2\log x
\big],
\end{eqnarray}
normalized such that $F_i^j(1)=1$. The U(1) and SU(2) gauge couplings
are given by $g_{1,2}=e/\{c_W,s_W\}$, such that the one-loop
contributions are of the order $\alpha=e^2/(4\pi)$. A class of large
two-loop logarithms can be taken into account by the replacement
$\alpha\to\alpha(M_{\rm SUSY})$ (see later for more details).

For discussing the one-loop contributions $\amu^{\chi^{0,\pm}}$ it is
noteworthy that the terms linear in $m_{\chi^{0,\pm}}$ are not
enhanced by a factor $m_{\chi^{0,\pm}}/m_\mu$ compared to the other
terms. Rather, these terms involve either an explicit factor of the
muon Yukawa coupling $y_\mu$ or of the combination
$\Usmu_{m1}\Usmu_{m2}/m_{\tilde{\mu}_m}^2$, which in turn is
proportional to $(M_\mu^2)_{12}$ and thus to $y_\mu$. Hence, all terms
are of the same basic order $m_\mu^2/M_{\rm SUSY}^2$, and the terms
linear in $m_{\chi^{0,\pm}}$ are enhanced merely by a factor
$\tan\beta$ from the muon Yukawa coupling.

It is instructive to close this subsection by deriving a simple
approximation of (\ref{amuchi0}), (\ref{amuchipm})  for
large $\tan\beta$ and the case that all SUSY mass parameters in the
smuon, chargino and neutralino mass matrices are equal to a common
scale $M_{\rm SUSY}$. In this case only the terms linear 
in $m_{\chi^{0,\pm}}$ have to be considered, and the loop functions
$F_i^j(x)$ can be approximated by a Taylor series around $x=1$. For
example, the factors $m_{\chi^\pm_k} c_k^L c_k^R F_2^C(x_k)$ appearing
in $\amucha$ can be approximated as
\begin{equation} 
\fl
-g_2 y_\mu \sum_k U_{k2} V_{k1}
m_{\chi^\pm_k} 
\left(\frac74-\frac{3}{4}
\frac{m_{\chi^\pm_k}^2}
{m_{\tilde{\nu}_\mu}^2}
\right) 
\approx
\frac{3 g_2 y_\mu}{4} \frac{ X_{22} (X^\dagger)_{21}
   X_{11}}{m_{\tilde{\nu}_\mu}^2}
\approx
\frac{3 g_2 y_\mu}{4} {\rm sign}(\mu M_2) X_{12}.
\end{equation}
Here terms that are suppressed by $1/\tan\beta$ or $M_W/M_{\rm
  SUSY}$ have been neglected.
Note that the factors on the right-hand side correspond directly to
the mass-insertion diagram (C) in figure
\ref{fig:massinsertiondiagrams} and the approximation
(\ref{Diagabcd}).
The factors appearing in $\amuneu$ can be similarly
approximated. Inserting these approximations, one obtains
\cite{Moroi1L}
\begin{eqnarray}
\amuneu&= \frac{g_1^2-g_2^2}{192\pi^2}\frac{m_\mu^2}{\MSUSY^2}
\ {\rm sign}(\mu M_2)\ \tan\beta\ 
\left[1+{\cal O}\left(\frac{1}{\tan\beta},
\frac{M_W}{M_{\rm SUSY}}\right)\right],\\
\amucha&= \ \frac{g_2^2}{32\pi^2}\ \ \frac{m_\mu^2}{\MSUSY^2}
\ {\rm sign}(\mu M_2)\ \tan\beta\ 
\left[1+{\cal O}\left(\frac{1}{\tan\beta},
\frac{M_W}{M_{\rm SUSY}}\right)\right],
\label{amucha}
\end{eqnarray}
where real parameters and equal signs of $M_1$ and $M_2$ have been
assumed.

\subsection{Two-loop contributions}

It is useful to classify the MSSM two-loop diagrams similar to the
one-loop diagrams, into
\begin{enumerate}
\item  two-loop corrections to SM one-loop diagrams, where the
  $\mu$-lepton number is carried only by $\mu$ and/or $\nu_\mu$.
\item  two-loop corrections to SUSY one-loop diagrams, where the
  $\mu$-lepton number is carried also by $\tilde{\mu}$ and/or
  $\tilde{\nu}_\mu$.
\end{enumerate}
The first class contains in particular SM-like diagrams with an
insertion of a loop of SUSY particles, e.g.\ of $\tilde{t}$,
$\tilde{b}$ or $\chi^\pm$. Such diagrams are particularly interesting
since they constitute SUSY two-loop contributions that involve other
particles and have a completely different parameter dependence than
the SUSY one-loop contributions. Most importantly, these two-loop
contributions can be large even if $\amuSUOL$ is suppressed. The
contributions of this class are exactly known \cite{HSW03,HSW04}.

The SUSY two-loop contributions of the second class involve the same
particles as the SUSY one-loop contributions (possibly plus additional
ones). Hence
they can be expected to have a similar parameter dependence as
$\amuSUOL$. The contributions of this class are known in the
approximation of leading QED-logarithms \cite{DG98}.

\begin{figure}
\null\hfill
\begin{picture}(125,130)(0,-12)
\Vertex(25,5){2}
\Vertex(100,5){2}
\Vertex(45,40){2}
\Vertex(80,40){2}
\Vertex(62.5,85.67){2}
\DashLine(25,5)(45,40){4}
\Photon(100,5)(80,40){4}{4}
\Photon(62.5,85.67)(62.5,114){4}{3}
\CArc(62.5,60)(25.67,0,360)
\Line(0,5)(25,5)
\Line(100,5)(125,5)
\Line(25,5)(100,5)
\Text(0,-6)[l]{$\mu$}\Text(125,-6)[r]{${\mu}$}
\Text(62.5,-6)[c]{${\mu},\nu_\mu$}
\Text(28,25)[r]{$H,G$}\Text(97,25)[l]{$V$}
\Text(33,65)[r]{$X$}
\Text(55,120)[rt]{$\gamma$}
\end{picture}
\hfill
\begin{picture}(125,130)(0,-12)
\Vertex(25,5){2}
\Vertex(100,5){2}
\Vertex(45,40){2}
\Vertex(80,40){2}
\Vertex(62.5,85.67){2}
\Photon(62.5,85.67)(62.5,114){4}{3}
\Photon(25,5)(45,40){4}{4}
\Photon(100,5)(80,40){4}{4}
\CArc(62.5,60)(25.67,0,360)
\Line(0,5)(25,5)
\Line(100,5)(125,5)
\Line(25,5)(100,5)
\Text(0,-6)[l]{$\mu$}\Text(125,-6)[r]{${\mu}$}
\Text(62.5,-6)[c]{${\mu},\nu_\mu$}
\Text(28,25)[r]{$V$}\Text(97,25)[l]{$V$}
\Text(33,65)[r]{$X$}
\Text(55,120)[rt]{$\gamma$}
\end{picture}
\hfill\null
\caption{\label{fig:chisfTL}
  Sample two-loop diagrams with closed chargino/neutralino or
  sfermion loop, contributing to
  $\amuchiTL$ and $\amusfTL$. The diagrams are
  categorized into classes $(XVH)$, $(XVG)$ and $(XVV)$, where
  $X=\chi^{\pm,0},\tilde{f}$.
  $V=\gamma,Z,W^\pm$ denotes gauge bosons, $H=h^0,H^0,A^0,H^\pm$
  denotes physical Higgs bosons, and $G=G^{\pm,0}$ denotes
  Goldstone bosons. See \cite{HSW03,HSW04} for more details on the
  possible diagram topologies.}
\end{figure}
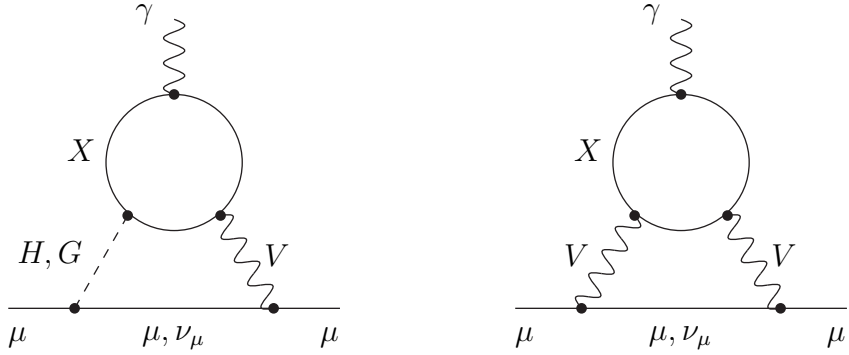

\subsubsection{Two-loop corrections to SM one-loop diagrams}
\label{sec:SUTLa}

\ 
The MSSM two-loop contributions of the first class can be decomposed
into a SM- and SUSY-part,
\begin{eqnarray}
 \amuSMTL+\amuSUTLa,
\end{eqnarray}
where $\amuSMTL$ denotes the SM two-loop
contributions. The genuine SUSY contributions of this class can be split
into four parts:
\begin{eqnarray}
\amuSUTLa &= \amuchiTL + \amusfTL + \amuSUTLferm + \amuSUTLbos.
\label{amuSUTLa}
\end{eqnarray}
The first two terms correspond to diagrams involving a closed
chargino/neutralino or sfermion loop, respectively. These
diagrams are further categorized according to the particles coupling
to the muon line,
\begin{eqnarray}
\amu^{X,\rm 2L} &= \amu^{(XVH)}+\amu^{(XVV)}+\amu^{(XVG)},
\qquad&
X=\chi,\tilde{f}.
\end{eqnarray}
Diagrams where one gauge boson and one physical Higgs
boson couple to the muon line are denoted as $(XVH)$ with
$V=\gamma,W,Z$ and $H=h^0,H^0,A^0,H^\pm$. Diagrams
where only gauge bosons or unphysical Goldstone bosons couple to the
muon are denoted as $(XVV)$, $(XVG)$.\footnote{%
 Diagrams of the form
  $(XHH)$, $(XHG)$ etc.\ in which two Higgs or Goldstone bosons couple
 to the muon line are suppressed by an additional muon Yukawa coupling
 and can be neglected.}
Sample diagrams are shown in figure
\ref{fig:chisfTL}. 

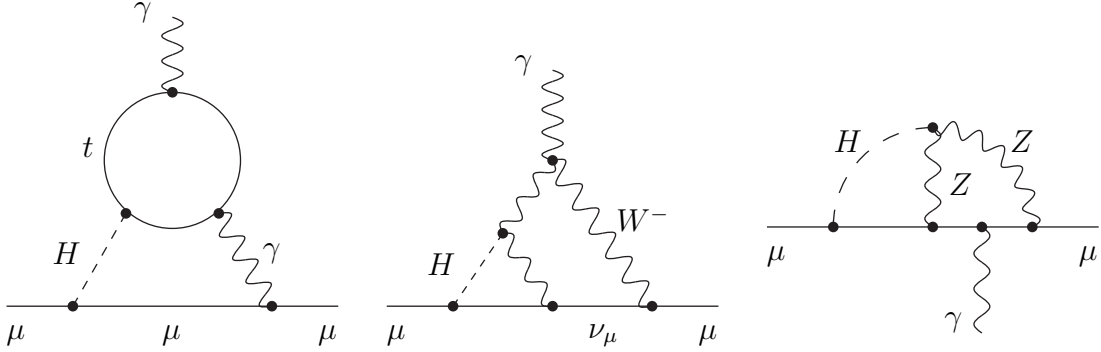
\begin{figure}
\null\hfill
\begin{picture}(125,130)(0,-12)
\Vertex(25,5){2}
\Vertex(100,5){2}
\Vertex(45,40){2}
\Vertex(80,40){2}
\Vertex(62.5,85.67){2}
\Photon(62.5,85.67)(62.5,114){4}{3}
\DashLine(25,5)(45,40){4}
\Photon(100,5)(80,40){4}{4}
\CArc(62.5,60)(25.67,0,360)
\Line(0,5)(25,5)
\Line(100,5)(125,5)
\Line(25,5)(100,5)
\Text(0,-6)[l]{$\mu$}\Text(125,-6)[r]{${\mu}$}
\Text(62.5,-6)[c]{${\mu}$}
\Text(28,25)[r]{$H$}\Text(97,25)[l]{$\gamma$}
\Text(33,65)[r]{$t$}
\Text(55,120)[rt]{$\gamma$}
\end{picture}
\hfill
\begin{picture}(125,130)(0,-12)
\Vertex(25,5){2}
\Vertex(100,5){2}
\Vertex(43.75,32.5){2}
\Vertex(62.5,5){2}
\Vertex(62.5,60){2}
\Photon(62.5,60)(62.5,94){4}{4}
\DashLine(25,5)(43.75,32.5){3}
\Photon(62.5,5)(43.75,32.5){3}{3}
\Photon(100,5)(62.5,60){4}{6}
\Photon(62.5,60)(43.75,32.5){3}{3}
\Line(0,5)(25,5)
\Line(100,5)(125,5)
\Line(25,5)(100,5)
\Text(0,-6)[l]{$\mu$}\Text(125,-6)[r]{${\mu}$}
\Text(82,-6)[c]{$\nu_\mu$}
\Text(26,21)[r]{$H$}\Text(87,38)[l]{$W^-$}
\Text(55,100)[rt]{$\gamma$}
\end{picture}
\hfill
\begin{picture}(125,130)(0,-42)
\Vertex(25,5){2}
\Vertex(100,5){2}
\Vertex(62.5,5){2}
\Vertex(62.5,42.5){2}
\Vertex(81,5){2}
\Photon(81,5)(81,-35){3}{3}
\Photon(62.5,5)(62.5,42.5){3}{3}
\PhotonArc(62.5,5)(37.5,0,90){3}{6}
\DashCArc(62.5,5)(37.5,90,180){6}
\Line(0,5)(25,5)
\Line(100,5)(125,5)
\Line(25,5)(100,5)
\Text(0,-6)[l]{$\mu$}\Text(125,-6)[r]{${\mu}$}
\Text(92,37.5)[l]{$Z$}
\Text(36,37.5)[r]{$H$}
\Text(68.75,21)[l]{$Z$}
\Text(74,-35)[rb]{$\gamma$}
\end{picture}
\hfill\null
\caption{\label{fig:TLSMlike}
Sample two-loop diagrams involving only SM- or
  two-Higgs-doublet model particles and either with or without fermion
  loop. These
  diagrams are different in the MSSM and the SM due to the 
  modified Higgs sector, and this difference   constitutes the SUSY
  contributions $\amuSUTLferm$ and $\amuSUTLbos$.}
\end{figure}

The remaining two terms in (\ref{amuSUTLa})
correspond to diagrams involving only SM- or two-Higgs-doublet model
particles and no SUSY particles. These diagrams are different in the
MSSM and the SM due to the additional Higgs bosons and the modified
Higgs boson couplings. $\amuSUTLferm$ denotes the difference between the
MSSM- and SM-evaluation of the diagrams involving a SM fermion (i.e.\
quark or lepton) loop; likewise, $\amuSUTLbos$ denotes the
corresponding difference of the diagrams without fermion loop, the
so-called bosonic contributions. Sample diagrams are shown in figure
\ref{fig:TLSMlike}.

The SUSY two-loop diagrams can be conveniently evaluated by first
applying a large mass expansion \cite{smirnov}, where the muon mass is
treated as small and all other masses as large. This results in a
separation of scales, and all remaining integrals are of one of two
types. One type are one-scale two-point integrals with external
momentum $p^2=m_\mu^2$ and all internal masses being either zero or
equal to $m_\mu$. The other type are integrals where all internal
masses are heavy but the external momentum can be neglected. All these
integrals and the corresponding prefactors can be evaluated analytically
\cite{HSW03,HSW04}. In addition to the genuine two-loop diagrams,
one-loop counterterm diagrams have to be evaluated. These contain
renormalization constants corresponding to charge, mass, and tadpole
renormalization, which are defined in the on-shell renormalization
scheme \cite{Aoki,BHS,HKRRSS}.

The diagrams of classes $(XVH)$,
$X=\chi,\tilde{f},f$ can be calculated in an alternative way. In these
so-called Barr-Zee diagrams \cite{BarrZee}, a closed loop generates an
effective $\gamma$--$V$--$H$ vertex, and this vertex can be
evaluated first by a one-loop computation. By inserting the result and
performing the second loop integral one obtains a simple integral
representation for the full two-loop diagram.  Barr-Zee diagrams were
first considered because they give rise to important contributions to
electric dipole moments in extensions of the SM (see e.g.\ Refs.\
\cite{BarrZee,ChangKP98,Pilaftsis98}). The contributions from
particular Barr-Zee diagrams of the classes $(f\gamma A^0)$,
$(f\gamma H)$, $(\tilde{f}\gamma H)$,  $(\tilde{f}W^\pm 
H^\mp)$ to $\amu$ were considered in Refs.\
\cite{ChangCCK00,CheungCK01,ArhribBaek,ChenGeng},
respectively.

In Refs.\ \cite{HSW03,HSW04} the numerical results of all
two-loop contributions in  (\ref{amuSUTLa}) were compared and analyzed in
detail, taking into account that the SUSY parameters are constrained
by experimental bounds on $b$-decays, $M_h$ and other quantities. It
turned out that the numerical values of the various 
contributions is very different:
\begin{itemize}
\item The by far largest contributions are the ones from the photonic
  Barr-Zee diagrams $(\chi\gamma H)$ and
  $(\tilde{f}\gamma H)$, where $H$ denotes the neutral physical
  Higgs bosons $h^0, H^0, A^0$. As explained in section
  \ref{sec:massinsertions} they are enhanced by a factor $\tan\beta$
  and, in the case of the sfermion loop diagrams, by the potentially
  large Higgs--sfermion coupling. They can have values up to 
\begin{eqnarray}
\amu^{(\chi\gamma H)}, \amu^{(\tilde{f}\gamma H)}\sim{\cal
  O}(10)\tunit.
\end{eqnarray}
  Barr-Zee diagrams with $Z$ or $W^\pm$ exchange have a
  similar parameter dependence but are typically smaller by a factor
  of about 3--5.
\item The diagrams of classes $(\chi VV)$ and $(\tilde{f}VV)$ and the
  corresponding Goldstone diagrams $(\chi VG)$ and
  $(\tilde{f}VG)$  involve no enhanced
  muon--Higgs Yukawa coupling and thus no  $\tan\beta$-enhancement,
  and they do not involve any other enhancement factors. Their
  numerical impact is tiny. For SUSY masses larger than 100 GeV these
  contributions are smaller than $0.1\tunit$.
\item The genuine SUSY contributions to the SM-like diagrams
  $\amuSUTLferm$ and $\amuSUTLbos$ depend only on $\tan\beta$ and
  $M_A$ and are small. Only for $M_A<200$ GeV they can reach
  $10^{-10}$, but for larger $M_A$ they are typically below
  $0.5\tunit$.
\end{itemize}
Hence, for the purpose of the present review, we only present the
analytical result for the dominant contributions from the photonic
Barr-Zee diagrams with physical Higgs bosons. They can be written as
%
\begin{eqnarray}
\label{amuchigammaphi}
\fl\qquad
\amu^{(\chi\gamma H)} = 
\frac{\alpha^2 m_\mu^2}{8\pi^2 M_W^2 s_W^2}\
\sum_{k=1,2}\Big[
{\rm Re}[ \lambda_\mu^{A^0} \lambda_{\chi^+_k}^{A^0}]
\ f_{PS}(m_{\chi^+_k}^2/M_{A^0}^2)
\nonumber\\
\qquad\qquad\qquad\qquad\qquad\ 
+
\sum_{S=h^0,H^0} {\rm Re}[\lambda_\mu^S \lambda_{\chi^+_k}^S ]
\ f_S(m_{\chi^+_k}^2/M_S^2)
\Big]
,\\
\label{amusfgammaphi}
\fl\qquad
\amu^{(\tilde{f}\gamma H)} =
\frac{\alpha^2 m_\mu^2}{8\pi^2 M_W^2 s_W^2}\
\sum_{\tilde{f}=\tilde{t},\tilde{b},\tilde{\tau}}\sum_{i=1,2}
\Big[
\sum_{S=h^0,H^0}
(N_c Q^2)_{\tilde{f}}\ 
{\rm Re}[ \lambda_\mu^S \lambda_{\tilde{f}_i}^S]
\ f_{\tilde{f}}(m_{\tilde{f}_i}^2/M_S^2)
\Big].
\end{eqnarray}
The Higgs--muon and Higgs--chargino coupling factors are given by
\begin{eqnarray}
\lambda_\mu^{\{h^0,H^0,A^0\}} &= \left\{
-\frac{s_\alpha}{c_\beta},\frac{c_\alpha}{c_\beta},t_\beta
\right\},\\
%
\lambda_{\chi^+_k}^{\{h^0,H^0,A^0\}} &= \frac{\sqrt2 M_W}{m_{\chi^+_k}}
\big(U_{k1}V_{k2}\big\{ c_\alpha, s_\alpha, -c_\beta\big\}
+U_{k2}V_{k1}\big\{ -s_\alpha,c_\alpha,-s_\beta \big\}\big)
.
\end{eqnarray}
In the Higgs--sfermion couplings we neglect terms that are
subleading in $\tan\beta$ and that give rise to negligible
contributions to $\amu$:
\begin{eqnarray}
\lambda_{\tilde{t}_i}^{\{h^0,H^0\}} &=
\frac{2m_t}
{m_{\tilde{t}_i}^2 s_\beta}
\big(+\mu^*\big\{s_\alpha,-c_\alpha\big\}+A_t\big\{c_\alpha,s_\alpha\big\}\big)
\  (U^{\tilde{t}}_{i1})^*\  U^{\tilde{t}}_{i2} 
,\\
\lambda_{\tilde{b}_i}^{\{h^0,H^0\}} &=
\frac{2m_b}
{m_{\tilde{b}_i}^2 c_\beta}
\big(-\mu^*\big\{c_\alpha,s_\alpha\big\}+A_b\big\{-s_\alpha,c_\alpha\big\}\big)
\  (U^{\tilde{b}}_{i1})^* \   U^{\tilde{b}}_{i2} 
,\\
\lambda_{\tilde{\tau}_i}^{\{h^0,H^0\}} &=
\frac{2m_\tau}
{m_{\tilde{\tau}_i}^2 c_\beta}
\big(-\mu^*\big\{c_\alpha,s_\alpha\big\}
+A_\tau\big\{-s_\alpha,c_\alpha\big\}\big)
\  (U^{\tilde{\tau}}_{i1})^* \   U^{\tilde{\tau}}_{i2} 
.
\end{eqnarray}
The loop integral function $f_{PS}$ can be given either as a
one-dimensional integral or in terms of
dilogarithms:
\begin{eqnarray}
\fl\qquad
f_{PS}(z) &= z\int_0^1 \frac{{\rm d}x\log\frac{x(1-x)}{z}}{x(1-x)-z}
=
\frac{2z}{y}\Big[{\rm Li}_2\Big(1-\frac{1-y}{2z}\Big)
                -{\rm Li}_2\Big(1-\frac{1+y}{2z}\Big) \Big]
\end{eqnarray}
with $y=\sqrt{1-4z}$. Note that $f_{PS}(z)$ is real and analytic even
for $z\ge1/4$. The other loop functions are related to
$f_{PS}$ as
\begin{eqnarray}
f_S(z) &= (2z-1)f_{PS}(z) - 2z(2+\log z),\\
f_{\tilde{f}}(z) &=\frac{z}{2}\Big[2+\log z-f_{PS}(z)\Big].
\end{eqnarray}

We remark that useful numerical estimates for the leading two-loop
contributions can be obtained by taking into account only the
$\tan\beta$-enhanced terms in the couplings and by 
approximating the loop functions. In the case of the sfermion-loop
contributions, simple approximations for the most important
Barr-Zee diagrams $(\tilde{t}HV)$, $(\tilde{b}HV)$ with stop or
sbottom loops can be derived \cite{HSW03}. The approximation for
$(\tilde{t}HV)$ agrees with the estimate (\ref{Diage}) discussed in
section \ref{sec:massinsertions}, and the approximation for
$(\tilde{b}HV)$ has a similar form. 
In the case of the chargino/neutralino-loop diagrams the
parameter dependence is simpler, and the diagrams with $W$- and
$Z$-exchange can be included in the approximation 
\cite{HSW04}. We will collect all these approximations below in
equations (\ref{chaneuapprox})--(\ref{sbotapprox}).

\subsubsection{Two-loop corrections to SUSY one-loop diagrams}
\label{sec:SUTLb}

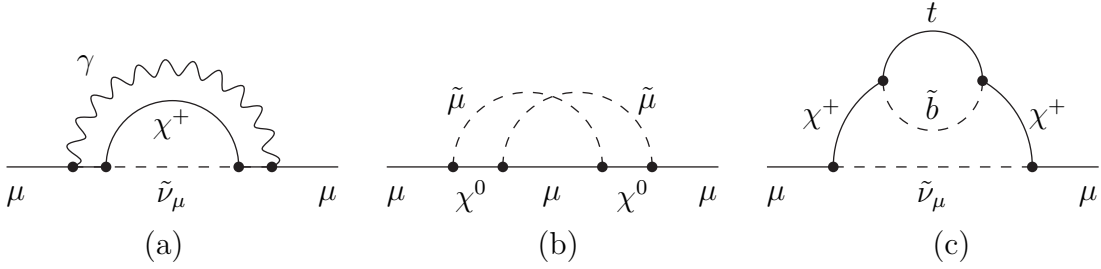
\begin{figure}
\null\hfill
\begin{picture}(125,75)(0,-12)
\Vertex(25,5){2}
\Vertex(100,5){2}
\Vertex(37.5,5){2}
\Vertex(87.5,5){2}
\PhotonArc(62.5,5)(37.5,0,180){3}{12}
\CArc(62.5,5)(25,0,180)
\Line(0,5)(37.5,5)
\Line(87.5,5)(125,5)
\DashLine(25,5)(100,5){4}
\Text(0,-6)[l]{$\mu$}\Text(125,-6)[r]{${\mu}$}
\Text(62.5,-6)[c]{$\tilde{\nu}_\mu$}
\Text(62.5,27)[t]{$\chi^+$}\Text(30,40)[b]{$\gamma$}
\end{picture}
\hfill
\begin{picture}(125,75)(0,-12)
\Vertex(25,5){2}
\Vertex(100,5){2}
\Vertex(43.75,5){2}
\Vertex(81.25,5){2}
\DashCArc(53.15,5)(28.15,0,180){4}
\DashCArc(71.875,5)(28.15,0,180){4}
\Line(0,5)(25,5)
\Line(100,5)(125,5)
\Line(25,5)(100,5)
\Text(0,-6)[l]{$\mu$}\Text(125,-6)[r]{${\mu}$}
\Text(38,-6)[r]{$\chi^0$}\Text(87,-6)[l]{$\chi^0$}
\Text(62.5,-6)[c]{${\mu}$}
\Text(30,30)[r]{$\tilde{\mu}$}\Text(95,30)[l]{$\tilde{\mu}$}
\end{picture}
\hfill
\begin{picture}(125,75)(0,-12)
\Vertex(25,5){2}
\Vertex(100,5){2}
\Vertex(43.75,37.5){2}
\Vertex(81.25,37.5){2}
\CArc(62.5,5)(37.5,0,60)
\CArc(62.5,5)(37.5,120,180)
\CArc(62.5,37.5)(18.75,0,180)
\DashCArc(62.5,37.5)(18.75,180,360){4}
\Line(0,5)(25,5)
\Line(100,5)(125,5)
\DashLine(25,5)(100,5){4}
\Text(0,-6)[l]{$\mu$}\Text(125,-6)[r]{${\mu}$}
\Text(62.5,-6)[c]{$\tilde{\nu}_\mu$}
\Text(28,25)[r]{$\chi^+$}\Text(99,25)[l]{$\chi^+$}
\Text(62.5,23)[b]{$\tilde{b}$}\Text(62.5,60)[b]{$t$}
\end{picture}
\hfill\null\\
\null\hfill(a)\hfill\hfill(b)\hfill\hfill(c)\hfill\null
\caption{\label{fig:SUTLb}
Sample two-loop diagrams contributing to $\amuSUTLb$, i.e.\ involving
a SUSY one-loop diagram. The external photon can be attached to all
charged internal lines. (a) shows a diagram with additional photon
loop, giving rise to large QED-logarithms. (b) shows a diagram of the
class computed in \cite{FengLM06}. (c) shows a diagram with an
additional fermion/sfermion loop.}
\end{figure}

The two-loop corrections to SUSY one-loop diagrams can be decomposed
into two pieces,
\begin{eqnarray}
\label{amuTLbdecomp}
\amuSUTLb &=
c_L^{\rm SUSY,2L(b)}\log \frac{M_{\rm SUSY}}{m_\mu}
+c_0^{\rm SUSY,2L(b)}
.
\end{eqnarray}
The first piece contains the large logarithm of the ratio
$M_{\rm SUSY}/m_\mu$, where $M_{\rm SUSY}$ is the generic SUSY mass
scale, and the second piece contains at most small logarithms of
ratios of different SUSY masses. Sample diagrams are shown in figure
\ref{fig:SUTLb}. The diagrams all involve the same particles and the
same couplings as the SUSY one-loop diagrams (possibly plus additional
ones). Hence the overall parameter dependence of $\amuSUTLb$ and of
$\amuSUOL$ can be expected to be similar, up to the additional
two-loop suppression of $\amuSUTLb$. The large logarithm is the
most relevant enhancement factor.

The decomposition (\ref{amuTLbdecomp}) is analogous to the one of the
bosonic two-loop contributions in the SM, which have been evaluated in
\cite{SMbosold,HSW04,SMbosnew}. In the case of the SM, the term
enhanced by $\log M_Z/m_\mu$ is roughly a factor 10 larger than the
non-logarithmic piece. 

In the case of the MSSM, the logarithmic term is known \cite{DG98},
but for the non-logarithmic remainder only a particular subclass of
diagrams has been computed recently \cite{FengLM06}. In the following
we first discuss the logarithmic term, which can be assumed to be
dominant like in the SM, and then we describe the computation of
\cite{FengLM06}. 

The large logarithms in (\ref{amuTLbdecomp}) are QED-logarithms and
arise from two-loop diagrams that involve a SUSY one-loop diagram and
an additional photon loop. The loop integrals of such diagrams have an
infrared singularity in the limit $m_\mu\to0$ and therefore give rise
to terms $\propto\log m_\mu$. As shown by \cite{DG98}, the appropriate
framework to evaluate these logarithms efficiently is the framework of
effective field theories. 

The relevant effective field theory is
obtained from the MSSM by integrating out all fields of mass $\geq
M_{\rm SUSY}$ and retaining only the muon and photon. All further
light or heavy SM fields are irrelevant in this analysis and can be
ignored. The 
resulting theory is QED with additional higher-dimensional terms,
described by
\begin{eqnarray}
{\cal L}_{\rm eff} &=-2\sqrt2 G_\mu \sum_i C_i(\mu){\cal O}_i
\end{eqnarray}
in the notation of \cite{DG98}. The ${\cal O}_i$ are
higher-dimensional operators. The analysis of \cite{DG98} shows that
in the MSSM, like in many new physics models, only one
higher-dimensional operator has to be considered, namely
the one corresponding to the muon anomalous magnetic moment,
\begin{eqnarray}
H_\mu &= -\frac{e}{16\pi^2}\, m_\mu\, \bar\mu\,\sigma^{\nu\rho}\,
\mu\, F_{\nu\rho}.
\end{eqnarray}
The prefactors $C_i(\mu)$ are renormalization-scale dependent Wilson
coefficients, which can be 
determined by matching the effective theory to the full MSSM at the
high scale $M_{\rm SUSY}$. Determining the Wilson coefficient
$C_{H_\mu}(M_{\rm SUSY})$ thus corresponds to the one-loop
computation of $\amuSU$.

By construction, the large logarithms are identical in the full MSSM
and the effective theory. However, in the effective theory the
logarithms can be obtained simply from the one-loop
renormalization-group running  of the Wilson coefficient
$C_{H_\mu}(\mu)$ from $\mu=M_{\rm SUSY}$ down to $\mu=m_{\mu}$. This
running is described by
\begin{eqnarray}
C_{H_\mu}(\mu) &= C_{H_\mu}(M_{\rm SUSY})-
\gamma(H_\mu,H_\mu)\, \frac{\alpha(\mu)}{4\pi}\,
\log\frac{M_{\rm SUSY}}{\mu}\, C_{H_\mu}(M_{\rm SUSY})
.
\label{DGFormula}
\end{eqnarray}
where $\gamma(H_\mu,H_\mu)$ is the anomalous dimension of
$H_\mu$. On a diagrammatic level, the correspondence of this formula
to the two-loop computation in the full MSSM is easy to see. In the
MSSM, the logarithms arise from diagrams like the one in
figure~\ref{fig:SUTLb}~(a). Corresponding diagrams in the effective
theory are obtained by contracting the insertion of the SUSY
one-loop diagram to a point. The resulting diagrams are one-loop 
contributions to $H_\mu$, involving the effective vertex
$H_\mu$. Their UV-divergence, and thus their $\log\mu$-terms,
determine the anomalous dimension $\gamma(H_\mu,H_\mu)$. 

The value of the anomalous dimension is
\begin{eqnarray}
\gamma(H_\mu,H_\mu)=16
.
\end{eqnarray}
As a result, the QED-logarithms in the two-loop contributions to
$\amuSU$ are given by
\begin{eqnarray}
\label{amutllogs}
\amuSUTLb &=
-\frac{4\alpha}{\pi}\,\log \frac{M_{\rm SUSY}}{m_\mu}
\,
\amuSUOL
+c_0^{\rm SUSY,2L(b)}
.
\end{eqnarray}

This logarithmic correction is negative, and it amounts to
$-7\%\ldots-9\%$ of the SUSY one-loop contributions for $M_{\rm SUSY}$
between 100 and 1000 GeV. This result can be compared to the case of
the bosonic SM two-loop contributions, where the logarithms amount to
$-19\%$ of the SM electroweak one-loop result.

As mentioned before, the non-logarithmic terms in $\amuSUTLb$,
$c_0^{\rm SUSY,2L(b)}$,  are not
known so far. A first evaluation of a subclass of diagrams has been
carried out in \cite{FengLM06}. The considered diagrams involve only
sleptons, charginos and neutralinos, and only topologies as in figure
\ref{fig:SUTLb}~(b) 
are taken into account that contain no self-energy
subdiagrams. These diagrams constitute a finite contribution to
$c_0^{\rm SUSY,2L(b)}$. However, the result of \cite{FengLM06} can
only be viewed 
as intermediate because there are more diagrams, e.g.\ containing
self-energy subdiagrams or $W$- or $Z$-exchange, that would involve
exactly the same coupling constants, such that non-trivial
cancellations could be possible.

Nevertheless, the investigation of \cite{FengLM06} provides a first
insight to the possible values of the remaining two-loop
contributions. The numerical values found in \cite{FengLM06} are
surprisingly large. In a range of SUSY parameters with SUSY
masses of the order 300\ldots500 GeV, the values are mostly below
$10^{-10}$ but can become up to $2\tunit$, which is significantly
larger than the corresponding non-logarithmic terms of the bosonic SM
two-loop contributions.

\subsection{Summary of known contributions and error estimate}

To summarize, the SUSY contributions to $\amu$ up to the two-loop
level, i.e.\ the difference of $\amu$ in the MSSM and the SM, are given
by
\begin{eqnarray}
\amuSUSY &=
\amuSUOL\left(1-\frac{4\alpha}{\pi}\log\frac{M_{\rm
    SUSY}}{m_\mu}\right)
+
\amu^{(\chi\gamma H)}+\amu^{(\tilde{f}\gamma H)}
\nonumber\\
&+\amu^{(\chi\{W,Z\} H)}+\amu^{(\tilde{f}\{W,Z\}H)}
+\amuSUTLferm + \amuSUTLbos
+\ldots,
\label{amuSUSYknown}
\end{eqnarray}
where the terms in the first line have been given analytically in
(\ref{amuSUOL}), (\ref{amuchigammaphi}),
(\ref{amusfgammaphi}), (\ref{amutllogs}). Note that the discussion of
the two-loop QED-logarithms and equation (\ref{DGFormula}) also show
that the one-loop result should be parametrized in terms of the running
$\alpha(M_{\rm SUSY})$. In practice, it is sufficiently accurate to
approximate $\alpha(M_{\rm SUSY})$ by $\alpha(M_Z)=1/127.9$, and we
define $M_{\rm SUSY}$ in the logarithm as the mass of the lightest
charged SUSY particle.

For many applications it 
should be sufficient to take into account the terms in the first line.
The explicitly written terms in the second line are known, and they
can be up to ${\cal O}(1)\tunit$, but in the largest part of the MSSM
parameter space they are much smaller. The dots denote the known but
negligible contributions of the type $(\chi VV)$, $(\tilde{f}VV)$,
the contributions evaluated in \cite{FengLM06}, and the remaining unknown
two-loop contributions.

Handy approximations for the dominant terms are given by
\begin{eqnarray}
\label{oneloopapprox}
\amuSUOL &\approx
\ \ 13\times10^{-10}\left(\frac{100\,\rm GeV}{M_{\rm SUSY}}\right) ^2
\ \tan\beta\ \mbox{sign}(\mu M_2),\\
\label{chaneuapprox}
\amu^{(\chi VH)} &\approx
\ \ 11\tunit\ \left(\frac{\tan\beta}{50}\right)
\left(\frac{100\ {\rm GeV}}{\MSUSY}\right)^2\
{\rm sign}(\mu M_2),\\
\label{stopapprox}
\amu^{(\tilde{t}\gamma H)} &\approx
-13\tunit\ \left(\frac{\tan\beta}{50}\right)
\left(\frac{m_t}{m_{\tilde{t}}}\right)
\left(\frac{\mu}{20 M_H}\right)
\ {\rm sign}(X_t)
,\\
\label{sbotapprox}
\amu^{(\tilde{b}\gamma H)} &\approx
-3.2\tunit\ \left(\frac{\tan\beta}{50}\right)
\left(\frac{m_b\tan\beta}{m_{\tilde{b}}}\right)
\left(\frac{A_b}{20 M_H}\right)
\ {\rm sign}(\mu)
.
\end{eqnarray}
The first two are valid if all SUSY masses are approximately equal
(note that the relevant masses are different in the two cases),
and the third  and fourth are valid if the stop/sbottom mixing is
large and the relevant stop/sbottom and Higgs masses are of similar
size. The result for the $(\tilde{b}\gamma H)$ contribution has not
been here discussed before, but it can be understood in the same way
as the $(\tilde{t}\gamma H)$ result \cite{HSW03}.

In the following we list the missing contributions and estimate the
theory error of the SUSY prediction (\ref{amuSUSYknown}).
\begin{itemize}
\item Two-loop QED-corrections beyond the leading logarithm
  (\ref{amutllogs}). The leading-log approximation does not exactly
  fix the scale $M_{\rm SUSY}$ in the logarithm and in $\alpha(M_{\rm
  SUSY})$ (the latter appears in the one-loop result). The exact form
  of the logarithms and of the non-logarithmic 
  terms can only be found by a complete computation of the two-loop
  diagrams with a SUSY one-loop diagram and additional photon
  exchange. The error of the approximation (\ref{amutllogs}) can be
  estimated by varying $M_{\rm SUSY}$ in the range 100\ldots1000 GeV
  to about $2\%$ of the SUSY one-loop contributions. If the SUSY
  contributions to $\amu$ are the origin of the observed deviation
  (\ref{expSM}), they are certainly smaller than roughly $50\tunit$,
  and then this error is below $1\tunit$.
\item Further electroweak and SUSY two-loop corrections to SUSY
  one-loop diagrams. These corrections include two-loop diagrams
  similar to figure \ref{fig:SUTLb}~(a) but with $W$-, $Z$-, Higgs-
  instead of photon-exchange, and like in figure \ref{fig:SUTLb}~(b)
  with purely SUSY particles in the loops. Given the result for the
  subclass evaluated in \cite{FengLM06}, we assign an error of
  $\pm2\tunit$ to these diagrams. Note that this is a factor of 10
  larger than the known result of the corresponding non-logarithmic
  bosonic two-loop contributions in the SM
  \cite{SMbosold,HSW04,SMbosnew}.
\item Two-loop corrections to SUSY one-loop diagrams with
  fermion/sfermion-loops (see figure \ref{fig:SUTLb}~(c) for an
  example). This class of diagrams involves in particular top/stop-
  and bottom/sbottom-loops, which are enhanced by the large 3rd
  generation Yukawa couplings. We estimate the numerical value of
  these diagrams to be less than $\pm0.5\tunit$ for not too light SUSY
  masses for the following
  reasons. SUSY relates these diagrams to SM   diagrams with
  top/bottom loops, which  amount to about $0.6\tunit$   but are not
  suppressed by  possibly heavy SUSY masses. SUSY
  also relates the fermion/sfermion-loop diagrams to pure
  sfermion-loop diagrams such as $\amu^{(\tilde{f}\gamma H)}$, see
  (\ref{stopapprox}), (\ref{sbotapprox}). However, this relation
  should be most accurate for rather small $A_{t,b}$ and $\mu$, since
  the fermion/sfermion-loop diagrams are not enhanced by making these
  parameters large. In that case, the approximations
  (\ref{stopapprox}), (\ref{sbotapprox}) lead to values 
  below $0.5\tunit$.
\item Three-loop contributions. In general, three-loop contributions
  can be expected to be significantly smaller than the two-loop
  contributions. Two potential exceptions are three-loop diagrams that
  correspond to the two-loop contributions of the types $(\chi\gamma H)$,
  $(\tilde{f}\gamma H)$ with subloop-corrections to the Higgs-boson
  masses or the $b$-quark Yukawa coupling. It is well-known that
  the one-loop corrections to the Higgs-boson masses, in particular
  to $M_h$, and to $y_b$ can be very large. Hence, in cases where the
  diagrams with $h$-exchange and/or sbottom loop are very large, the
  missing three-loop contributions could amount to ${\cal
  O}(1)\tunit$. Fortunately, however, the
  influence of the lightest Higgs boson mass and $y_b$ on 
  the $(\chi\gamma H)$,   $(\tilde{f}\gamma H)$ diagrams is
  small in the largest part of the MSSM parameter space. Hence we
  neglect the theory error associated with the missing three-loop
  contributions. 
\end{itemize}
To summarize, we estimate the theory error associated with
(\ref{amuSUSYknown}) to
\begin{eqnarray}
\delta\amuSUSY(\mbox{unknown}) &= 0.02\ \amuSUOL + 2.5\tunit,
\label{amuSUSYerror}
\end{eqnarray}
where the errors associated with the individual classes of missing
diagrams have been added linearly. If $\amuSUSY$ is approximated by
only the first line of (\ref{amuSUSYknown}), the error
increases by the neglected contributions in the second line. An upper
limit of these can be  well approximated by \cite{HSW03,HSW04}
\begin{eqnarray}
\delta\amuSUSY(\mbox{2nd line}) &= 0.3\ 
\left(\amu^{(\chi\gamma H)}+\amu^{(\tilde{f}\gamma H)}\right)
+0.3\tunit.
\label{amuSUSYerror2}
\end{eqnarray}
It should be noted that the error estimate is deliberately
conservative. The later numerical analysis, see e.g.\ table
\ref{SPSTable}, shows that often already the known two-loop
contributions are much smaller than $10^{-10}$. In these cases, it is
reasonable to assume that the theory error due to the unknown
higher-order corrections is also much smaller than the estimate
(\ref{amuSUSYerror}). In any case, the theory error of the SUSY
contributions is smaller than both the current SM theory error
and the experimental uncertainty.

\section{Numerical behaviour of the SUSY contributions}
\label{sec:Numerics}

The SUSY contributions to $a_\mu$ depend on the MSSM parameters in a
complicated way. In this section we analyze the parameter
dependence of the numerical results in detail. For each of the one-
and two-loop contributions to $a_\mu$ we will describe which
parameters are more and which are less influential and 
explain the dominant behaviour. We discuss the
parameter choices for which each contribution can become particularly
large and the numerical values that can be expected for various typical 
parameter choices. Finally, we provide the values of the SUSY
contributions to $\amu$ for the benchmark ``SPS'' reference points
\cite{SPSDef}.

\subsection{One-loop contributions}

\begin{figure}
\mbox{\epsfbox{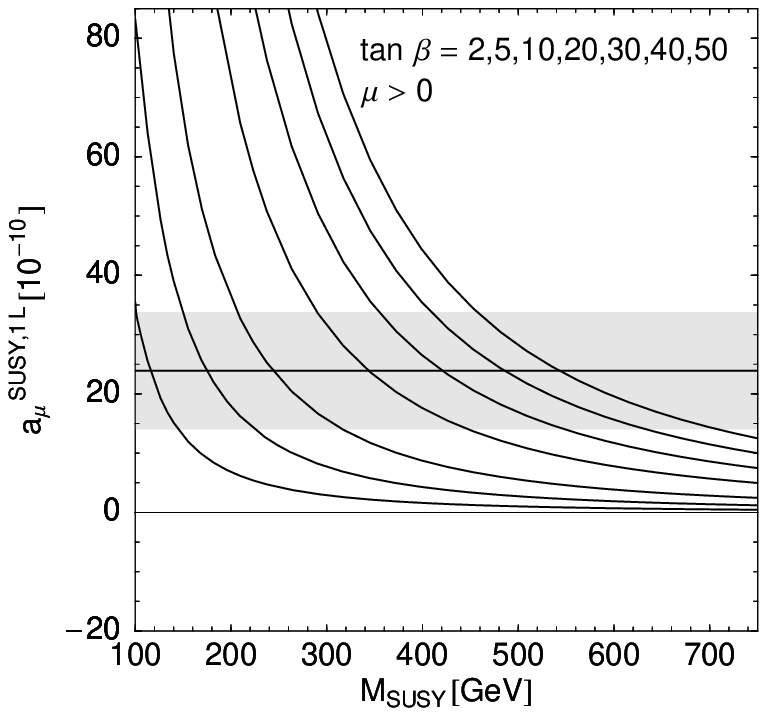}
\epsfbox{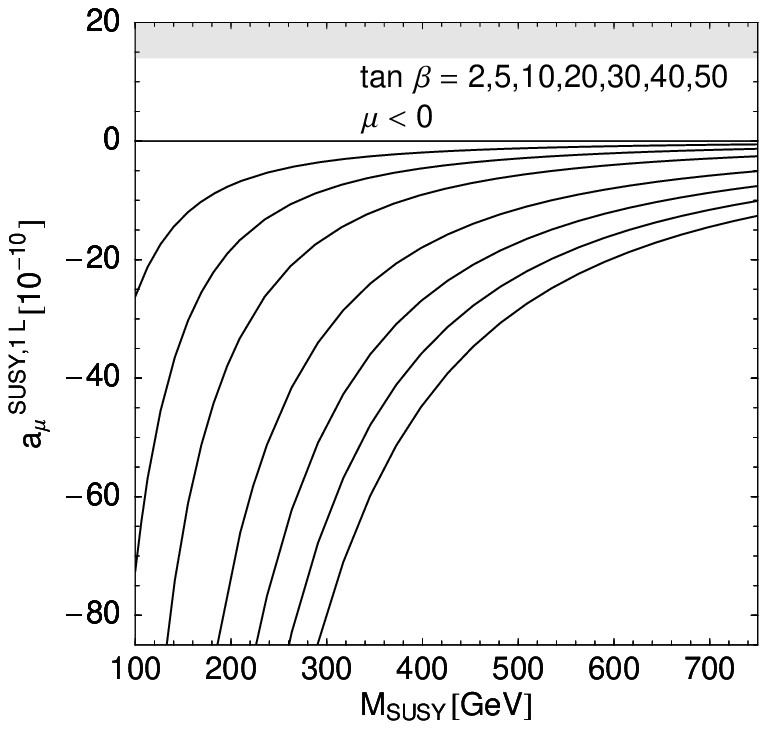}}
\null\qquad\hfill(a)\quad\hfill\hfill\qquad(b)\ \hfill\null
\caption{\label{fig:oneloopTBMUE}
$\amuSUOL$ as a function of a common mass scale $M_{\rm
    SUSY}=|\mu|=M_2=m_{L,\tilde{\mu}}=m_{R,\tilde{\mu}}$ for various
  values of $\tan\beta$ and $\mu>0$ (panel (a)), $\mu<0$ (panel
  (b)). The smaller values of $\tan\beta$ correspond to smaller
  $|\amuSUOL|$. The preferred $1\sigma$ region (\ref{expSM}) is
  indicated in grey.}
\end{figure}

The generic behaviour of the SUSY one-loop contributions to $a_\mu$ is
well described by (\ref{oneloopapprox}). The suppression by
$1/M_{\rm SUSY}^2$, the enhancement $\propto\tan\beta$ and the
dependence on the sign of $\mu$ has been explained in section
\ref{sec:massinsertions} using mass insertion diagrams. This generic
$\tan\beta\,$sign$(\mu)$ behaviour is illustrated by figure
\ref{fig:oneloopTBMUE}, which
shows $\amuSUOL$ for various values of $\tan\beta$ and both signs of
$\mu$. A common mass scale
$|\mu|=M_2=m_{L,\tilde{\mu}}=m_{R,\tilde{\mu}}\equiv M_{\rm SUSY}$ has
been chosen, except that $M_1$ is fixed by the GUT relation  $M_1/M_2=
5g_1^2/3g_2^2$  and $A_\mu=0$. Only for small $\tan\beta$ there
are visible deviations from this simple behaviour due to formally
$1/\tan\beta$-suppressed terms. The figure also demonstrates that
$\amuSUOL$ can be well in agreement with the observed deviation
(\ref{expSM}), but the small uncertainty in (\ref{expSM}) results in
stringent constraints on SUSY parameters. In the
following we consider the $\tan\beta\,$sign$(\mu)$ dependence as
understood and fix $\mu$ to be positive and $\tan\beta$ to be large.

\subsubsection{Dependence on mass parameters}

In general,
$\amuSUOL$ depends on $\tan\beta$ and six mass parameters $\mu$, $M_{1,2}$,
$m_{L,\tilde{\mu}}$, $m_{R,\tilde{\mu}}$, $A_\mu$.
In the following we will first concentrate on the behaviour of
$\amuSUOL$ as a function of the four parameters
\begin{eqnarray}
\mu,\ M_2,\ m_{L,\tilde{\mu}},\ m_{R,\tilde{\mu}}
\end{eqnarray}
and fix $M_1$ by the GUT relation $M_1/M_2= 5g_1^2/3g_2^2\approx
1/2$  and set $A_\mu=0$.

\begin{figure}
\mbox{\epsfbox{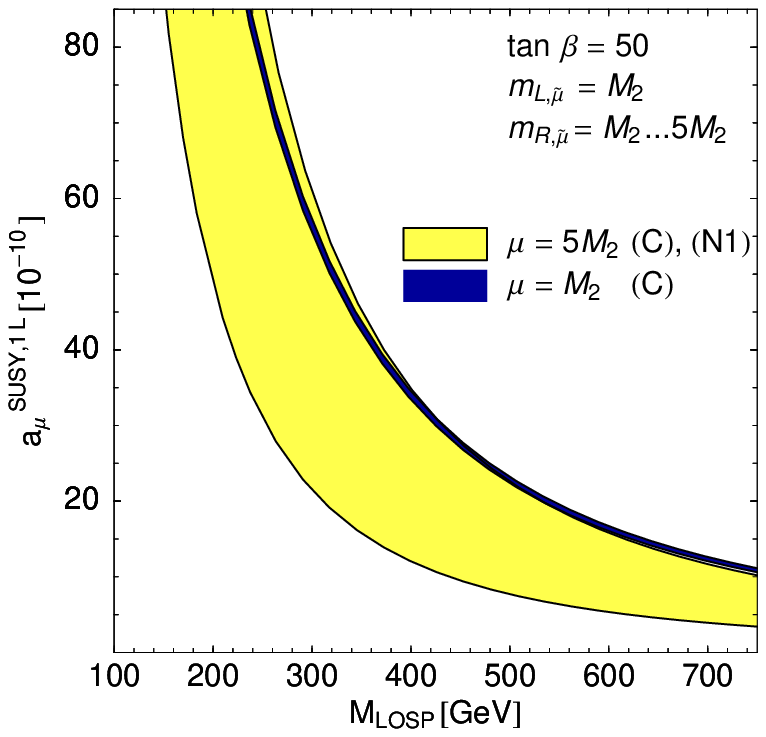}
\epsfbox{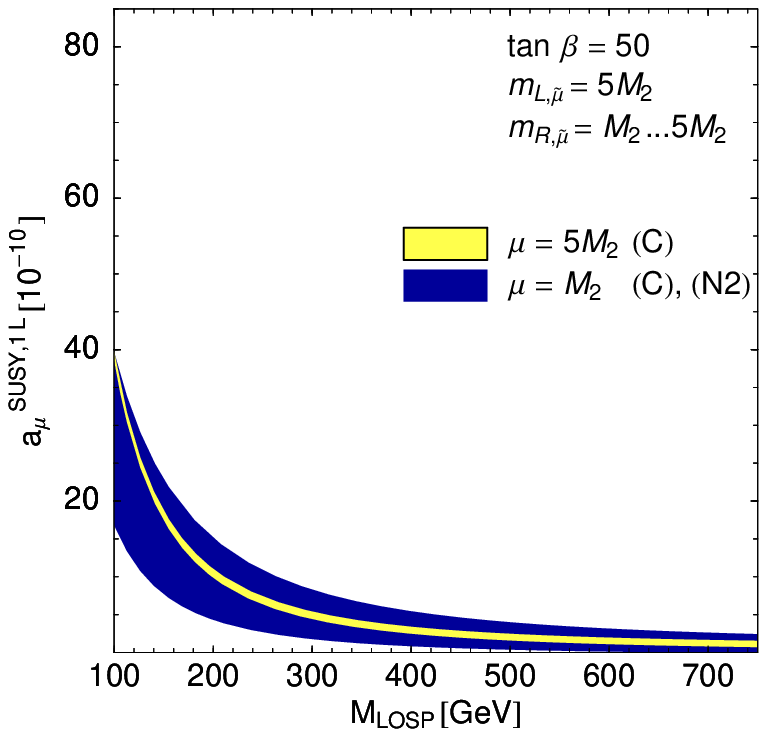}}
\null\qquad\hfill(a)\quad\hfill\hfill\qquad(b)\ \hfill\null
\caption{\label{fig:oneloopplot}
$\amuSUOL$ as a function of the mass of the lightest observable supersymmetric 
particle $M_{\rm LOSP}$, for $\tan\beta=50$
and the mass parameters varied in the range $M_2\ldots5M_2$. The blue
(yellow) regions correspond to 
$\mu=M_2$ ($\mu=5M_2$) and $m_{R,\tilde{\mu}}=(1\ldots5)M$. In panel (a),
  the left-handed smuon mass is small, $m_{L,\tilde{\mu}}=M_2$, in panel
(b) $m_{L,\tilde{\mu}}=5M_2$. For each case it is indicated which of
  the mass-insertion diagrams (C), (N1), (N2) are dominant.}
\end{figure}

We visualize the behaviour of $\amuSUOL$ for $\tan\beta=50$ in this
four-dimensional parameter space with the help of figure
\ref{fig:oneloopplot}. In this figure, $M_2$ is chosen to be the
smallest of the four mass parameters, and $\mu$, $m_{L,\tilde{\mu}}$,
$m_{R,\tilde{\mu}}$ are independently varied in the range
$M_2\ldots5M_2$. The left and right panels correspond to the choices
$m_{L,\tilde{\mu}}=M_2$ (panel (a)) and $m_{L,\tilde{\mu}}=5M_2$
(panel (b)). The different colours correspond to the values $\mu=M_2$
(dark blue) and $\mu=5M_2$ (light yellow). In all regions, the
right-handed smuon mass is varied between
$m_{R,\tilde{\mu}}=(1\ldots5)M_2$. Keeping these parameter ratios
fixed, $M_2$ is varied, and the resulting $\amuSUOL$ is plotted as a
function of the mass $M_{\rm LOSP}$ of the lightest observable SUSY
particle \cite{FengMatchev}, defined here as min($m_{\tilde{\mu}_1}$,
$m_{\chi^\pm_1}$, $m_{\chi^0_2}$). 

The two panels show that the $\mu$-dependence is quite different for
small and for large $m_{L,\tilde{\mu}}$. This intricate interplay
between the parameters has been first studied in \cite{Moroi1L}, where
it has been shown that the mass-insertion diagrams of figure
\ref{fig:massinsertiondiagrams} provide an intuitive understanding of
the parameter-dependence. In general, the chargino-diagram (C)
dominates due to its large numerical prefactor in (\ref{amucha}). In
special parameter regions, the diagrams (N1) with bino-exchange and
(N2) with $\tilde{\mu}_R$-exchange can become important.
\begin{itemize}
\item {\em Panel (a), dark blue:}
In the simple case $\mu=m_{L,\tilde{\mu}}=M_2$,
the chargino diagram (C) dominates. There is hardly any suppression
for heavy $m_{R,\tilde{\mu}}$, which enters only via the smaller
neutralino diagrams. Therefore the corresponding dark blue region in
panel (a) is very narrow.
\item {\em Panel (a), light yellow:}
If $\mu$ is increased, the chargino and most neutralino diagrams are
suppressed. For large $\mu$ the largest contribution can come from the
bino-diagram (N1), which is linear in $\mu$. 
However, (N1) is only large if both left- and
right-handed smuons are light. The large spread of the yellow region
in panel (a) corresponds mainly to this $m_{R,\tilde{\mu}}$-dependence
of (N1). The upper border corresponds to small $m_{R,\tilde{\mu}}$ (i.e.\
to $\mu=5M_2$, $m_{L,\tilde{\mu}}=m_{R,\tilde{\mu}}=M_2$) and leads to
values of $\amuSUOL$ that are even larger than the
ones in the blue region. This shows that the largest $\amuSUOL$ for a
given value of $M_{\rm LOSP}$ is obtained for $\mu\gg
m_{L,R,\tilde{\mu}},M_2$ via diagram (N1). 
The values at the lower border are suppressed
by the large $m_{R,\tilde{\mu}}=5M_2$ and are roughly by a factor 3
smaller.
\item {\em Panel (b), light yellow:}
The situation is reversed in panel (b) for large
$m_{L,\tilde{\mu}}$. In this case large $\mu$ does not lead to an
enhancement of diagram (N1) or any other diagram, since then all
diagrams are suppressed either by the large $\mu$ or the large
$m_{L,\tilde{\mu}}$. Consequently the result for
$\mu=m_{L,\tilde{\mu}}=5M_2$ is dominated by diagram (C) and is almost
independent of $m_{R,\tilde{\mu}}$. Hence the yellow region in panel
(b) is narrow. 
\item {\em Panel (b), dark blue:}
For large $m_{L,\tilde{\mu}}$ but small $\mu$, however, diagram (N2)
with $\tilde{\mu}_R$-exchange becomes important. It is the unique
diagram that is not suppressed by the large $m_{L,\tilde{\mu}}$, but it
has the opposite sign and a smaller numerical prefactor than diagram
(C). 
The lower border of the blue region in panel (b) corresponds to the
case of small $m_{R,\tilde{\mu}}$, i.e.\ to
$m_{R,\tilde{\mu}}=\mu=m_{L,\tilde{\mu}}/5$, where the contributions
of (C) and (N2) almost cancel. For 
larger ratios than displayed in the plot,
$m_{L,\tilde{\mu}}/m_{R,\tilde{\mu}}>5$, $\amuSUOL$ can even change
sign. The upper border of the same region
corresponds to the case of large $m_{R,\tilde{\mu}}$, where (C)
dominates and (N2) is suppressed.
\end{itemize}

\begin{figure}
\mbox{\epsfbox{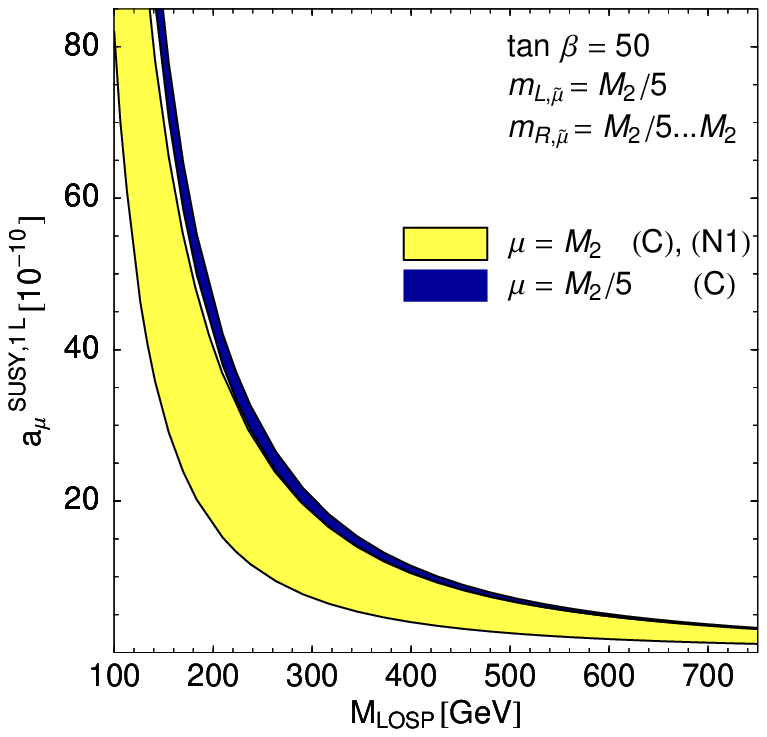}
\epsfbox{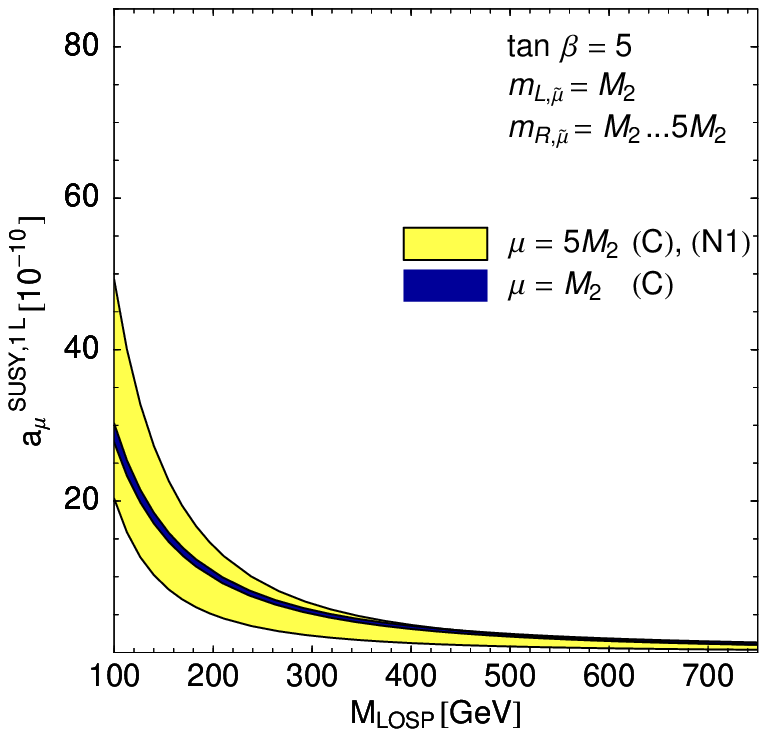}}
\null\qquad\hfill(a)\quad\hfill\hfill\qquad(b)\ \hfill\null
\caption{\label{fig:oneloopplotalternative}
$\amuSUOL$ as a function of the mass of the lightest observable supersymmetric 
particle $M_{\rm LOSP}$, for the same
  parameters as in figure \ref{fig:oneloopplot}~(a), except $M_2\to5M_2$
  (left panel), $\tan\beta=5$ (right panel).}
\end{figure}

Figure \ref{fig:oneloopplotalternative} shows how the results are
modified for different choices of $M_2$ and $\tan\beta$. Generally, 
larger values of $M_2$ lead to a suppression of the results without
dramatic change of the qualitative behaviour, and smaller values of
$\tan\beta$ lead to a suppression of the results due to the almost
linear $\tan\beta$-dependence. Figure
\ref{fig:oneloopplotalternative}~(a) shows the same as figure
\ref{fig:oneloopplot}~(a) except that  
$M_2$ has been replaced by $5M_2$, i.e.\ $M_2$ is the largest of the
four mass parameters and the other three are varied between
$M_2/5\ldots M_2$. We find that the qualitative features are
essentially unchanged. The larger value of $M_2$ leads to a
suppression by a factor of roughly 3, and the largest contributions
are now obtained for small $\mu=M_2/5$, corresponding to the dark blue
region. Panel (b) shows the same again but for $\tan\beta=5$ (and small
$M_2$). Here the results are reduced by almost a factor 10 compared to
the case with $\tan\beta=50$, as expected.


The two mass parameters we have not explicitly discussed yet are
$A_\mu$ and $M_1$. $M_1$ has generally less influence than $M_2$ since
the $M_1$-independent chargino diagrams dominate in most of the
parameter space. However, as exemplified by the discussion of figure
\ref{fig:oneloopplot}~(a) there are situations where the
bino-exchange diagram (N1) becomes important. If $M_1$ is not tied to
$M_2$ by the GUT relation, it is possible to choose $M_2,\mu\gg
M_1$. In this case one can obtain large contributions to $\amuSUOL$
from diagram (N1) even if both charginos are very heavy
\cite{MartinWells}. 

The dependence of $\amuSUOL$ on $A_\mu$ arises only
via the smuon mixing matrix, where $A_\mu$ is accompanied by the term
$\mu\tan\beta$, which is typically much larger. Therefore, $A_\mu$ is
quite insignificant for the SUSY contributions to $a_\mu$. The small
influence of $A_\mu$ was verified in \cite{FengMatchev} by means of a
scan of the SUSY parameter space, where $|A_\mu|$ was varied up to 100
TeV.

\subsubsection{CP violation and flavour violation\label{sec:LFV}}

So far, we have neglected the possibilities of complex phases and
generation mixing in the SUSY parameters. The influence of complex
phases has been studied in
\cite{IbrahimNath00a,IbrahimNath00b,MartinWells}. Clearly, 
the phase of the $\mu$-parameter has the most significant impact,
corresponding to the sign($\mu$)-dependence in the real case. At the
same time, this phase is strongly restricted by negative results for
electric dipole moment (EDM) measurements. Nevertheless,
non-negligible effects of the complex phases can be obtained
\cite{IbrahimNath00a,IbrahimNath00b} even if only small phases that do
not violate the 
EDM-bounds are considered. On the other hand, CP invariance is not a
critical symmetry for the magnetic moment, in contrast to e.g.\ chiral
invariance. Therefore, CP-violating phases do not lead to new
enhancement factors, and the largest SUSY contributions are
naturally obtained for real parameters \cite{MartinWells}. 

The situation is different for generation mixing in the slepton
sector. If the smuons can mix with staus, it is possible to obtain
contributions where the chirality is flipped at a stau-line instead of
a smuon-line and thus contributes a factor $m_{\tau}$ instead of
$m_\mu$ \cite{Moroi1L}. An example is provided by diagram (N1), where
the $\tilde{\mu}_L$--$\tilde{\mu}_R$ insertion is replaced by a
$\tilde{\tau}_L$--$\tilde{\tau}_R$ insertion. Due to the large
enhancement factor $m_\tau/m_\mu$, even a small smuon--stau mixing can
lead to sizable corrections to $\amuSUOL$ \cite{Moroi1L}.

\subsection{Two-loop contributions}

\subsubsection{Chargino/neutralino-loop contributions}

The two-loop contributions from closed chargino/neutralino loops have
the same linear $\tan\beta$-dependence and $1/M_{\rm
  SUSY}^2$-suppression as the one-loop contributions, see
(\ref{chaneuapprox}). The detailed parameter-dependence, however,
shows interesting differences, as shown in \cite{HSW04}. 

In contrast to the one-loop contributions, the two-loop
chargino/neutralino diagrams are independent of the smuon mass
parameters. They only depend on $\mu$, $M_{1,2}$, and the Higgs boson
mass $M_A$. Moreover, the dependence on these parameters is quite
straightforward, while the $\mu$-dependence at the one-loop level is
very complicated due to the different behaviour of diagrams (C), (N1)
in figure \ref{fig:massinsertiondiagrams}, for example.

\begin{figure}[tb!]
\mbox{\epsfbox{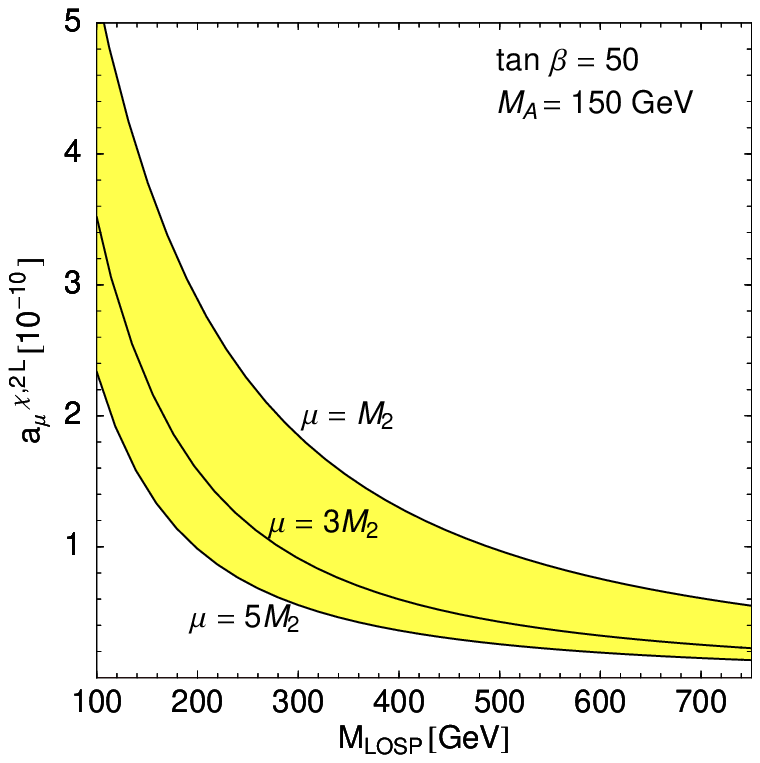}
\psfrag{MSl1000TB50}{\footnotesize \hspace{-1em}$m_{L,R,\tilde{\mu}}
  = 1 $ TeV, $\tan\beta = 50$}
\begin{picture}(220,230)(0,-10)\epsfxsize=7.7cm\epsfysize=7.7cm
 \put(000,000){\epsfbox{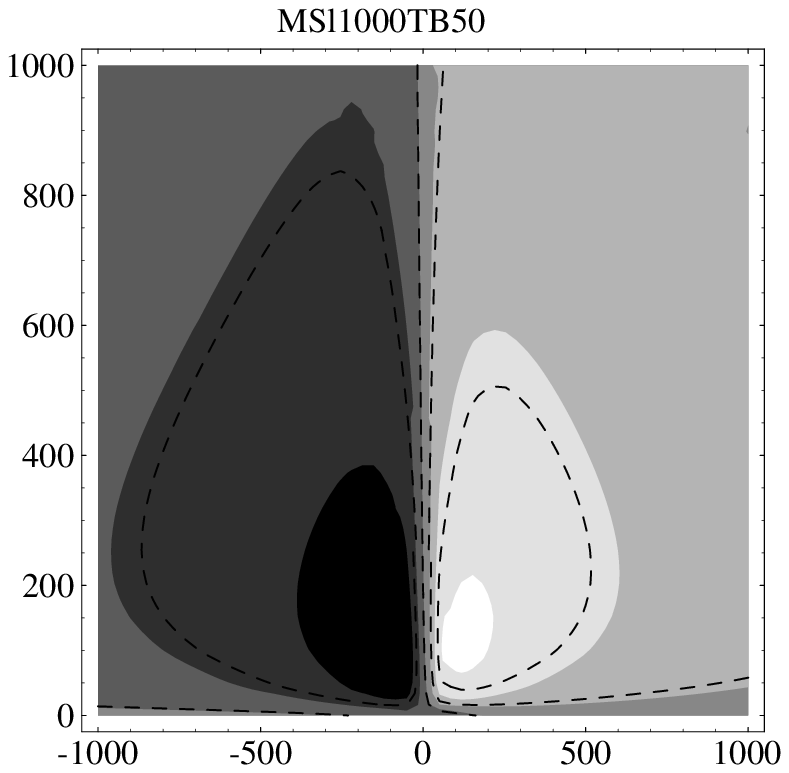}}
\put(005,212){\footnotesize $M_2$ [GeV]}
\put(105, -5){\footnotesize $\mu$ [GeV]}
\put(133,075){\footnotesize $<1\sigma$}
\put(160,160){\footnotesize $1-2\sigma$}
\put(119,188){\footnotesize $\longleftarrow2-3\sigma$}
\put(045,180){\footnotesize $3-4\sigma$}
\end{picture}}
\null\qquad\hfill(a)\quad\hfill\hfill\qquad(b)\ \hfill\null
\caption{\label{fig:chaneuplot}
Panel (a): $\amu^{(\chi VH),(\chi VV)}$ as a function of 
the mass of the lightest observable supersymmetric 
particle $M_{\rm LOSP}=$min($m_{\chi^\pm_1}$, $m_{\chi^0_2}$), for
$\tan\beta=50$ and $M_A=150$ GeV and for three different ratios
$\mu/M_2=1$, $3$, $5$ (from top to bottom). Panel (b): Contours of
$\amuSUOL+\amu^{(\chi VH),(\chi VV)}$ (solid border) and $\amuSUOL$
alone (dashed line) in the $(\mu$--$M_2)$-plane for $M_A=200$ GeV,
$m_{L,R,\tilde{\mu}}=1000$ GeV and
  $\tan\beta=50$. The contours are at $(24.5,15.5, 6.5, -2.5, -11.5,
  -20.5)\tunit$. Likewise, the colours of the fully draws areas
  correspond to the following values: white: $>24.5\tunit$, lightest
  grey: $(15.5\ldots24.5)\tunit$,\ldots, black: $<-20.5\tunit$. The
  plot has been taken from \cite{HSW04}, and the contours and regions
  correspond to $\Delta\amu(\mbox{exp}-\mbox{SM})$ as taken in that
  reference.}
\end{figure}

Figure \ref{fig:chaneuplot}~(a) shows the result of the two-loop
chargino/neutralino contributions $\amu^{(\chi VH),(\chi VV)}$  
for three different ratios $\mu/M_2=1$, $3$, $5$ as a
function of $M_{\rm LOSP}$, defined here as min($m_{\chi^\pm_1}$,
  $m_{\chi^0_2}$). The other parameters are chosen as $\tan\beta=50$
and $M_A=150$ GeV, and
the bino mass parameter $M_1$ is determined by $M_2$ via the GUT
relation. For $M_{\rm LOSP}>100$ GeV, contributions of up to $5\tunit$
can be obtained. This is a lot smaller than the largest possible
one-loop contributions, but it can be a significant correction if the
one-loop contributions are suppressed by heavy smuons and sneutrinos.

This can be seen immediately by comparing the two-loop
contributions in figure~\ref{fig:chaneuplot}~(a) with the one-loop
contributions in the blue region in figure~\ref{fig:oneloopplot}~(b),
where $m_{L,\tilde{\mu}}=5M_2=5\mu$. In this case the two-loop
correction can amount to more than 20\% of the one-loop result.

The result for a given $M_{\rm LOSP}$ is
largest if $\mu$ and $M_2$ are equal. There is no enhancement for
larger $\mu$ similar to the one-loop diagram (N1). Instead, increasing
$\mu$ to $\mu=5M_2$ leads to a suppression of the result by a factor
$2\ldots3$; for even larger ratios than the ones displayed in the
figure, the result becomes even smaller. 

For larger values of $M_A$, the result is suppressed as well. We have
checked that the results in figure \ref{fig:chaneuplot}~(a) would be
smaller by a factor $1.5$ (for large $M_{\rm LOSP}$) to $2.5$ (for
small $M_{\rm LOSP}$) for $M_A=300$ GeV instead of 150 GeV. This is in
agreement with the analysis of the $M_A$-dependence in \cite{HSW04}
for a wider selection of values for $\mu$ and $M_2$.

Figure  \ref{fig:chaneuplot}~(a) does not contain the case $\mu<M_2$.  It
is straightforward to show from the exact expression
(\ref{amuchigammaphi}) that the dominant contribution $\amu^{(\chi
   \gamma H)}$ is symmetric
under the exchange $\mu\leftrightarrow M_2$. Therefore the behaviour
for $\mu<M_2$ is very similar to the one for $\mu>M_2$ and does not
need to be analyzed separately.

The importance of the two-loop corrections is also visible in figure
\ref{fig:chaneuplot}~(b), which has been taken from \cite{HSW04}. 
Here the one-loop contributions and the sum of one- and two-loop
corrections are shown in a contour plot in the
$(\mu$--$M_2)$-plane. The  smuon mass parameters are
heavy, $m_{L,R,\tilde{\mu}}=1000 $ GeV, and $\tan\beta=50$, $M_A=200$
GeV. The white and lightest grey regions correspond to the
$1\sigma$-region around the experimentally preferred value (the value
$24.5(9.0)\tunit$ used for the plot is very similar to the one quoted
in (\ref{expSM})). For the chosen
parameters, this region extends up to $\mu$, $M_2\leq600$ GeV. However,
it is clearly visible that if the two-loop contributions were
neglected, the contours would shift considerably, and the $1\sigma$
region would extend only up to $\mu$, $M_2\leq500$ GeV.

\subsubsection{Sfermion-loop contributions}

The two-loop contributions from sfermion-loop diagrams can be even larger
than the ones from chargino/neutralino-loop diagrams. As discussed in
\cite{ArhribBaek,ChenGeng,HSW03}, however, and as
the discussion in section \ref{sec:massinsertions} and 
formulas (\ref{stopapprox}) , (\ref{sbotapprox}) show, these large
contributions arise only in special corners of parameter space. The
stop-loop diagrams become large if the stop mixing is large, one stop
is light,  $\mu$ is very large and $M_A$ is small. Similarly,
sbottom-loop diagrams become large for large sbottom mixing and if
$A_b/M_A$ is very large.

Such parameter regions are quite restricted for several
reasons. Primarily, small stop or sbottom masses in conjunction with
large $\tan\beta$ can have dramatic effects on the SUSY predictions
for the lightest Higgs boson mass and $b$-decays and are therefore
constrained by experimental limits on these observables. Furthermore,
large $\mu$ is restricted by the requirement that the sbottom mass
eigenvalues are positive, and by naturalness arguments due to its
appearance in the Higgs potential. 

\begin{figure}[tb!]
\mbox{\epsfbox{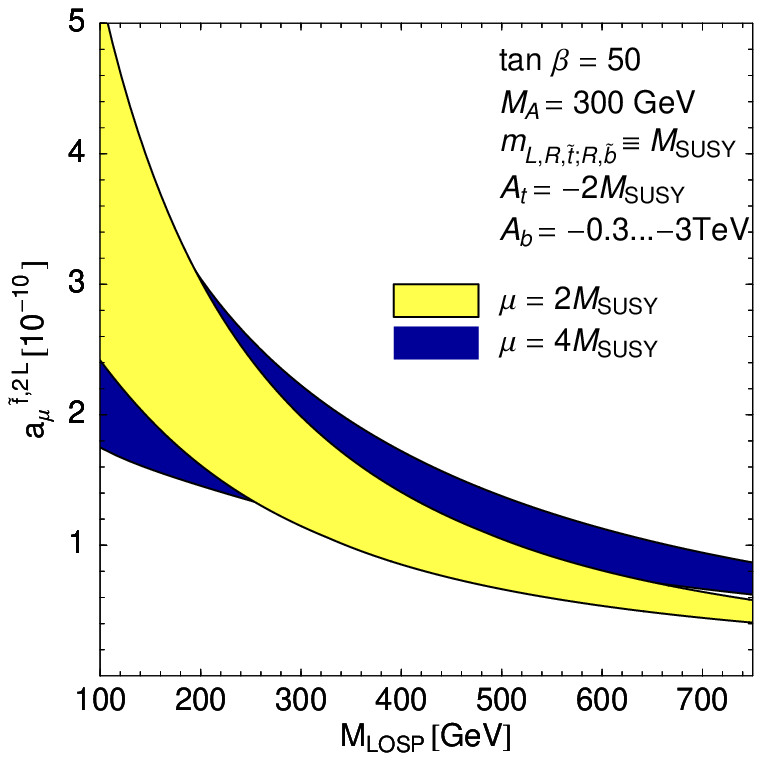}
\begin{picture}(220,220)(0,-3)
\put(0,-10){\epsfxsize=7.7cm\epsfysize=7.7cm\epsfbox{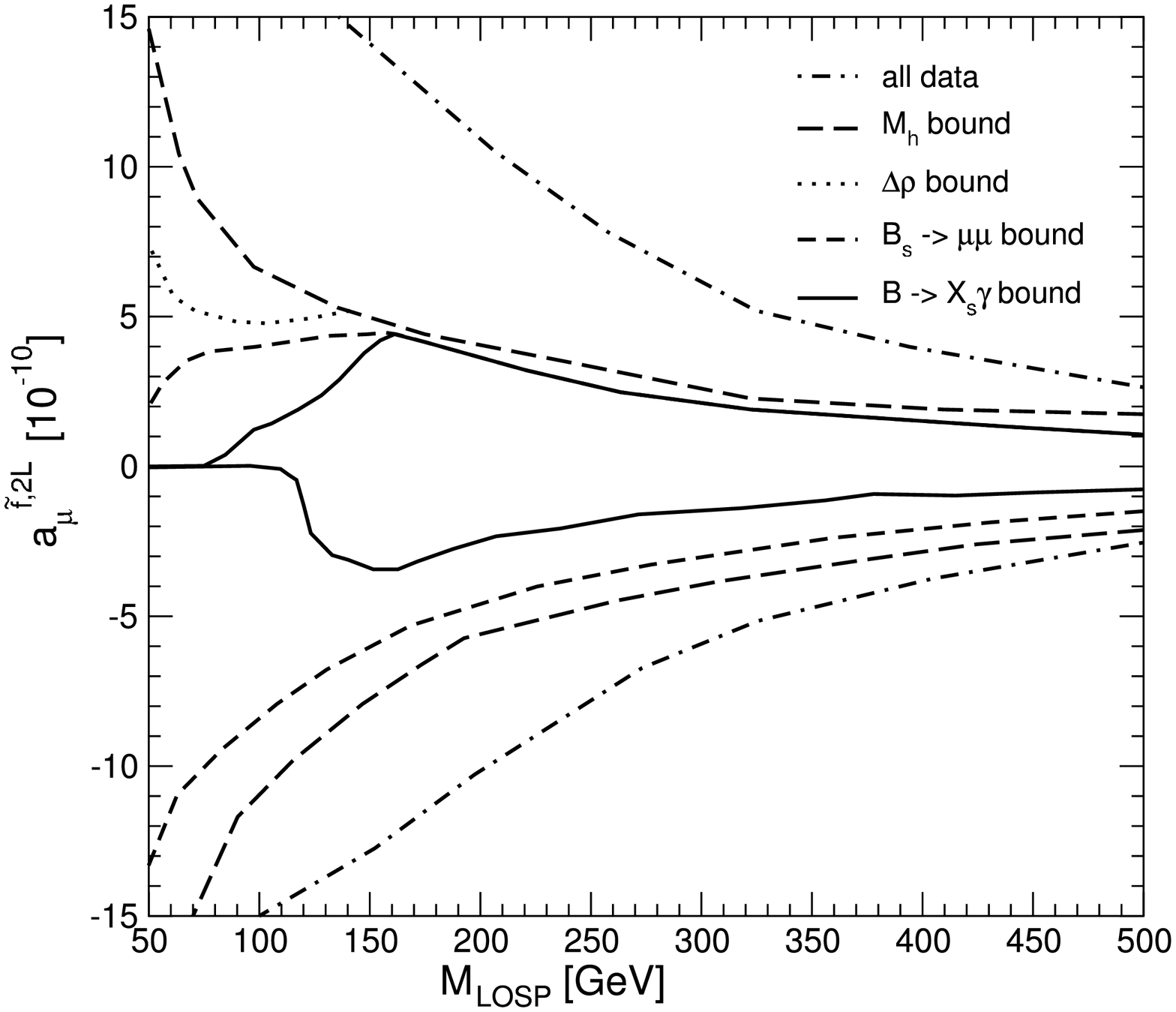}}
\end{picture}}
\null\qquad\hfill(a)\quad\hfill\hfill\qquad(b)\ \hfill\null
\caption{\label{fig:sfplot}
Panel (a): $\amusfTL$ as a function of 
the mass of the lightest observable supersymmetric 
particle $M_{\rm LOSP}=$min($m_{\tilde{t}_1}$, $m_{\tilde{b}_1}$), for
the parameter values given in the plot. The light yellow (dark blue)
region corresponds to $\mu=-A_t=2M_{\rm SUSY}$ ($\mu=-2A_t=4M_{\rm
  SUSY}$), where $M_{\rm SUSY}$ is the common sfermion mass parameter,
and $A_b=A_\tau=300\ldots3000$ GeV.
Panel (b): Maximum values of $\amusfTL$ obtained in a parameter scan
in the region (\ref{parameterscan}) if the experimental constraints
described in the text are incrementally applied. The plot has been
taken from \cite{HSW03}.}
\end{figure}

Figure \ref{fig:sfplot}~(a) shows the results of the two-loop
contributions from sfermion-loop diagrams as a function of the
lightest sfermion mass in two particular parameter scenarios.  The 
parameters have been chosen such that they are in agreement with
experimental bounds
and that the results are sizable. The Higgs boson
mass $M_A=300$ GeV, $\tan\beta=50$, the sfermion mass parameters
$m_{L,\tilde{t}}$, $m_{R,\tilde{t}}$, $m_{R,\tilde{b}}$,
$m_{L,\tilde{\tau}}$, $m_{R,\tilde{\tau}}$
have been chosen equal to 
a common scale $M_{\rm SUSY}$. The trilinear parameter $A_t=-2M_{\rm
SUSY}$, which leads to maximal stop mixing and values of $M_h$,
evaluated using {\em FeynHiggs}
\cite{FH1,Heinemeyer:1998np,Degrassi:2002fi,FH3} including
higher-order 
corrections, which are in agreement with current bounds. 

The light yellow region of figure \ref{fig:sfplot}~(a) corresponds to
$\mu=-A_t=2M_{\rm SUSY}$ and thus to a situation where the lightest
stop and sbottom masses are approximately equal. In the
dark blue region,  $\mu=-2A_t=4M_{\rm SUSY}$, and thus the lightest
sbottom is significantly lighter than the lightest stop. In both
regions, $A_b=A_\tau=-300\ldots-3000$ GeV. The signs of the 
parameters are such that the sbottom and stop contributions add up
constructively. The lower borders of the yellow and blue bands
correspond to small $|A_b|$ and thus mainly to the pure stop-loop
contributions. The width of the bands is essentially due to the
sbottom-loop contributions, which are approximately linear in $A_b$. 

The larger value of $\mu$ in the blue region has two effects. On the
one hand, the lightest sbottom is lighter than the lightest stop, and
thus the sbottom-loop contributions have more relative
importance. Hence the blue band is wider than the yellow band.
On the other hand, since the stop-loop contributions increase linearly
with $\mu$, they are enhanced as well in spite of the heavier stops,
and hence the blue band lies at higher values of $\amusfTL$. 

The upper borders of the two bands almost exhaust the limits found in
\cite{HSW03} for the largest possible values of $\amusfTL$.
Figure \ref{fig:sfplot}~(a), taken from \cite{HSW03}, shows these
largest possible values depending on which experimental constraints
are taken into account. The allowed parameter range is
\begin{eqnarray}
\fl\quad
\tan\beta=50,\quad M_{\rm SUSY}\leq1
\mbox{ TeV},\quad
|\mu|, |A_{t,b}|\leq3 \mbox{ TeV},\quad
 150 \mbox{ GeV} \leq M_A \leq1
\mbox{ TeV}
\label{parameterscan}
\end{eqnarray}
with common sfermion mass parameters
$m_{L,\tilde{t}}=m_{R,\tilde{t}}=m_{R,\tilde{b}}=m_{L,\tilde{\tau}}=
m_{R,\tilde{\tau}}=M_{\rm SUSY}$
and $A_\tau=A_b$.
The experimental constraints are $M_h>106.4$ GeV, $\Delta\rho^{\rm
  SUSY}<4\times10^{-3}$, BR$(B_s\to\mu^+\mu^-)<1.2\times10^{-6}$,
$|\mbox{BR}(B\to X_s\gamma)-3.34\times10^{-4}|<1.5\times10^{-4}$,
corresponding to conservative bounds taking into account experimental
and theoretical uncertainties (see \cite{HSW03} and references therein). 

If all experimental constraints are ignored,
$\amusfTL>15\times10^{-10}$ is possible, in agreement with the results
of \cite{ArhribBaek,ChenGeng}. Taking into account the Higgs
boson mass limit reduces the maximum contributions drastically, and
if all constraints are taken into account $\amusfTL$ turns out to be
smaller than $5\times10^{-10}$ in the parameter region
(\ref{parameterscan}).\footnote{For very low sfermion masses the
  parameter scenarios used in figure \ref{fig:sfplot}~(a) violate the
  experimental bounds, which is why the results displayed
  in figure \ref{fig:sfplot}~(a) are larger than the limits found in
  figure \ref{fig:sfplot}~(b) at $M_{\rm LOSP}\approx100$ GeV.}

Significantly larger values of these contributions can be obtained if
the sfermion mass parameters are non-universal. In particular, if the
ratio of $m_{R,\tilde{b}}$ and $m_{R,\tilde{t}}$ is either very large
or very small,
values of $\amusfTL>15\times10^{-10}$ can be in agreement with
all experimental constraints \cite{HSW03}. We will come back to this
possibility in section \ref{sec:scans}, where general scans of the
MSSM parameter space are described.

\subsection{Results for SPS benchmark points}

After having discussed the general parameter dependence of the
individual SUSY contributions to $\amu$, we present here the results
obtained in the Snowmass Points and Slopes (SPS) benchmark points
\cite{SPSDef}. These results provide an 
overview and reference of the SUSY contributions that can be expected
in various well-motivated and often considered parameter
scenarios. Furthermore, we use these results to assess the relative
importance of the individual one- and two-loop contributions.

\begin{table}
\caption{\label{SPSTable}Results of the SUSY contributions to $\amu$
  in units of $10^{-10}$
  for the SPS benchmark parameter points. The one-loop contributions
  include the two-loop QED-logarithms (\ref{amutllogs}). The SUSY
  two-loop corrections to SM one-loop diagrams  have
  been split up as 
$\amuSUTLa=\amu^{(\chi\gamma H)}+\amu^{(\tilde{f}\gamma H)}+
\amu^{\rm SUSY,\chi+\tilde{f},rest}+\amu^{\rm SUSY,ferm+bos,2L}$
  into the photon-loop contributions, the remaining
  chargino/neutralino and sfermion contributions, and the   bosonic
  and fermionic contributions.
}
\lineup
\begin{tabular}{@{}llcccc}
\br
SPS Point \ \null & $\amuSUOL$(improved)\hspace{-1em} 
&
\null\quad$\amu^{(\chi\gamma H)}$\quad\null
&
\null\ $\amu^{(\tilde{f}\gamma H)}$\ \null
&
$\amu^{\rm SUSY,\chi+\tilde{f},rest}$
&
$\amu^{\rm SUSY,ferm+bos,2L}$
\\
\mr
SPS 1a& \qquad 29.29 &  0.168 &  0.029 &  0.056 & 0.267 \\
SPS 1b& \qquad 31.84 &  0.273 &  0.044 &  0.106 & 0.222 \\
SPS 2& \qquad \01.65 &  0.032 &  \-0.002 &  0.027 & 0.068 \\
SPS 3& \qquad 13.55 &  0.078 &  0.009&  0.029 & 0.187 \\
SPS 4& \qquad 49.04 &  0.786 &  0.085 & 0.288 & 0.349 \\
SPS 5& \qquad \08.59 &  0.029 &  0.135 &  \-0.046 & 0.153 \\
SPS 6& \qquad 16.87 &  0.125 &  0.015 &  0.044 & 0.230 \\
SPS 7& \qquad 23.71 &  0.236 & 0 &  0.089 & 0.282 \\
SPS 8& \qquad 17.33 &  0.163 &  \-0.001 &  0.062 & 0.211 \\
SPS 9& \qquad \0\-8.98 &  \-0.046 &  \-0.002 & \-0.018 &  0.115 \\
\br
\end{tabular}
\end{table}

Table \ref{SPSTable} shows the results of the known contributions to
$\amuSU$, split up according to (\ref{amuSUSYknown}). The QED-improved
one-loop and the two-loop contributions with photon exchange 
in the first line of (\ref{amuSUSYknown}) are listed explicitly.
The remaining contributions are combined into the ones from diagrams
with charginos, neutralinos or sfermions and the purely SM-like ones.

The SPS points 1a, 1b, 3, 6 correspond to
various minimal supergravity (mSUGRA)-scenarios with $\tan\beta$
between 10 and 30 and SUSY 
masses in the range $100\ldots1000$ GeV. They lead to predictions of
$\amu$ very close to the observed value (\ref{expSM}). The same is
true for points 7, 8, which correspond to gauge-mediated SUSY
breaking. The point SPS~2  does not fit so well to (\ref{expSM})
because it corresponds to the focus-point region in which sfermions
are very heavy. SPS~4,~5 involve very large/small $\tan\beta$,
respectively, and therefore yield too high/low values for $\amuSU$,
and SPS~9, which corresponds to anomaly-mediated SUSY breaking,
involves negative $(\mu M_{1,2})$ and thus leads to negative $\amuSU$.

The two-loop corrections are very small in all cases. In general, the
chargino- or sfermion-loop  contributions with photon exchange can be
the largest two-loop contributions, but in all SPS points they are
suppressed by moderate values of $\tan\beta$ and/or rather high
values of $M_A$. Particularly the sfermion-loop contributions are
suppressed in addition by the heavy stops and sbottoms in all points
except SPS~5.  The results show, however, that the photon-exchange 
contributions are larger than $\amu^{\rm SUSY,\chi+\tilde{f},rest}$ by
a factor $\approx3$, in agreement with the discussion in section
\ref{sec:SUTLa} and the error estimate (\ref{amuSUSYerror2}). 
Likewise, the two-loop corrections from SM-like diagrams with
fermionic or bosonic loops are in the expected range
$\approx(0.1\ldots0.3)\tunit$.

\section{Impact on SUSY phenomenology}
\label{sec:Impact}

In this section we assume that the MSSM is the correct description of
physics at the weak scale and above and interpret the muon $g-2$
measurement within this framework. We discuss the impact of the
result for $\amu$ on the SUSY parameter space and its relation to
other relevant observables.

The deviation $\Delta\amu(\mbox{exp}-\mbox{SM})$  is positive and
larger than the pure SM weak contribution, see (\ref{expSM}),
(\ref{gapweak}), (\ref{SMweak}). We have
seen before that the MSSM can easily accommodate this deviation,
preferably for a rather small SUSY mass scale and/or large $\tan\beta$
and a positive $\mu$-parameter. 
In the following, we  first focus
on the general MSSM and show that specific, quantitative upper and
lower bounds on SUSY masses can be derived from $\amu$. 
Then we compare $\amu$ to $b$-decays, the Higgs
boson mass and electroweak observables, and neutralino dark
matter. Several of these observables exhibit correlations with
$\amu$, while others lead to complementary information on SUSY
parameters. 

Parameter bounds as well as correlations between
observables become much more severe in more constrained models than
the MSSM. We discuss here the cases of minimal supergravity,
gauge-mediated SUSY breaking and anomaly-mediated SUSY breaking.
We  conclude the section with a full parameter scan in the general
MSSM that summarizes the current status of $\amu$ in SUSY.

\subsection{Constraints from $\amu$ on the general MSSM}

Already before the Brookhaven $g-2$ experiment, the muon magnetic
moment was an important observable in SUSY phenomenology. Since the SM
theory prediction agreed with the older CERN measurement of $\amu$
\cite{CERNfinal}, it was possible to derive lower bounds on SUSY
masses
\cite{Fayet1L,Grifols1L,Ellis1L,Barbieri1L,Kosower1L,Yuan1L,Romao1L,Vendramin1L,Abel1L,Lopez1L,Chatt1L}.
Even after many of these bounds were superseded by LEP bounds, taking
into account $\amu$ remained complementary in corners of parameter
space where light SUSY particles could escape LEP detection
\cite{Carena96}.

In 2001, the publication of the Brookhaven result \cite{BNL3} caused a
lot of excitement since it showed a deviation of $43(16)\tunit$
($2.6\sigma$) from the SM theory prediction at the
time (which involved a sign error in the light-by-light
contributions). Many authors interpreted this result as a possible
signal for 
supersymmetry, and $\amu$ was studied in general SUSY models 
\cite{Czarnecki:2001pv,EverettKRW,FengMatchev,ChattNath01,KomineMY01,IbrahimCN01,EllisNO01,ArnowittDHS01,Choi01,MartinWells,BaerBFT01,Djouadi:2001yk},
with emphasis on correlations and implications for particular other
observables
\cite{Baltz:2001ts,Hisano:2001qz,Kim:2001eg,Cho:2001nf,Arnowitt:2001pm,Belanger:2001am,Dedes:2001fv},
and special scenarios or extended models were considered
\cite{Komine:2001hy,Baek:2001nz,Baek:2001kh,Enqvist:2001qz,Cerdeno:2001aj,Graesser:2001ec,Chacko:2001xd,Blazek:2001zm}.
A general conclusion was that significant constraints, in particular
upper mass limits could be derived even in the general MSSM.

The current deviation (\ref{expSM}) 
between the corrected SM result and the final experimental value is
$23.9\,(9.9)\tunit$. This is smaller but statistically almost as significant as
the one in 2001, and it still allows to stringently constrain the MSSM
parameter space. Without assuming any specific scenario of SUSY
breaking it is possible to derive both {\em lower} and {\em upper}
bounds on  the masses of SUSY particles. It is particularly
encouraging that not only one but several SUSY masses can be
bounded from above. 

\begin{figure}[tb]
\setlength{\unitlength}{1cm}
\begin{picture}(15.4,11.8)
\put(0,0){
\epsfxsize=7.6cm
\epsfbox{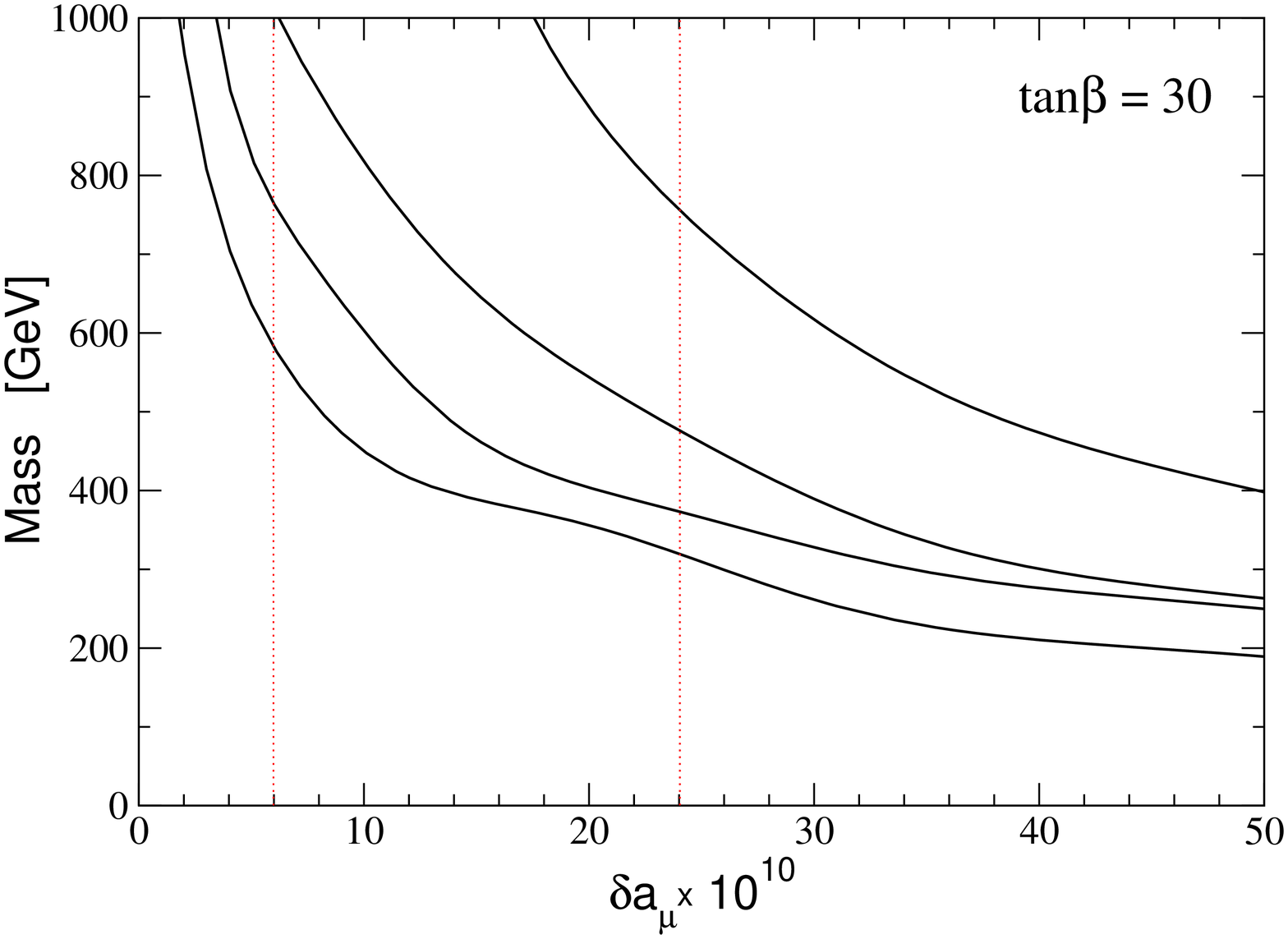}}
\put(7.9,0){
\epsfxsize=7.6cm
\epsfbox{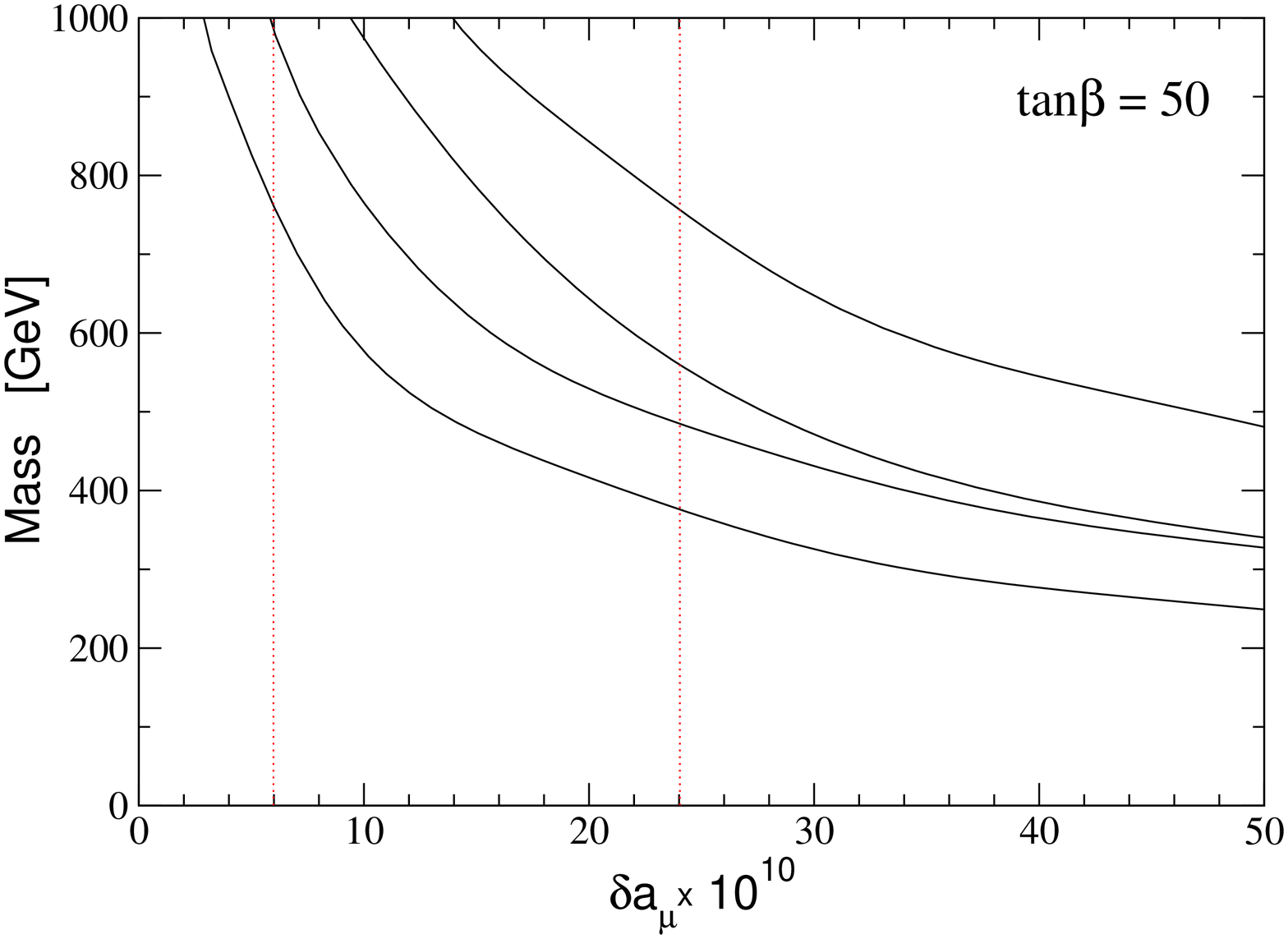}}
\put(0,5.8){
\epsfxsize=7.6cm
\epsfbox{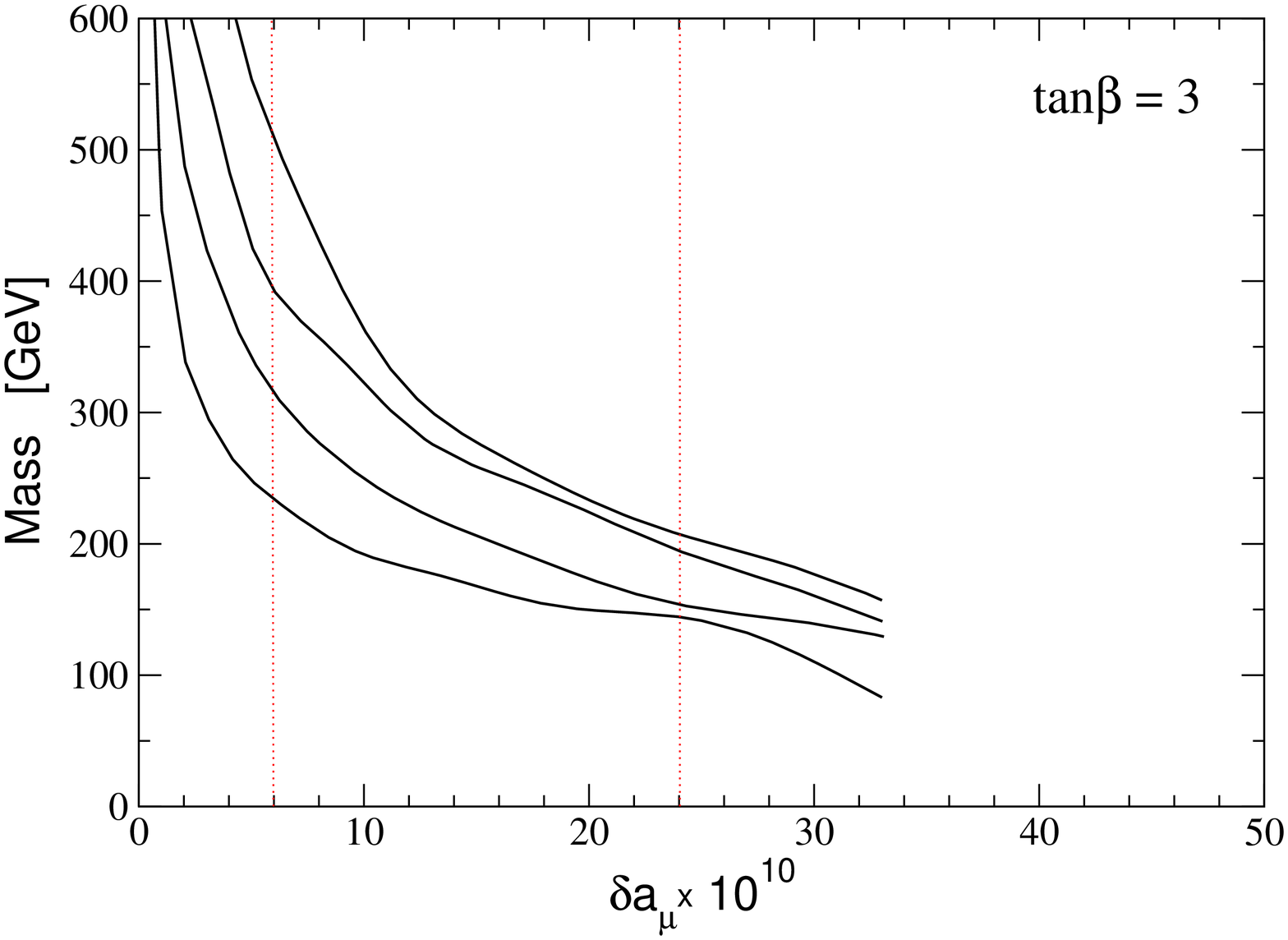}}
\put(7.9,5.8){
\epsfxsize=7.6cm
\epsfbox{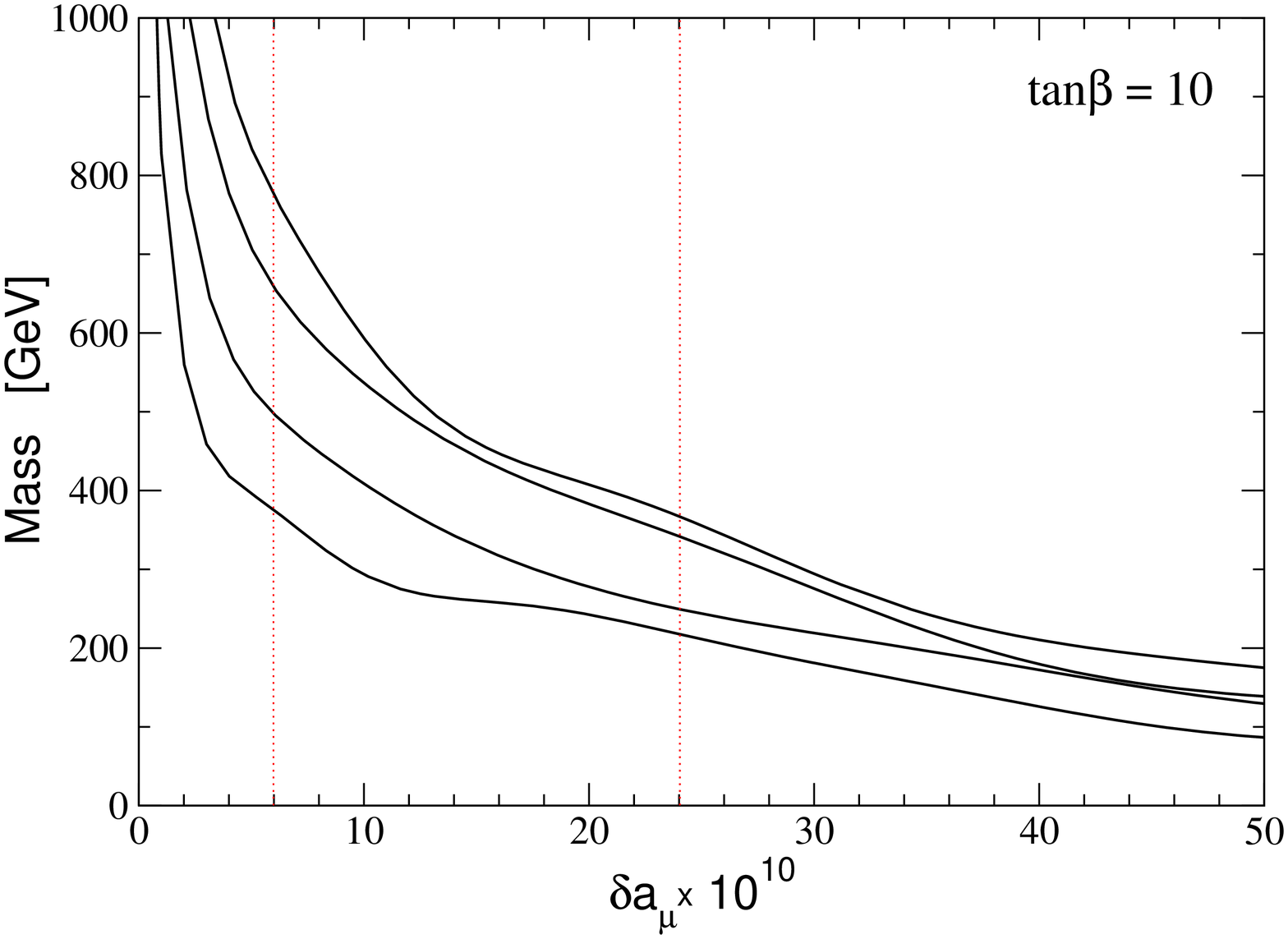}}
\end{picture}
\caption{\label{fig:ByrneKL02}
Upper bounds on the masses of the four lightest SUSY particles as a
function of $\delta\amu\equiv\amuSU$ for $\tan\beta=3,\ 10,\ 30,\
50$. The figure has been taken from \cite{ByrneKL02}, and the dotted
vertical lines at $\amuSU=6,24\tunit$ correspond to the $-1\sigma$
lines according to two different SM
theory evaluations (using or not using $\tau$-decay data) available at
the time of publication of \cite{ByrneKL02}. Incidentally, they are
close to the currently 
preferred value $23.9\tunit$ and the lower $2\sigma$ contour
$4.1\tunit$.}
\end{figure}

Figure \ref{fig:ByrneKL02} from \cite{ByrneKL02} shows the maximum
values of the four lightest SUSY particle masses, depending on the
value of $\amuSU$ and $\tan\beta$. These values are obtained
from a scan of the parameters $M_2$, $m_{L,\tilde{\mu}}$,
$m_{R,\tilde{\mu}}$, $\mu$ up to 2 TeV, assuming the GUT relation for
$M_1$. The bounds are independent of the identity of the SUSY
particles but they mainly restrict the chargino/neutralino and
smuon/sneutrino masses as these are most relevant for $\amuSU$. The
figure shows meaningful bounds even for the fourth lightest 
SUSY particle. For example, for $\tan\beta=10$, requiring that
$\amuSU>14\tunit$ ($\amuSU>4.1\tunit$), corresponding to a $1\sigma$
($2\sigma$) band in (\ref{expSM}), implies that four SUSY particles
are lighter than about 500 GeV (1 TeV) and two of them are even
lighter than 350 GeV (600 GeV).
The upper mass limits are tighter for smaller values of $\tan\beta$,
but all values of $\tan\beta$ allow for contributions $\amuSU$ larger
than $20\tunit$, so $\amu$ alone does not place a lower bound on
$\tan\beta$ \cite{ByrneKL02,MartinWells}. 

The upper mass limits are very promising for the search for SUSY
particles at the LHC and ILC. At a linear $e^+e^-$ collider, SUSY
particles can be pair-produced if they are lighter than half of the
center-of-mass energy. At the LHC, weakly interacting particles like
charginos, neutralinos and smuons can be best studied in cascade
decays of squarks, provided they are lighter than the squarks. In view
of figure~\ref{fig:ByrneKL02}, combined with lower squark mass limits
from Tevatron, these criteria might be satisfied for a number of SUSY
particles.

One might argue that it is too aggressive to require that $\amuSU$ is
within the $1\sigma$ or $2\sigma$ band of (\ref{expSM}), especially
given the difficulty in assessing the theoretical errors of the SM
hadronic contributions. Clearly, if e.g.\ a $3\sigma$ band is
admitted, zero SUSY contributions are allowed, and the upper limits on SUSY
masses disappear. However, interestingly even in a
``super-conservative'' approach significant parameter constraints can
be derived \cite{MaWe2}. The only requirement made in \cite{MaWe2} was
that $\amuSU$ falls into the interval 
\begin{eqnarray}
-36.8\tunit<\amuSU<89.9\tunit \qquad\mbox{\cite{MaWe2}},
\label{MaWeBand}
\end{eqnarray}
which was supposed to be a region that nobody could seriously argue
with. At the time of publication, it corresponded to the $5\sigma$
allowed region, now it corresponds roughly to the $6\sigma$
region. The bounds derived in \cite{MaWe2} from (\ref{MaWeBand}) can
therefore be regarded as definite bounds, independent of any future
theoretical or experimental developments. In addition to
(\ref{MaWeBand}), only the following assumptions have been made: all
SUSY parameters are real, $|\mu|>M_2$, $M_1=M_2/2$, smuon masses are
greater than 95 GeV, and $|A_\mu|/m_{\tilde{\mu}_2}<3$ in order to
avoid electric charge-violating minima. 

\begin{figure}[tb]
\mbox{\epsfxsize=7.7cm
\epsfbox{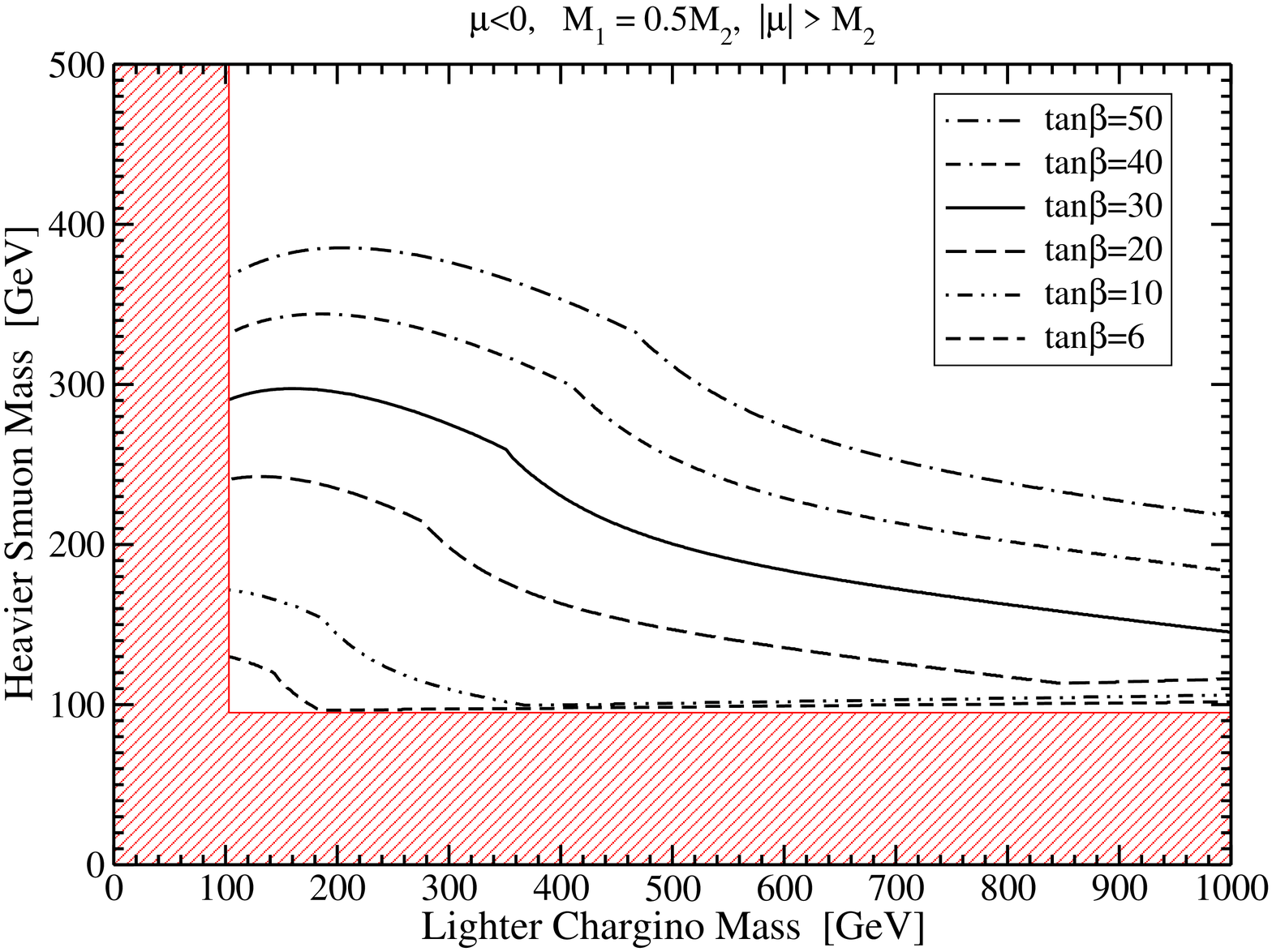}
\epsfxsize=7.7cm
\epsfbox{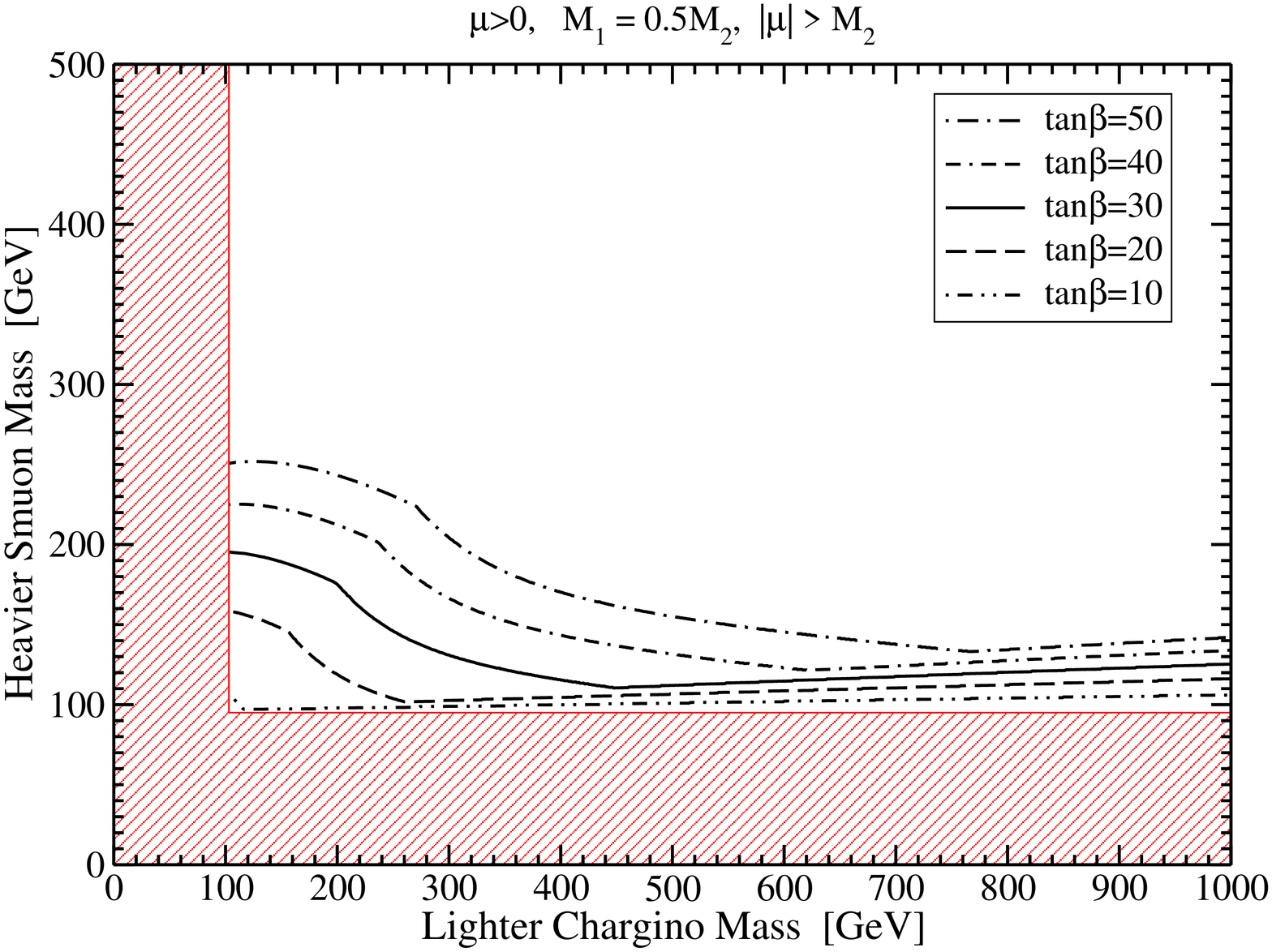}}
\caption{\label{fig:MaWe2}
Lower mass bounds derived from the requirement that $\amuSU$ lies in
the super-conservative interval (\ref{MaWeBand}). For each $\tan\beta$
and sign($\mu$), the region below the corresponding line is
excluded. The region constrained by direct searches is shaded. The
figure has been taken from \cite{MaWe2}.}
\end{figure}

Figure \ref{fig:MaWe2} shows an example of the (lower) mass bounds
that can be derived in this super-conservative approach in the plane
of the lighter chargino and heavier smuon mass. For each
$\tan\beta$, the region below the corresponding line is excluded. The
excluded regions are larger for $\mu<0$, reflecting the fact that
$\amu$ favours positive $\mu$. The excluded regions grow with
$\tan\beta$, but even for small $\tan\beta$,  $\amu$ excludes regions
of parameter space that are not excluded by any other experiment.

\subsection{General correlations with other observables}

Many observables can be related to $\amu$ in a meaningful way. Here we
discuss four particularly interesting examples: rare $b$-decays,
the neutralino dark matter density and detection rate, the Higgs boson
mass, and electroweak precision observables. Since the signs of the
parameters play an important role, we fix the convention $M_2>0$ for
simplicity in this and the following subsections. 

We do not
discuss in detail the possibility of lepton flavour violation. As
mentioned in section \ref{sec:LFV}, $\tilde{\mu}$--$\tilde{\tau}$
mixing can lead to substantial effects in $\amuSU$, and conversely the
measurement of $\amu$ implies bounds on lepton flavour-violating
parameters. Furthermore, the diagrams contributing to $\amu$ and
processes like $\tau\to\mu\gamma$ and $\mu\to e\gamma$ have a
similar structure and are correlated. Corresponding detailed studies
can be found in
\cite{Baek:2001kh,Graesser:2001ec,Chacko:2001xd,Baek:2002cc}.

\subsubsection{$B$-decays}

The rare $b$-decays $b\to s\gamma$ and $B_s\to \mu^+\mu^-$ are similar
to $\amu$ in two respects. They are loop-induced, and they involve a
chirality flip in the $b$--$s$ transition. Correspondingly, the SUSY
contributions to both branching ratios are enhanced by $\tan\beta$, in
the second case even $\propto\tan^6\beta$. In
the case of $b\to s\gamma$ also sign($\mu$) plays a crucial role as it
determines whether diagrams with $H^\pm$ exchange (always positive)
and diagrams with chargino or gluino exchange (the leading terms are
$\propto$ sign($\mu A_t$), sign($\mu M_3$), respectively) interfere
constructively or destructively. The current experimental results for
the two branching ratios are \cite{HFAG,CDFBSmumu}
\begin{eqnarray}
{\rm BR}(b\to s\gamma)&= (3.39^{+0.30}_{-0.27})\times10^{-4},\\
{\rm BR}(B_s\to\mu^+\mu^-)&<1.0\times10^{-7}\quad (95\% \mbox{ C.L.}).
\label{CDFBmumu}
\end{eqnarray}
Since both results are well in agreement with the SM theory
predictions, the possible SUSY 
contributions are bounded from above. As long as minimal flavour
violation is assumed, this implies for example that
destructive interference between the $H^\pm$ and $\chi^\pm$
contributions to $b\to s\gamma$, and thus the 
specific sign $\mu A_t<0$ is favoured (see
\cite{Hurth:2003vb,Dedes:2003kp} for reviews and references and
\cite{Carena06} for a recent thorough analysis of $b$-physics and
SUSY).

The interplay between rare $b$-decays and $\amu$ is particularly
strong in the framework of specific models such as minimal
supergravity or gauge-mediated SUSY breaking. Such models relate
squark and slepton masses and make specific 
predictions on the signs of $A_t$ and $M_3$ and thus also the 
sign of $\mu$ favoured by $b\to s\gamma$. Therefore an interesting
tension can arise between $b\to s\gamma$ and $\amu$. Both observables
favour a particular sign of $\mu$, but this might or might not be the
same, depending on the values of $A_t$ and $M_3$. And both depend on
$\tan\beta$ and SUSY masses, but $\amu$ prefers rather
large and $b\to s\gamma$ rather small SUSY contributions. 

\subsubsection{Neutralino dark matter density and detection rate}

\begin{figure}[tb]
\begin{center}
\mbox{\epsfxsize=9cm
\epsfbox{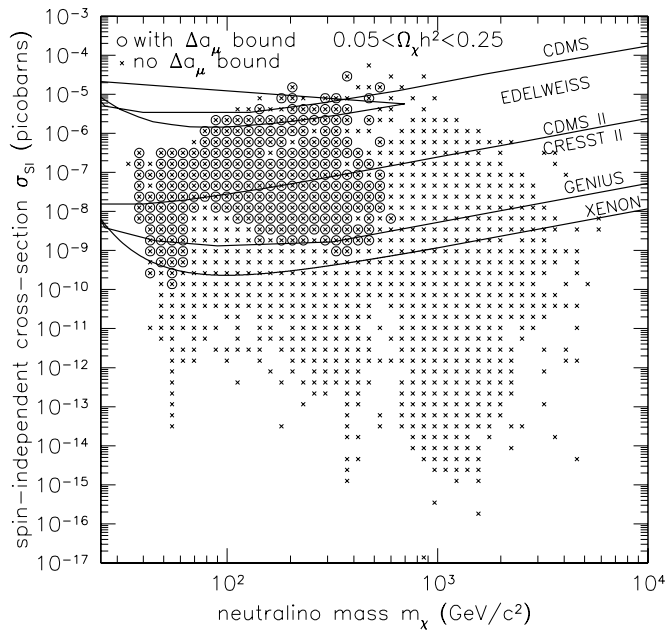}}
\end{center}
\caption{\label{fig:Baltz}
Neutralino--nucleon cross section versus neutralino mass, for a large
range of MSSM parameters (see text), taken from
\cite{Baltz:2002ei}. Only those points are shown for 
which the neutralino relic density could account for all dark matter,
but dropping this constraint does not affect the conclusions
\cite{Baltz:2002ei}. The circles denote points that satisfy the
constraint $7\tunit<\amuSU<51\tunit$, corresponding to a $2\sigma$
range at the time of publication.}
\end{figure}

If the lightest neutralino is stable it provides a promising candidate
for cold dark matter. In this case, two observables are of particular
interest: the relic density and the detection rate, governed by the
neutralino-nucleon cross section (for reviews and references see
\cite{Jungman:1995df,Bertone:2004pz}). The spin-independent contribution 
$\sigma_{\rm SI}$ to this cross section is yet another
chirality-flipping quantity, and 
it is also enhanced by large $\tan\beta$. Moreover, for positive
$\mu$ this cross section is typically larger than $10^{-10}$~pb, which
should be accessible to future detectors, while negative $\mu$ would
allow for cancellations that could strongly suppress
$\sigma_{\rm SI}$, such that neutralino dark matter detection would
be out of reach. Hence, the preference of $\amu$ for positive $\mu$
and not too small $\tan\beta$ has significant impact on the prospect
of dark matter detection
\cite{Drees:2000bs,Baltz:2002ei,Kim:2002cy,Ellis:2005mb,Baltz:2001ts,ChattNath01,ArnowittDHS01}.
Figure \ref{fig:Baltz} from \cite{Baltz:2002ei} shows the
spin-independent neutralino--nucleon cross section versus the
neutralino mass for a wide range of the MSSM parameters $\mu$, $M_2$,
$\tan\beta$, $M_A$, $A_{b,t}$, $m_0$, where $m_0$ is a common sfermion
mass parameter.
The circles denote points that satisfy the $\amu$
constraint. Imposing this constraint increases the minimum value of
$\sigma_{\rm SI}$ from $10^{-16}$~pb to $10^{-10}$~pb, significantly
improving the prospect for direct detection of galactic neutralinos. 

Under the assumption that standard cosmology (involving a fixed
cosmological constant and a certain amount of cold dark matter) is
valid and that neutralinos constitute the only component of 
cold dark matter, the neutralino relic density $\Omega_\chi$ is fixed
by astrophysical observations, in particular from WMAP \cite{WMAP3}. 
Fixing $\Omega_\chi$ effectively selects a one-dimensional
hyper-surface from the MSSM parameter space. However, the dependence
of $\Omega_\chi$ on the MSSM parameters is rather uncorrelated with
the one of $\amu$. Hence, the measurement of $\amu$ and the
increasingly precise determination of the cold dark matter density are
complementary probes of the MSSM parameter space.

\subsubsection{Lightest Higgs boson mass}

The lightest Higgs boson mass $M_h$ is an important observable since
it is tightly constrained in the MSSM. At tree level, $M_h$ must be
smaller than $M_Z$, but the current lower LEP-bound is $M_h>114.4$
GeV (95\% C.L.) \cite{LEPHiggs}.\footnote{This bound applies only if
  the MSSM Higgs boson $h$ is SM-like, which however is the case in
  the largest part of parameter space, in particular for $M_A\gsim150$
  GeV.}
Large quantum corrections can reconcile the MSSM prediction for $M_h$
with the lower bound, but this leads to severe constraints on the MSSM
parameter space. The main quantum corrections are enhanced by
$m_t^4\log(m_{\tilde{t}_1}m_{\tilde{t}_2}/m_t^2)$, and large
$\tan\beta$ and large stop mixing can lead to further enhancements
(see \cite{HHW} for a review and references). Therefore, like $\amu$,
the $M_h$-constraint favours larger $\tan\beta$, but unlike $\amu$ it
also favours larger SUSY masses, at least in the stop sector. 

Hence there
can be a certain tension between the $\amu$- and $M_h$-constraints,
particularly in models that connect stop and smuon masses.
This tension has become stronger recently owing to the latest
determination $m_t=171.4(2.1)$~GeV \cite{MTOP,LEPEWWG}. This value is lower
than previous ones, and it increases the tendency of the
$M_h$-constraint to favour rather large stop masses.
In \cite{Endo:2001ym}, the tension is illustrated for a class of SU(5) GUT
models by plots of the largest possible $\amuSU$ as a function of
$M_h$. The higher the Higgs boson mass, the lower the possible values
for $\amuSU$. However, the current bounds on $\amu$ and $M_h$ can
still be simultaneously satisfied in considerable parameter regions.

\subsubsection{Electroweak precision observables}

Finally, we briefly comment on electroweak precision observables,
particularly on $M_W$ and the effective weak mixing angle
$\sin^2\theta_{\rm eff}$. The present experimental values are
\cite{EWPOexp,LEPEWWG} 
\begin{eqnarray}
M_W^{\rm exp} &= 80.392(29)\ \mbox{GeV},\\
\sin^2\theta_{\rm eff}^{\rm exp} &= 0.23153(16).
\end{eqnarray}
The SM predictions for these observables sensitively depend on the
values of the SM input parameters, in particular on $m_t$ and $M_h$. 
For the experimentally preferred values of the input parameters,
the SM predictions agree with quite well $M_W^{\rm exp}$ and
$\sin^2\theta_{\rm eff}^{\rm exp}$. It has been noted
that the agreement with experiment can be even better in the MSSM (see
e.g.\ \cite{Arne} for an analysis that takes into account the most
recent experimental data). The SUSY contributions to both observables
are dominated by the quantity $\Delta\rho$, which is sensitive to the
breaking of isospin invariance and thus e.g.\ to the mass splittings in
the stop/sbottom sector. These SUSY contributions are enhanced for
smaller stop/sbottom masses and also depend on the chargino and
neutralino masses, but they are not particularly sensitive to
$\tan\beta$. Therefore, $\amu$ as well as the electroweak precision
observables have a tendency to favour not too heavy SUSY masses, and
it is fruitful to combine analyses of both kinds of quantities
\cite{ChoH01,Ellis:2004tc,Ellis:2006ix}.

\subsection{MSUGRA}

In the general MSSM, supersymmetry breaking is parametrized by a large
set of free parameters. In specific scenarios of supersymmetry
breaking these parameters find an explanation in terms of some
underlying physical mechanism. Typically, then, the MSSM parameters
can be related to a much smaller set of more fundamental
quantities. Such supersymmetry breaking scenarios are very predictive,
and parameter bounds implied by experimental constraints as well as
correlations between observables are much more stringent than in the
general MSSM.

The first scenario we discuss here is minimal supergravity (mSUGRA)
and the very similar constrained MSSM (CMSSM) (see e.g.\ \cite{Abel:2000vs}
and references therein). In this scenario supersymmetry breaking is
assumed to take place in a hidden sector and to be transmitted to the
observable sector via gravitational interactions. The K\"ahler
potential of the underlying supergravity theory is assumed to be
minimal, i.e.\ in particular to involve no generation-dependent
couplings. Furthermore, at the GUT scale $M_{\rm GUT}\approx
2\times10^{16}$~GeV the SM gauge interactions unify. The free
parameters of this model are
\begin{eqnarray}
m_0, m_{1/2}, A_0, \tan\beta, \mbox{sign}(\mu),
\end{eqnarray}
where $m_0$, $m_{1/2}$ and $A_0$ are the universal values of all
scalar mass,  gaugino mass, and $A$ parameters of the MSSM at the GUT
scale. The low-energy values of the MSSM soft-breaking parameters are
determined by renormalization-group running, and the value of $|\mu|$
is determined by the requirement that electroweak symmetry breaking
leads to the correct value of $M_Z$. (The mSUGRA scenario
also leads to a prediction of the gravitino mass, but in the present
review we will not make use of this prediction.)

The phenomenology of mSUGRA has been studied extensively, in
particular in view of the $\amu$ constraint (see
\cite{ChattNath01,EllisNO01,ArnowittDHS01,Djouadi:2001yk,Dedes:2001fv,Komine:2001rm,Chattopadhyay:2001mj,Chattopadhyay:2002jx}
for mSUGRA studies focussing on $\amu$ and
\cite{Baer:2003yh,Chattopadhyay:2003xi,Ellis:2003si,Baer:2004xx,Ellis:2004tc,Allanach:2005kz,Allanach:2006jc,Djouadi:2006be,Ellis:2006ix}
for recent general analyses of mSUGRA).

\begin{figure}[tb]
\begin{center}
\parbox[b]{7.7cm}{\epsfxsize=7.7cm
\epsfbox{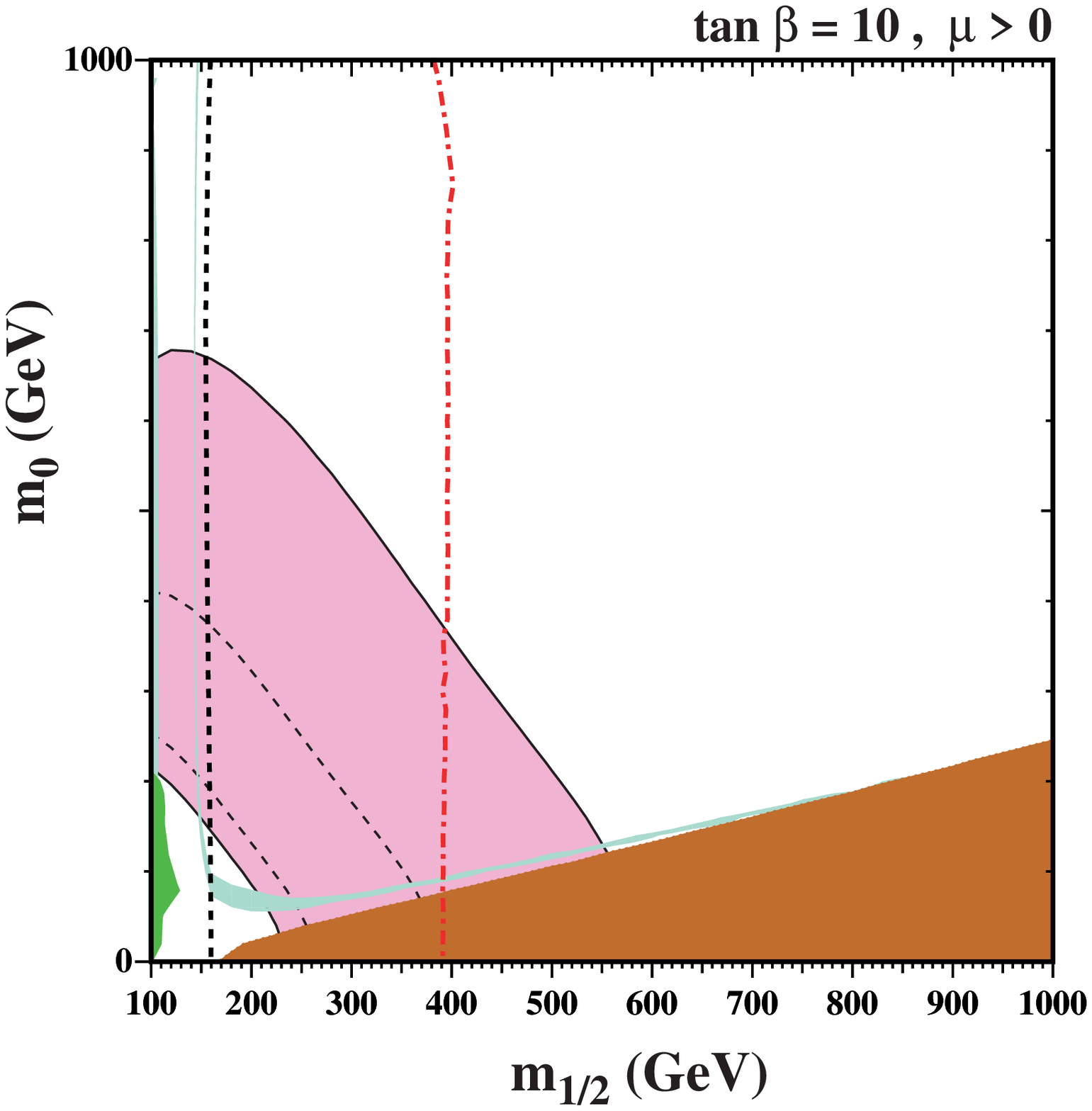}}
\parbox[b]{7.7cm}{
\epsfxsize=7.7cm\epsfysize=7.5cm
\epsfbox{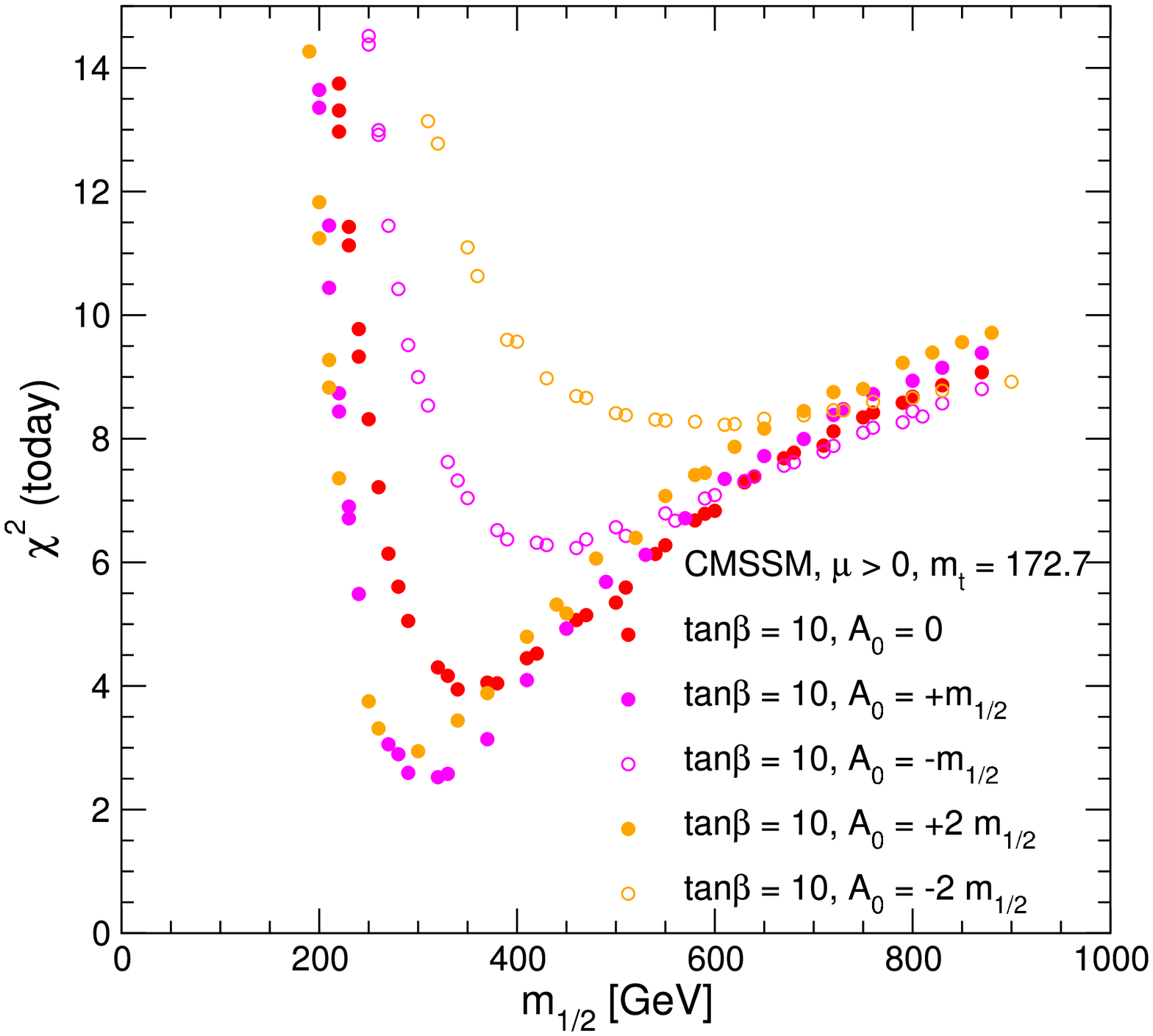}}
\null\qquad\hfill(a)\quad\hfill\hfill\qquad(b)\ \hfill\null
\end{center}
\caption{\label{fig:mSUGRA}
(a) The $m_{0}$--$m_{1/2}$ plane of the CMSSM for $\mu>0$,
  $\tan\beta=10$, $A_0=0$, taken from \cite{Ellis:2006vc}. The medium
  pink band and the thin black dashed lines show the $2\sigma$ and
  $1\sigma$ contours for $\amu$. The near-vertical lines correspond to
  $M_h=114$~GeV (red, dot-dashed), $m_{\chi^\pm_1}=104$~GeV (black,
  dashed); the medium green area is excluded by $b\to s\gamma$; and
  the light turquoise band satisfies the relic density constraint. In
  the dark red area the LSP is charged.\\
(b) Likelihood function $\chi^2_{\rm tot}$ for the observables $\amu$,
  $b\to s\gamma$, $M_h$, $M_W$, $\sin^2\theta_{\rm eff}$ in the CMSSM
  for $\tan\beta=10$ and various values of $A_0$. $m_0$ is chosen to
  yield the central value of the relic density constraint. This figure
  has been taken from \cite{Ellis:2006ix}.
}
\end{figure}

As an example, figure
\ref{fig:mSUGRA}~(a) from \cite{Ellis:2006vc} shows the mSUGRA
$m_0$--$m_{1/2}$ plane for $A_0=0$, $\tan\beta=10$, $\mu>0$ including
contours corresponding to the most 
important observables. The region preferred by $\amu$ at the $2\sigma$
and $1\sigma$ level is shown as the medium pink band and thin black
dashed lines. The Higgs boson mass bound $M_h>114$~GeV is satisfied to
the right of the near-vertical dot-dashed red line at
$m_{1/2}\approx400$~GeV, and to the right of the black dashed line
at $m_{1/2}\approx150$~GeV $m_{\chi^\pm}>104$~GeV. The plot displays
clearly the tension between $M_h$, which is increased by larger
$m_{1/2}$, and $\amu$, which prefers smaller SUSY masses, but there is
a non-vanishing region where both constraints are satisfied. For
larger $\tan\beta$ this overlap region grows. 

The small green region around $m_{1/2}\approx100$~GeV is excluded by
the decay $b\to s\gamma$. As mentioned before, the good agreement
between SM theory and experiment favours destructive interference
between the various SUSY contributions. In mSUGRA, the dominant
contributions are the ones with charged Higgs or chargino
exchange, and destructive interference requires $\mu A_t<0$. Since
mSUGRA predicts $A_t<0$ almost independently of $A_0$, positive $\mu$
is preferred by $b\to s\gamma$. It is a non-trivial success of mSUGRA
that both $\amu$ and $b\to s\gamma$ prefer the same sign of $\mu$, and
this is the reason why the green region in figure \ref{fig:mSUGRA}~(a)
is so small. However, for larger $\tan\beta$ the $b\to
s\gamma$-constraint excludes a larger part of the mSUGRA parameter space,
and this counterbalances the preference of $M_h$ and $\amu$ for larger
$\tan\beta$. Contours for the decay $B_s\to \mu^+\mu^-$ are not shown
in figure \ref{fig:mSUGRA}, but for large $\tan\beta\approx50$ this decay
becomes relevant, too. The current CDF limit (\ref{CDFBmumu}) already
begins to constrain the mSUGRA parameter space
\cite{Dedes:2001fv,Baek:2002wm}.

The constraint imposed
by the requirement that the lightest neutralino relic density
coincides with the cold dark matter density preferred by WMAP
\cite{WMAP3} at the $2\sigma$ level is shown by the light
turquoise band. This band is very narrow and thus allows to
fix e.g.\ $m_{0}$ as a function of $m_{1/2}$. But since the relic
density band is essentially orthogonal to the other regions
it is possible to satisfy all constraints simultaneously. 

All observables considered in figure \ref{fig:mSUGRA}~(a) have been
combined with the electroweak precision observables $M_W$ and
$\sin^2\theta_{\rm eff}$ in \cite{Ellis:2004tc,Ellis:2006ix}. 
In these references, $\chi^2$ fits within mSUGRA have been
performed, and remarkably there are mSUGRA parameter points that are
consistent with all constraints. Figure \ref{fig:mSUGRA}~(b) shows an
example of the total $\chi^2$ as a function of $m_{1/2}$ for
$\tan\beta=10$ and various values of $A_0$ ($m_0$ is always fixed by
the relic density constraint). The minimum $\chi^2_{\rm tot}=2.55$ is
obtained for $A_0=m_{1/2}=320$~GeV. 

In fact, a similarly good fit quality with a
$\chi^2_{\rm tot}<3$ can be achieved for all values of $\tan\beta$
between 10 and 50, and the preferred mass range for $m_{1/2}$ is always
between $300$ and $600$ GeV. For lower $\tan\beta$, the $M_h$- and
$\amu$-constraint are difficult to reconcile. For higher $\tan\beta$,
$b\to s\gamma$ is a serious constraint. It should be noted that the
significance of these tensions has grown recently since the
experimental value of the top quark mass has gone down
\cite{Djouadi:2006be,Ellis:2006ix}.
Nevertheless, the fact that mSUGRA fits well to all observables and
that the preferred mass range $m_{1/2}=300\ldots600$~GeV is rather low
is very encouraging also in view of forthcoming SUSY searches at
colliders.

\subsection{Gauge-mediated SUSY breaking}

Gauge-mediated SUSY breaking (GMSB) assumes that supersymmetry
breaking is mediated from a hidden sector to the observable sector by
gauge fields. In the simplest case there is an integer number
$N_{\rm mess}$ of such messenger gauge fields, and these gauge fields
have mass $M_{\rm mess}$ and form vector like {\boldmath{$5+\bar{5}$}}
representations of SU(5). A major advantage of GMSB over the mSUGRA
framework is that flavour universality naturally follows from the
symmetries of the messenger interactions (see \cite{Giudice:1998bp}
for a review of GMSB). 

The low-energy properties of GMSB are described
by the following parameters:
\begin{eqnarray}
M_{\rm mess}, N_{\rm mess}, \Lambda, \tan\beta, \mbox{sign}(\mu),
\end{eqnarray}
where the mass scale $\Lambda$ is related to the SUSY breaking scale
$\sqrt{F}$ by $\Lambda=F/M_{\rm mess}$. $\Lambda$ determines the
overall scale of the soft breaking parameters and can be traded for
one of them, e.g.\ $M_1$. At $M_{\rm mess}$ boundary conditions are
imposed on the MSSM soft parameters, and renormalization group running
is used to determine the MSSM spectrum at the electroweak scale.

\begin{figure}[tb]
\begin{center}
\mbox{\epsfxsize=7.7cm
\epsfbox{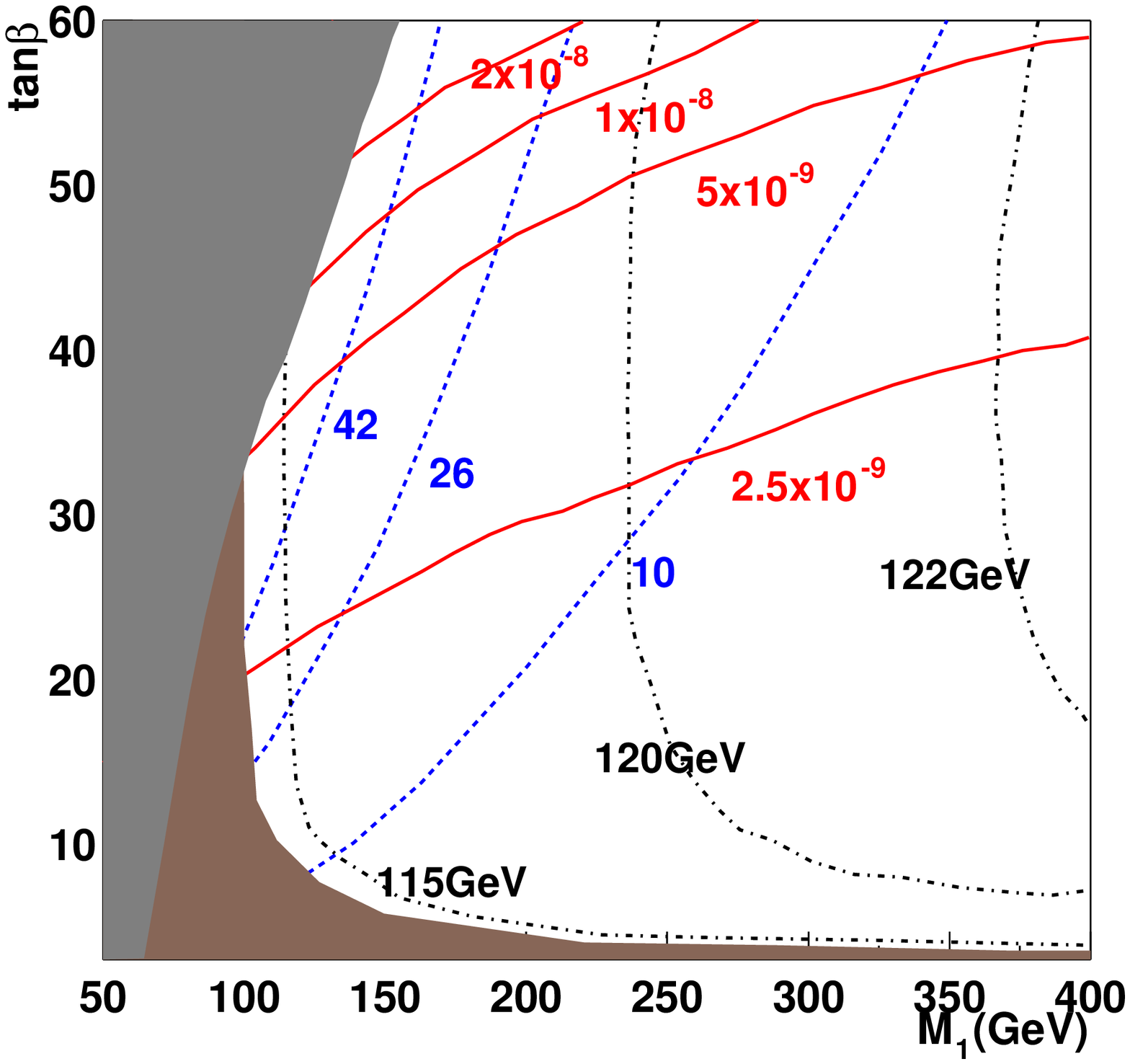}
\epsfxsize=7.7cm
\epsfbox{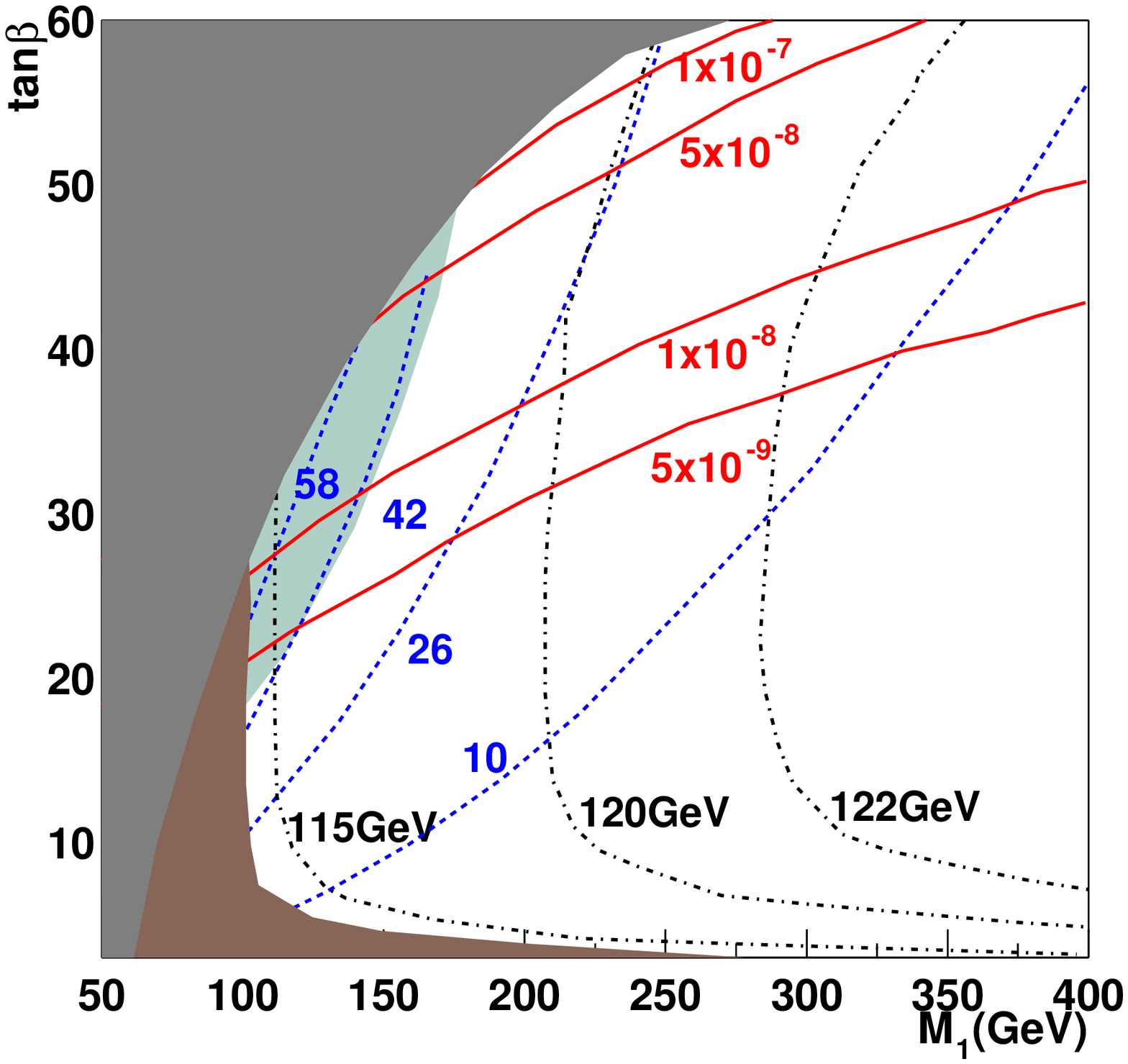}}
\null\qquad\hfill(a)\quad\hfill\hfill\qquad(b)\ \hfill\null
\end{center}
\caption{\label{fig:GMSB}
The $M_1$--$\tan\beta$ plane in GMSB for $\mu>0$ and for (a) $M_{\rm
  mess}=10^6$~GeV, $N_{\rm mess}=1$ and (b) $M_{\rm mess}=10^{15}$~GeV,
$N_{\rm mess}=5$, taken from \cite{Baek:2002wm}. 
The shown contours correspond to $\amu\times10^{10}=10,26,42,58$
(blue, dashed), to ${\rm BR}(B_s\to\mu^+\mu^-)$ (red, solid), and to
$M_h$ (black, dot-dashed). The light green region visible only in
panel (b) is excluded by $b\to s\gamma$. They grey and dark brown
regions are excluded by mass bounds on SUSY particles and the lightest
Higgs boson.
}
\end{figure}

The GMSB predictions for $\amu$ and related observables have been
studied in
\cite{Carena96,Gabrielli:1997jp,Gabrielli:1998sw,Mahanthappa:1999ta}
and compared to the predictions of mSUGRA and other 
models in \cite{Baek:2002wm,MartinWells,BaerBFT01}. Figure
\ref{fig:GMSB} from \cite{Baek:2002wm} shows 
two contour plots in the $M_1$--$\tan\beta$ plane for $\mu>0$ and for
$M_{\rm mess}=10^6$~GeV, $N_{\rm mess}=1$ and $M_{\rm mess}=10^{15}$~GeV,
$N_{\rm mess}=5$. In both cases, the
experimental result for $\amu$ can be easily accommodated by GMSB. For
any given $M_1$ and $\tan\beta$, $\amu$ is larger in panel (b) mainly
due to the relative suppression of the sfermion masses by
$1/\sqrt{N_{\rm mess}}$. However, curiously due to the other
constraints the highest values for $\amuSU$ can be obtained for
smaller $N_{\rm mess}$ \cite{MartinWells}.

Like in mSUGRA $\amu$ and $b\to s\gamma$ both favour the same,
positive sign of $\mu$. Moreover, in GMSB both stops and
charged Higgs bosons are typically rather heavy, especially for small
$M_{\rm mess}$. 
As a result, GMSB is not very significantly
constrained by the $b$-decays \cite{Baek:2002wm}. Conversely,
observation of non-standard effects in $b$-decays would seriously
constrain GMSB models.

\subsection{Anomaly-mediated SUSY breaking}

In anomaly-mediated SUSY breaking (AMSB) SUSY breaking takes place in
a hidden sector and is transmitted to the observable sector via the
anomalous breaking of superconformal (or super-Weyl) invariance
\cite{Randall:1998uk,Giudice:1998xp}. In this scenario the gaugino and
scalar mass soft parameters are related to the breaking of scale
invariance, expressed in terms of the gauge $\beta$ functions and the
anomalous dimensions of the matter fields. Most notably, the
AMSB contributions to the gaugino masses are
$M_i\propto-{b_i\alpha_i}$ with the one-loop coefficients
$(b_i)=(3,-1,-33/5)$  of the SU(3), SU(2) and U(1) $\beta$
functions. While these AMSB contributions to the soft SUSY breaking
terms are always present in hidden sector models, they are usually
subdominant. In the absence of any larger contributions, AMSB leads to
a very predictive framework that is qualitatively very different from
mSUGRA or GMSB. 

The parameters of minimal AMSB (mAMSB) are 
\begin{eqnarray}
M_{\rm aux}, m_0, \tan\beta, \mbox{sign}(\mu).
\end{eqnarray}
$M_{\rm aux}$ is a common scale for all soft parameters, and $m_0$ is
a universal additional scalar mass term that does not originate from
the super-Weyl anomaly. This term is necessary in order to avoid
tachyonic sleptons. 

\begin{figure}[tb]
\begin{center}
\mbox{\epsfxsize=7.7cm
\epsfbox{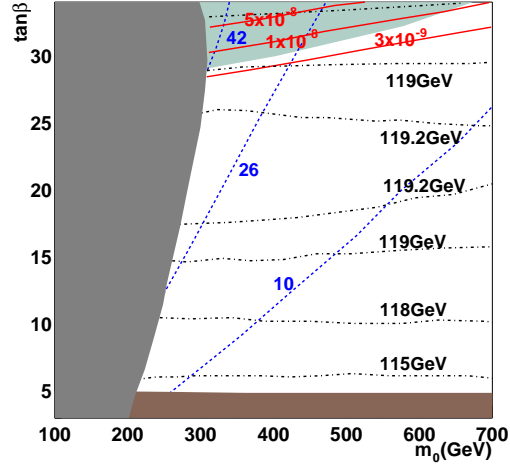}}
\end{center}
\caption{\label{fig:AMSB}
The $m_0$--$\tan\beta$ plane in mAMSB for $\mu>0$ and $M_{\rm
  aux}=50$~TeV, taken from \cite{Baek:2002wm}. 
The shown contours correspond to $\amu\times10^{10}=10,26,42$
(blue, dashed), to ${\rm BR}(B_s\to\mu^+\mu^-)$ (red, solid), and to
$M_h$ (black, dot-dashed). The light green region  is excluded by
$b\to s\gamma$. They grey and dark brown regions are excluded by mass
bounds on SUSY particles and the lightest Higgs boson.
}
\end{figure}

A crucial feature of AMSB is that the signs of $M_3$ and $A_t$
(in the convention that $M_2>0$) are reversed compared to the mSUGRA
and GMSB cases. As a result, the $b\to s\gamma$ constraint favours
negative $\mu$ and thus negative $\amuSU$. If the observed deviation
of the experimental from the SM theory value of $\amu$ would have
turned out to have a negative sign, AMSB would be favoured over
virtually all other models of SUSY breaking
\cite{FengMoroi,Chattopadhyay:2000ws}. However, since the deviation is
positive AMSB is disfavoured compared to other scenarios such as
mSUGRA or GMSB \cite{FengMatchev}. This clearly shows how combining
information on low-energy physics can discriminate between different
well-motivated SUSY breaking mechanisms.

Figure \ref{fig:AMSB} shows an example from \cite{Baek:2002wm} of the
$m_0$--$\tan\beta$ plane of mAMSB for $M_{\rm aux}=50$ TeV and
$\mu>0$. The region for $\tan\beta>30$ is almost entirely excluded by
the $b\to s\gamma$ constraint (light green). For low $\tan\beta$, the
Higgs boson mass bound and $\amu$ restrict the parameter
space. Nevertheless, for moderate values of $\tan\beta$ and $m_0$ it
is possible to consistently accommodate $b$-decays, $\amu$ and $M_h$
within mAMSB.

\subsection{MSSM parameter scan}
\label{sec:scans}

In this subsection we present a general,
model-independent MSSM parameter scan that summarizes the current
status of $\amu$ in SUSY. This scan shows the maximum results for
$\amu$ in the MSSM if all parameters are independently varied and 
the range of SUSY masses for which the MSSM can accommodate the
experimental result. The full set of one- and two-loop contributions
in (\ref{amuSUSYknown}) are taken into account, and the results are
compared with the current deviation between the experimental and SM
result (\ref{expSM}). The scan 
can be viewed as an update of the one presented in
\cite{FengMatchev}.

\begin{figure}[tb!]
\epsfbox{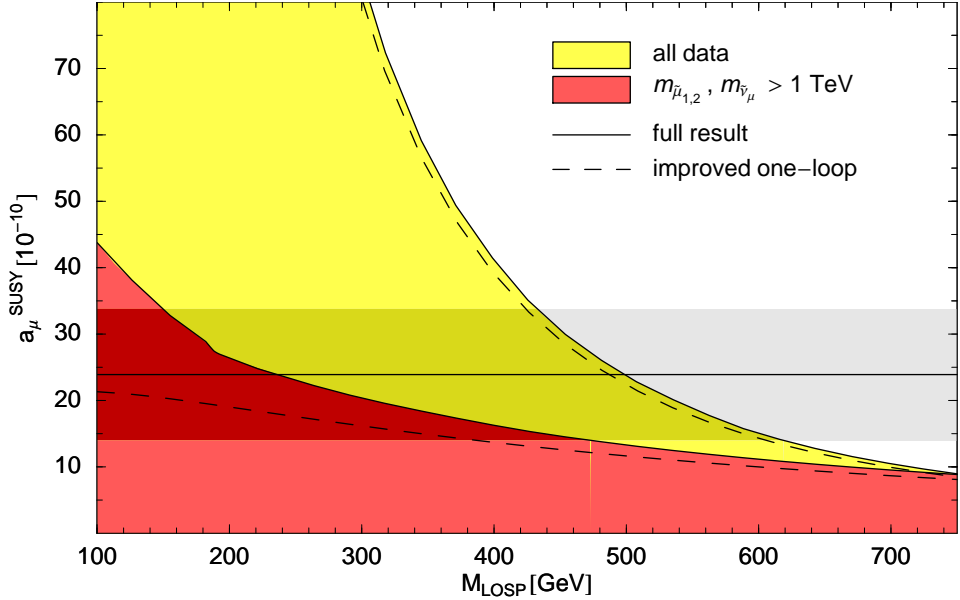}
\caption{\label{fig:scanTB50}
Allowed values of $\amuSU$ as a function of the mass of the lightest observable supersymmetric 
particle $M_{\rm
  LOSP}=$min$(m_{\tilde{\chi}^\pm_1}, m_{\tilde{\chi}^0_2},
m_{\tilde{f}_i})$, from a scan of the MSSM parameter space in the range
(\ref{scanrange}) and for $\tan\beta=50$. The $1\sigma$ region
corresponding to the deviation between experimental and SM values
(\ref{expSM}) is indicated. The light yellow region corresponds to all
data points that satisfy the experimental constraints from $b$-decays,
$M_h$ and $\Delta\rho$. In the red region, the smuons and sneutrinos
are heavier 
than 1 TeV. The dashed lines correspond to the contours that arise
from ignoring the two-loop corrections from chargino/neutralino- and
sfermion-loop diagrams.
}
\end{figure}

The MSSM parameters have been varied in the ranges
\begin{eqnarray}
\fl\quad
|\mu|, M_2, m_{L,\tilde{f}}, m_{R,\tilde{f}}, 
|A_f|
 \leq 3000\mbox{ GeV},
\quad
M_A=90\ldots3000\mbox{ GeV},
\quad
\tan\beta=50,
\label{scanrange}
\end{eqnarray}
where the upper limit is motivated by naturalness arguments, and the
lower limit on $M_A$ corresponds to the experimental limit. The
parameter $\tan\beta$ has not been varied because the essentially
linear $\tan\beta$ dependence of $\amuSU$ has been sufficiently
established before. All other
parameters have been varied independently, except that $M_1$ has been fixed
via the GUT relation, $A_\mu=0$,
$m_{L,\tilde{b}}=m_{L,\tilde{\tau}}$, $m_{R,\tilde{b}}=m_{R,\tilde{\tau}}$ and $A_b=A_\tau$. These
constraints do not have much impact on the maximum contributions
\cite{FengMatchev}. The 3rd generation sfermion parameters are
significantly restricted by experimental constraints on  $M_h$,
$\Delta\rho$ and 
$b$-decays. Only parameter points have been considered that
are in agreement with these constraints, according to the same
criteria as in the ``weak bounds'' of \cite{HSW03}. Note that the
precise values used in these constraints have not much influence on
the maximum results in figure   \ref{fig:scanTB50} as they affect
essentially only the two-loop sfermion contributions.

Figure \ref{fig:scanTB50} shows the results of the scan. The light
yellow region corresponds to the possible 
results for $\amuSU$, given by (\ref{amuSUSYknown}), as a 
function of $M_{\rm LOSP}$, if all parameters are varied in the range
(\ref{scanrange}). The red region corresponds to the situation that
the smuons and sneutrinos are heavy, 
\begin{eqnarray}
m_{\tilde{\mu}_{1,2}},\ m_{\tilde{\nu}_\mu}>1000\mbox{ GeV},
\end{eqnarray}
while charginos, neutralinos and stops and sbottoms can still be
light. A dedicated analysis of this situation is of interest since
heavy 1st and 2nd generation sfermions are sometimes considered as a
possible explanation of the absence of observable SUSY contributions to
flavour-changing neutral currents and CP-violating observables.

The dashed lines in figure \ref{fig:scanTB50} correspond to the
results if the genuine two-loop diagrams
are neglected and only the one-loop contributions and the logarithmic
QED-corrections (\ref{amutllogs}) are taken into account. 

The yellow region corresponds to all data points and thus to the
maximum possible values of $\amuSU$ compatible with the parameter
range (\ref{scanrange}) and the experimental constraints from
$b$-decays, $M_h$ and $\Delta\rho$. The SUSY contributions can
accommodate the observed result (\ref{expSM}) within 
$1\sigma$ for LOSP masses below about 600 GeV. For LOSP masses below
about 440 GeV, the SUSY contributions can be even too large, and 
thus the $\amu$-measurement significantly restricts the MSSM
parameter space in this low-mass region.

The red region of figure \ref{fig:scanTB50} shows that heavy smuons
and sneutrinos significantly suppress the maximum SUSY contributions
to $\amu$. Nevertheless, the contributions can be in the $1\sigma$
region (\ref{expSM}) for $M_{\rm LOSP}$ between about 150 and 470 GeV. 

We close the section with a couple of instructive but more technical
remarks. On the one hand we analyze which parameter regions give rise
to the maximum contributions found in figure \ref{fig:scanTB50}. On
the other hand we explain the reasons for the different behaviour of
the two-loop contributions in the red and yellow regions.

In the yellow region, corresponding to all data points, the maximum
values are obtained for $\mu=3000$ GeV, the upper border of the
allowed range, and for $M_2\approx m_{L,R,\tilde{\mu}}$. The reason
is that the one-loop bino-exchange diagram (N1) of
figure~\ref{fig:massinsertiondiagrams} is linear in $\mu$ and 
therefore dominant for large $\mu$.
In comparing our results with the ones of \cite{FengMatchev}, one
should note that since the maximum contributions are obtained at the
border of the allowed multi-dimensional parameter region, an unbiased
random generation of parameter points  as the one used in
\cite{FengMatchev} has difficulties in finding the maximum
contributions. Therefore the results displayed in figure 1 and
equation (3) of \cite{FengMatchev} slightly underestimate the maximum
contributions allowed by the employed parameter ranges. 
Since the maximum contributions are obtained for large $\mu$, the
two-loop chargino/neutralino contributions are negligible, but the
two-loop sfermion contributions shift the contour by about
5\%. Another consequence of the dominance of diagram (N1) 
is that the border of the yellow contour scales linearly with the
maximum value for $\mu$ used in scanning over the SUSY parameters.

The one-loop contributions in the red region are maximized for smuon
masses of 1 TeV and 
$\mu=M_2$. Since therefore $\mu$, $M_2$ and $M_A$ can be
simultaneously small, substantial two-loop
chargino/neutralino contributions are possible. Furthermore, the
two-loop sfermion contributions can be large as well since the stop
and sbottom mass parameters are not required to be universal. The
largest results in the red region are obtained if both
$m_{L,\tilde{b}}$ and $m_{R,\tilde{b}}$ are small but
$m_{R,\tilde{t}}$ is large. In this situation, the sbottom-loop
contribution can be larger than the results displayed in figure
\ref{fig:sfplot} without any violation of the experimental constraints
on $M_h$, $\Delta\rho$ or $b$-decays.

The significance of the two-loop corrections in the red region is
reflected in the fact that the contour shifts by more than 30\% in the
low-mass region if the two-loop corrections are neglected.

\section{Concluding remarks and outlook}

The final result of the Brookhaven $\amu=(g_\mu-2)/2$ measurement
shows a deviation of $23.9\,(9.9)\tunit$, corresponding to 2 ppm and
more than $2\sigma$, from the corresponding SM prediction based on
$e^+e^-$ data. The Brookhaven experiment  is the first magnetic dipole
moment measurement that is sensitive to physics at 
the electroweak scale, just like the previous CERN experiment was the
first to be sensitive to hadronic effects. But unlike the CERN
experiment, which confirmed the SM prediction of the hadronic effects,
the Brookhaven experiment would prefer electroweak effects that are
about $2.5$ times larger than predicted by the SM.

The SUSY contributions to $\amu$ are essentially proportional to
$\tan\beta$~sign$(\mu)/M_{\rm SUSY}^2$. Hence for moderate or large
$\tan\beta$ these contributions can easily be larger than the
electroweak SM contributions and thus constitute the origin of the
discrepancy between the experiment and the SM prediction. Furthermore,
in this case $\amu$ strongly favours positive $\mu$, which is a very
important piece of information for SUSY phenomenology. The qualitative
behaviour of $\amuSU$ can be well understood using the mass insertion
technique. The $\tan\beta$-enhancement arises in diagrams where the
necessary chirality flip occurs at a muon Yukawa coupling, either to a
Higgsino or Higgs boson, because this coupling is enhanced by
$1/\cos\beta\approx\tan\beta$ compared to its SM value. The
$\mu$-parameter mediates the transition between the two Higgs/Higgsino
doublets $H_{1,2}$, and this transition enhances diagrams because only
$H_1$ couples to muons while $H_2$ has the larger vacuum expectation
value. 

For a quantitative analysis, the MSSM prediction of $\amu$ has to be
known with an accuracy that matches the one of the SM prediction. This
goal has been achieved with the calculation of the full one-loop and
leading two-loop contributions, as given in
(\ref{amuSUSYknown}). Since the remaining theory uncertainty
is dominated by one particular class of unknown two-loop diagrams, an
even more satisfactory MSSM prediction could be obtained by evaluating
this missing class of diagrams.

If supersymmetry is assumed to exist, many non-trivial details about
the SUSY spectrum can be derived from $\amu$. For example, for
$\tan\beta=10$ four SUSY particles must be lighter than 500(1000)~GeV
at the $1\sigma(2\sigma)$ level. And even in a very conservative
interpretation, where a $5\sigma$ deviation in $\amu$ is tolerated,
significant lower bounds on the SUSY masses can be derived in a
model-independent way. These bounds cannot be obtained from any other
observable and establish $\amu$ as one of the most important indirect
probes of SUSY. 

There are several other observables that are relevant for SUSY
phenomenology and provide constraints on SUSY parameters, although
$\amu$ is unique in its largest discrepancy between experimental and SM
values. On a qualitative level, the relation between $\amu$ and these
other observables is the following. $\amu$ favours rather large
$\tan\beta$ and/or small SUSY masses and positive $\mu$.
If neutralinos constitute a component of dark
matter, the dark matter detection rate is enhanced by large
$\tan\beta$ and positive $\mu$. Therefore, the $\amu$ result is
encouraging for future dark matter detection experiments.

The preference of $\amu$ for rather light SUSY masses is supported by
the experimental results for the electroweak precision
observables $M_W$ and $\sin^2\theta_{\rm eff}$, which favour small but
non-zero SUSY contributions.
The preference of $\amu$ for not too small $\tan\beta$ is supported by
the constraint derived from the lower bound on $M_h$. However, the bound
on $M_h$ also favours large SUSY masses (particularly stop masses)
over light ones. Furthermore, the MSSM
parameter space for large $\tan\beta$ is significantly constrained by
the rare $b$-decays $b\to s\gamma$ and $B_s\to\mu^+\mu^-$. The decay
$b\to s\gamma$ shows a similar dependence on sign$(\mu)$ as $\amu$. 
This already constitutes a crucial test of minimal models. For
instance, in mSUGRA or GMSB, but not in AMSB, $\amu$ and
$b\to s\gamma$ favour the same sign of $\mu$.

Given the tension between all these observables it is non-trivial that
all constraints can be simultaneously satisfied, even in the simplest
but well-motivated models such as mSUGRA or GMSB. Mainly driven by
$\amu$, the parameter points satisfying all constraints prefer a
number of (though not necessarily all) SUSY particles to be
light. This result is clearly encouraging for the SUSY search at the
LHC and the ILC.

We have summarized the current status of $\amu$ and SUSY in figure
\ref{fig:scanTB50}, which shows the possible values of $\amuSU$,
compared with the observed deviation between experiment and the SM
prediction. The MSSM parameters have been scanned over in the range
allowed by all relevant constraints from other collider experiments,
and in the evaluation of $\amuSU$ all known one- and two-loop
contributions have been taken into account. The result confirms again
that SUSY can accommodate the $\amu$ result consistently with all other
constraints and that the preferred mass range is rather low. On a more
technical level, the figure also shows that two-loop SUSY effects can
be important.

In spite of the impressive current status, progress on $\amu$ is
important and will come both from the experimental and the 
theoretical side. Already a small improvement of the precision
could be sufficient to increase the discrepancy between the
experimental and SM values of $\amu$ to $4$--$5\sigma$ and thus to
establish the existence of non-SM contributions. Currently the SM
theory prediction has a larger uncertainty than the experimental
value, and within the SM prediction the main uncertainty is
related to the hadronic vacuum polarization. However, this uncertainty 
can be expected to be significantly reduced as a result of
ongoing measurements of $e^+e^-\to\,$hadrons by CMD2 and SND
in Novosibirsk and, using radiative return, by $B$ factories and
KLOE. Another important source of theoretical uncertainty is related
to the hadronic light-by-light scattering contribution. It is a
very challenging theoretical task to evaluate this contribution,
and current error estimates vary between $2.5\ldots4\tunit$. However,
further progress in the understanding of this contribution can be
expected, and one can hope that the full SM theory uncertainty can be
reduced to below $4\tunit$ in the foreseeable future
\cite{HMNT}. Finally, progress 
can be expected from the experimental determination of $\amu$. Since
the current measurement precision is statistics limited, a new
experiment using similar methods with only straightforward
improvements could reduce the uncertainty by a factor of $2.5$ or
more, and an according plan  has already been outlined
\cite{BNL6,LeeRoberts:2005uy}.

After more than 40 years of experimental and theoretical progress, the
observable $\amu$ has become a sensitive probe of physics at the
electroweak scale. Already today it is one of the most important
constraints of physics beyond the SM. In the near future, particle
physics will enter a new era where the detailed structure of physics at
the electroweak scale and above will be unravelled by
experiments at the LHC and possibly later the ILC. Not only new
particles like smuons or charginos might be discovered, but also the
other indirect constraints from the observables mentioned above will
become more stringent. The magnetic moment of the muon will provide a
crucial cross-check and complement of the forthcoming experiments.

\ack{
It is a pleasure to thank A.~Martin for the invitation to write this
review and S.~Heinemeyer, W.~Hollik,
J.~Illana, S.~Rigolin and G.~Weiglein for discussions and
collaborations. Furthermore, I am
grateful to A.~Arhrib, W.-F.~Chang, A.~Czarnecki, A.~Dedes, T.~Jones,  S.~Martin, K.~Matchev, T.~Moroi, A.~Nyffeler,
M.~J.~Ramsey-Musolf and 
T.~Teubner for discussions on various aspects of the muon magnetic
moment and to C.~Fischer, S.~Heinemeyer, T.~Plehn and
H.~St\"ockinger-Kim for critical comments on the manuscript.
}

\end{document}